\documentclass[twocolumn,aps]{revtex4}
\usepackage{amsthm, amssymb, graphicx, verbatim, tikz, hyperref}

\usepackage{framed, color}
\definecolor{shadecolor}{gray}{.8}

\usetikzlibrary{arrows,decorations.pathmorphing}

\def\u{{\mathcal U}}
\def\G{{\mathcal G}}
\def\g{{\mathfrak{g}}}
\def\H{{\mathcal H}}
\def\h{{\mathfrak{h}}}
\def\tr{{\rm tr}}
\def\C{{\mathbb{C}}}

\def\m{\mathcal{M}}
\def\n{\mathcal{N}}
\theoremstyle{plain} 
\newtheorem{theorem}{Theorem}
\newtheorem{lemma}[theorem]{Lemma}

\newtheorem{corollary}[theorem]{Corollary}
\newtheorem{protocol}[theorem]{Protocol}
\newtheorem{definition}[theorem]{Definition}
\newtheorem{example}[theorem]{Example}
\def\R{{\mathbb{R}}}

\begin{document}

\title{Three-dimensionality of space and the quantum bit:\\ an information-theoretic approach}
\author{Markus P.\ M\"uller}
\affiliation{Perimeter Institute for Theoretical Physics, 31 Caroline Street North, Waterloo, ON N2L 2Y5, Canada.}
\author{Llu\'is Masanes}
\affiliation{H.\ H.\ Wills Physics Laboratory, University of Bristol, Tyndall Avenue, Bristol, BS8 1TL, UK.}

\date{May 3, 2013}

\begin{abstract}
It is sometimes pointed out as a curiosity that the state space of quantum two-level systems, i.e.\ the qubit,
and actual physical space are both three-dimensional and Euclidean. In this paper, we suggest an information-theoretic analysis
of this relationship, by proving a particular mathematical result:
suppose that physics takes place in $d$ spatial dimensions, and that some events happen probabilistically
(not assuming quantum theory in any way). Furthermore, suppose there are systems that carry ``minimal amounts of
direction information'', interacting via some continuous reversible time evolution. We prove that this uniquely determines spatial dimension $d=3$
and quantum theory on two qubits (including entanglement and unitary time evolution), and that it allows observers to
infer local spatial geometry from probability measurements.
\end{abstract}

\maketitle

\section{Introduction}
The fact that the state space of quantum two-level systems -- the Bloch ball -- and physical space are both
three-dimensional and Euclidean has been regarded as a remarkable coincidence for many years, provoking interesting
ideas and lines of research. Building on this observation, von Weizs\"acker~\cite{Weizsaecker,Lyre} constructed his ``ur theory''
as an attempt to derive spacetime from quantum mechanics.
Similarly, Penrose's twistor theory~\cite{Penrose}
was built on the idea that the geometry of physical and quantum state space are fundamentally related,
which was elaborated further by Wootters~\cite{WoottersThesis} pointing out
the relation between quantum state distinguishability and geometry.

The idea that the quantum bit state space and physical space are somehow logically intertwined has become
a widespread paradigm, cf.~\cite{Brody}. But what is the exact relationship -- which one of the two determines the other?
Could a similarly nice relationship also exist in other dimensions, or is there something special about $d=3$?

The goal of this paper is to offer a particular information-theoretic analysis of these questions:
we show that a certain natural interplay between geometry and probability is only possible if space has three dimensions,
and if outcome probabilities of measurements are exactly as predicted by quantum theory.
This result suggests to explore the idea that neither quantum theory nor spacetime are separately fundamental,
but that both might have a common information-theoretic origin.

Our approach rests on some natural background assumptions.
Suppose that physics takes place in $d$ spatial dimensions (and one
time dimension), and some of the physical processes involve probabilities. That is, there exist experiments with random
outcomes -- we can imagine that physicists, or nature, prepare physical systems in certain \emph{states},
and later on, \emph{measurements} on the systems reveal outcomes with certain probabilities. We do \emph{not} assume
that those probabilities are necessarily described by quantum theory.

\begin{figure*}[!hbt]
\begin{center}
\includegraphics[angle=0, width=16cm]{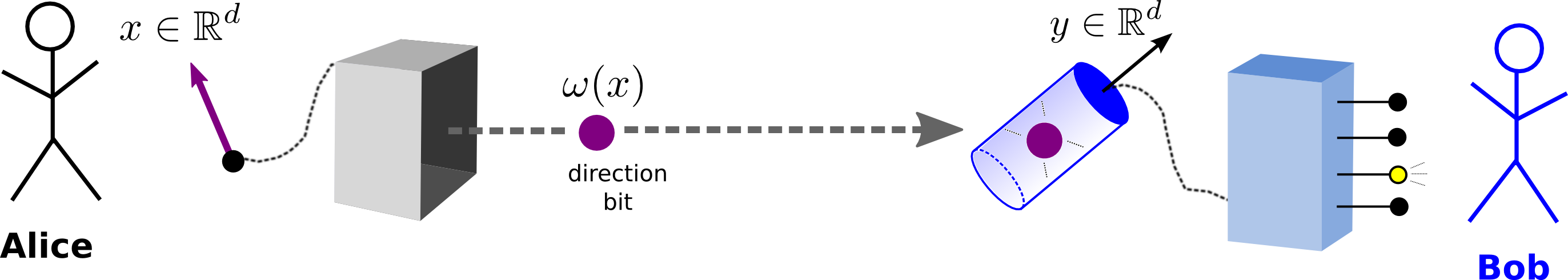}
\caption{The situation considered in this paper. Bob holds a macroscopic measurement device that he can rotate in $d$-dimensional space; its
orientation in space is thus described by a unit vector (``direction'') $y\in\R^d$. Alice's goal is to send a spatial direction $x\in\R^d$, $|x|=1$, to Bob,
by encoding it into a suitable state $\omega(x)$. After obtaining the state, Bob measures it with his device, obtaining one of several possible
measurement outcomes with some probability (indicated by a flashing lightbulb in the picture). After obtaining many identical copies of $\omega(x)$,
and measuring it in many different directions $y\in\R^d$, Bob is supposed to estimate Alice's direction $x$, such that his guess becomes
arbitrarily close to Alice's actual choice in the limit of infinitely many copies. We assume that Alice and Bob have agreed on an arbitrary protocol
beforehand, but they do not share a common coordinate system, such that Alice cannot simply send a classical description of $x$.
}
\label{fig_alicebob}
\end{center}
\end{figure*}

Then we consider the situation depicted in Fig.~\ref{fig_alicebob}: we have two agents, traditionally called Alice and Bob. Alice's goal is to send
some spatial direction -- that is, a unit vector $x\in\R^d$ -- to Bob, by encoding it into some state $\omega(x)$ and sending a physical system
that carries this state. We assume that they do not share a common coordinate system, such that Alice cannot simply send
a classical description of $x$. Bob holds a measurement
device that can be rotated in space, which he can apply to the state that he received, obtaining one of finitely many possible outcomes.
The outcome probabilities depend continuously on the device's spatial orientation.
Furthermore, suppose that the following four postulates are satisfied:
\begin{itemize}
\item[1.] Alice can encode any spatial direction $x\in\R^d$ into some state
such that Bob is able to retrieve $x$ in the limit of many copies.
\item[2.] It is impossible for Alice to encode any further information into the state without adding noise to the direction information.
\item[3.] There is a unique way to add up single-system observables on pairs of systems.
\item[4.] The state-carrying systems can interact pairwise by continuous reversible time evolution.
\end{itemize}
As we show below, these postulates can only be satisfied if $d=3$
and if these systems, and pairs of them, are described by quantum theory. That is, we derive the three-dimensionality
of space, two-qubit quantum state space and unitary time evolution as the unique solution.

These postulates declare some actions as \emph{possible} or \emph{impossible}: it is possible to let
two systems interact, but impossible to encode more than a spatial direction into one system (we define below
what this means in detail). This approach is in line with other recent developments like information causality~\cite{informationcausality},
where postulates of impossibility of certain information-theoretic tasks are exploited to derive properties of physical theories.
These approaches also have successful historical examples, like the postulate
of impossibility to build a perpetuum mobile of the second kind in thermodynamics.

The approach in this paper may be interpreted as the application of novel mathematical tools to the old question
of the relation between geometry and probability.
These tools have their origin in the recent wave of axiomatizations
of quantum theory~\cite{Fivel,Hardy2001,DakicBrukner,MasanesMueller,Chiribella,Hardy2011}, in particular in Hardy's seminal work~\cite{Hardy2001},
and are inspired by recent work on quantum reference frames~\cite{BartlettRS2006,GourSpekkens,Chiribella2004,Chiribella2006,Angelo}.

The first part of this paper consists of an introduction to one of these tools, which is the framework of convex state spaces,
generalizing quantum theory in a natural way. Then, the first two postulates will be defined in more detail, and will be used to
derive the state space of a single system. Finally, joint state spaces and the remaining two postulates will be
discussed in detail, yielding our main result. Throughout the paper, only the main ideas and proof sketches are given; the full proofs
are deferred to the appendix.

\section{Setting the stage: convex state spaces}
\label{SecConvexStateSpaces}
The framework of convex state spaces -- also called general probabilistic theories -- has proven useful in the context of quantum information
theory~\cite{Hardy2001,Wootters,BarnumWilce,Barrett,OppenheimWehner,ChiribellaPurification,MuellerUdudec},
but dates back much further~\cite{Mackey,Ludwig,AlfsenShultz,Araki,Holevo}. We now give a brief introduction; other useful introductory sources
include~\cite{Brassard,Brukner,Pfister,PostClassical}, in particular Chapters 1 and 2 in the paper by Mielnik~\cite{Mielnik1973}.

\begin{figure}[!hbt]
\begin{center}
\includegraphics[angle=0, width=7cm]{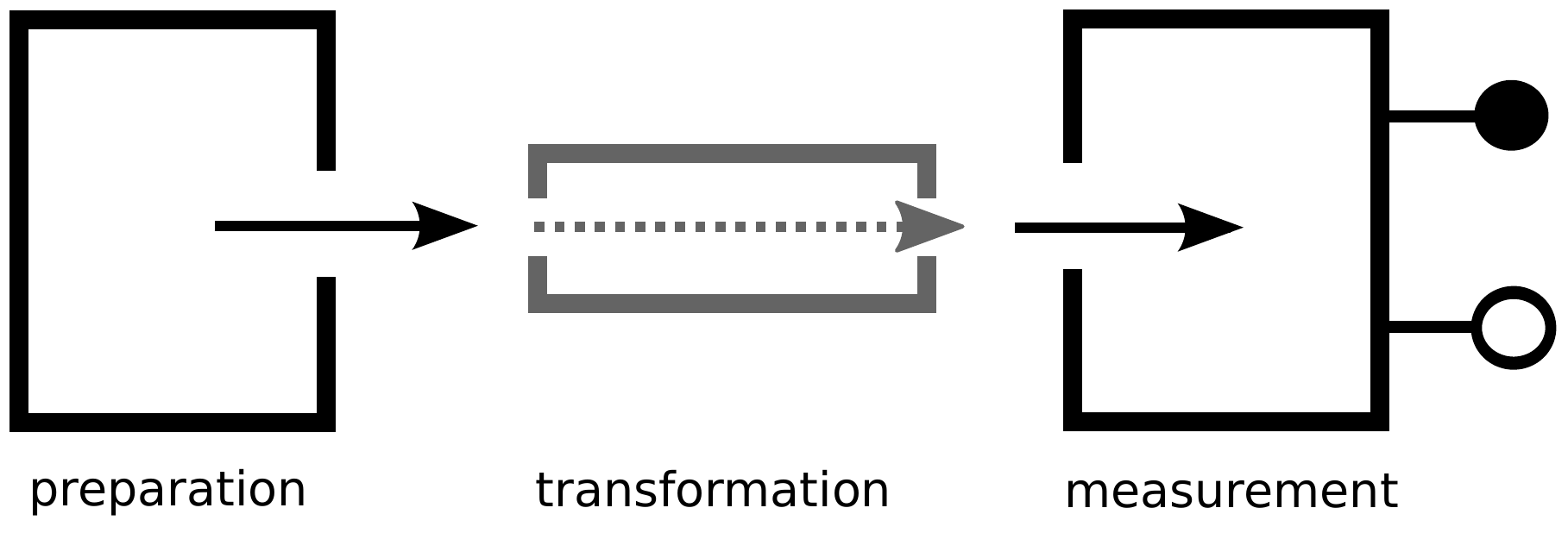}
\caption{Schematic of the physical setup underlying the framework of convex state spaces.
}
\label{fig_preptrafomeas}
\end{center}
\end{figure}

Consider the simple physical setup in Fig.~\ref{fig_preptrafomeas}. We have a \emph{preparation device}, which, whenever it is operated, generates an instance of
a physical system (for example, a particle). We assume that we can operate the preparation device as often as we want (say, by pressing a button on the device, or
by waiting until a periodic physical process has completed another cycle).
In the end, the system can be measured, by applying one of several possible \emph{measurement devices} with a finite number of outcomes.

The intuition is that the device prepares the system in a certain fixed state $\omega$; operating the preparation device several times
produces many independent copies of $\omega$. To define exactly what we mean by that, consider any fixed measurement device $\m$. If $\m$ is applied
to the preparation device's output, we assume that we get one of $k$ different measurement outcomes probabilistically, where $k\in\mathbb{N}$ is an arbitrary
natural number (in Fig.~\ref{fig_preptrafomeas}, we have $k=2$, represented by the two dots).
The probability to obtain the $i$-th outcome (where $1\leq i \leq k$) is denoted $\m^{(i)}(\omega)$, such that $\sum_i \m^{(i)}(\omega)=1$.

Suppose we have two devices, both preparing the same type of physical system, but in two different states $\varphi$ and $\omega$.
Then we can use them to build a new device that performs a random preparation: it prepares state
$\omega$ with probability $p$, and state $\varphi$ with probability $1-p$. The resulting state will be denoted $p\omega+(1-p)\varphi$.
This is a \emph{convex combination} of $\omega$ and $\varphi$. If we apply measurement $\m$ to that state, we will get the $i$-th
outcome with probability
\[
   \m^{(i)}\left(\strut p\omega+(1-p)\varphi\right)=p \m^{(i)}(\omega) + (1-p) \m^{(i)}(\varphi)
\]
by the basic laws of probability theory. In summary, we see that states $\omega$ of some physical system are elements of a real affine space which we denote by some capital letter, say $A$;
single measurement outcomes are described by affine maps $\m^{(i)}:A\to\R$ which yield values between $0$ and $1$ for every state.
Maps of this kind will be called \emph{effects}.
Full measurements are described by a collection of effects $\{\m^{(i)}\}_{i=1}^k$ that sum to unity if applied to any state.
The \emph{set of all possible states} of the corresponding physical
system will be denoted $\Omega_A$, the state space. It is a bounded subset of $A$. We have just seen that $\omega\in\Omega_A$ and $\varphi\in\Omega_A$ imply that
$p\omega+(1-p)\varphi\in\Omega_A$ for all $0\leq p\leq 1$; this means that $\Omega_A$ is \emph{convex}. We will only consider finite-dimensional state spaces in this paper.
Since outcome probabilities can only ever be determined to finite precision, we may (and will) assume that $\Omega_A$ is topologically closed.

\begin{figure}[!hbt]
\begin{center}
\includegraphics[angle=0, width=8.5cm]{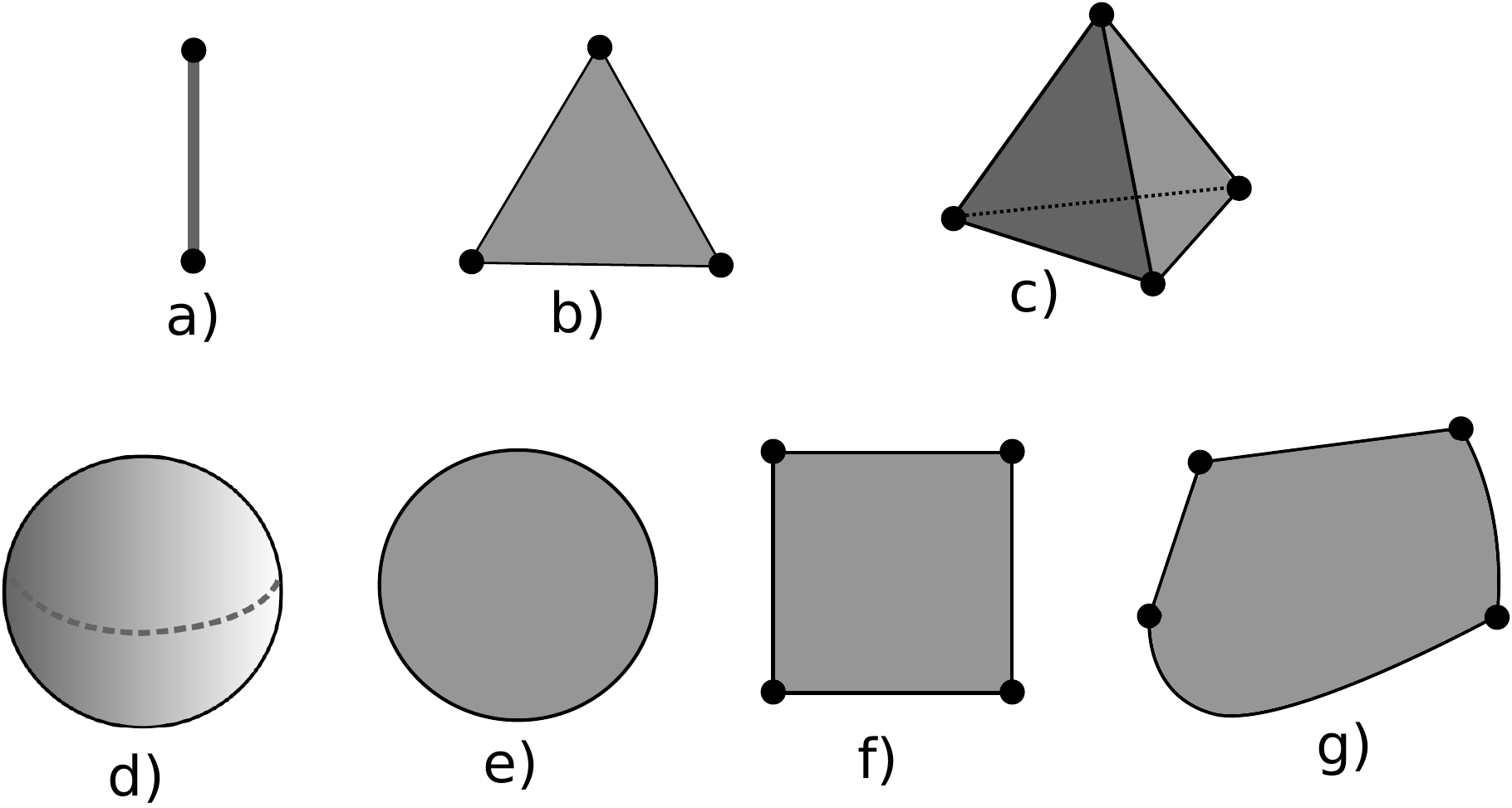}
\caption{Examples of convex state spaces: a) is a classical bit, b) and c) are classical $3$- and $4$-level systems, d)  is a quantum bit, e), f) and g)
are neither classical nor quantum, even though e) can naturally be embedded in a qubit.
Note that quantum $n$-level systems for $n\geq 3$ are \emph{not} balls~\cite{Weis}.}
\label{Fig2}
\end{center}
\end{figure}

As a simple example, consider a physical system which resembles a classical bit, or coin. We can perform a measurement by looking whether the coin
shows heads or tails; think of a two-outcome device which yields the first outcome if the coin shows heads, and the second otherwise.
The possible states are then characterized by the probability $p\in[0,1]$ of obtaining heads. The state space becomes a line segment,
with all states being probabilistic mixtures of two pure states that yield either heads or tails deterministically, see Fig.~\ref{Fig2}a).

The state spaces of a classical three- and four-level system are also shown in Fig.~\ref{Fig2}, b) and c): they are an equilateral triangle, resp.\ a tetrahedron.
In general, the state space of a classical $n$-level system is the set of all probability distributions $(p_1,\ldots, p_n)$, which is an $(n-1)$-dimensional
simplex.

Quantum state spaces look quite different. Quantum bits, the states of spin-$1/2$ particles, can be described by $2\times 2$ complex
density matrices $\rho$. These can always be written in the form $\rho=(\mathbf{1}+\vec r \cdot \vec \sigma)/2$, where $\vec r$ is an ordinary real vector in $\R^3$ with $|\vec r|\leq 1$,
and $\sigma=(\sigma_x,\sigma_y,\sigma_z)$ denotes the Pauli matrices~\cite{NielsenChuang}. We can consider $\vec r=(r_x,r_y,r_z)$ as the state of the qubit. Thus, the state space 
is a three-dimensional unit ball as shown in Fig.~\ref{Fig2}d). A spin measurement in the $z$-direction may be described by the two effects
$\m^{(1)}(\vec r)=(1+r_z)/2$ and $\m^{(2)}(\vec r)=(1-r_z)/2$, for example, where the two outcomes correspond to ``spin up'' and ``spin down'', respectively.

However, the state space of a quantum $n$-level system is only a ball for $n=2$; for $n\geq 3$, quantum state spaces are \emph{not} balls,
but intricate compact convex sets of dimension $n^2-1$~\cite{Zyczkowski,Weis}.

Given any state space $\Omega_A$, all effects, i.e.\ affine maps $\m:A\to\R$
with $\m(\omega)\in[0,1]$ describe outcomes of conceivable measurement devices. We can work out the set of these maps from a description of $\Omega_A$.
In general, some of these measurements might be physically impossible to implement; in order to describe a physical system, we have to specify which ones
are possible and which ones are not.

From the effects, we can construct expectation values of observables, simply called \emph{observables} in the following. These
are arbitrary affine maps $h:A\to\R$; in quantum theory, they are maps of the form $\rho\mapsto \tr(\rho H)$, where $H=H^\dagger$ is any
self-adjoint matrix. One way to measure an observable (on many copies of a state) is to write it as a linear combination of effects,
$h=\sum_i h_i\m_i$,  $h_i\in\R$, and to measure the effects $\m_i$ on different copies (in general, they may not be jointly
measurable on a single copy and thus be outcomes of different measurement devices).

Similarly, we can describe \emph{reversible transformations} of a physical system: these are physical processes that take a state to another state,
and may be inverted by another physical process (in quantum theory, these are the \emph{unitaries}, mapping $\rho$ to $U\rho U^\dagger$). Since they must respect
probabilistic mixtures, they must also be affine maps. Due to reversibility, they map the state space $\Omega_A$ onto itself -- they are symmetries of the
state space. The set of reversible transformations on $A$ is a closed subgroup $\G_A$ of all symmetries of $\Omega_A$.

\section{Single systems: Postulates 1 and 2}
We consider a particular situation where measurements take place in $d$-dimensional space, with one time dimension.
For simplicity, we assume that there is a fixed flat background space, such that there is a unique way to
transport vectors from one laboratory $A$ to another distant laboratory $B$ (however, we think that our results may apply to more
general situations). We will also assume that all physical operations considered in the following, such as measurements, are
performed locally in a way such that all parties (particles, measurement devices etc.) are relative to each other at rest~\footnote{In
the usual vocabulary of special relativity, if we imagine that direction bits are internal degrees of freedom of particles,
this assumption implies that these particles must be massive.
}. Thus, we do not have to consider conceivable relativistic effects.

In general, there may be many different kinds of physical systems described by convex state spaces. We now assume that there exists a particular type
of physical system which, in a sense to be made precise, behaves like a ``unit of direction information''. We will call these systems
``direction bits'' (later on, we show that they are effectively $2$-level systems,
therefore ``bits'', cf.\ Lemma~\ref{LemBits} in the appendix).
We will not specify by what type of physical object they are carried -- a direction bit could,
for example, correspond to the internal degrees of freedom of a particle, or it could be something completely different.
We will only assume that a direction bit may come in different states (matching the framework described above), with
a state space denoted $\Omega_d$.
\begin{figure}[!hbt]
\begin{center}
\includegraphics[angle=0, width=7cm]{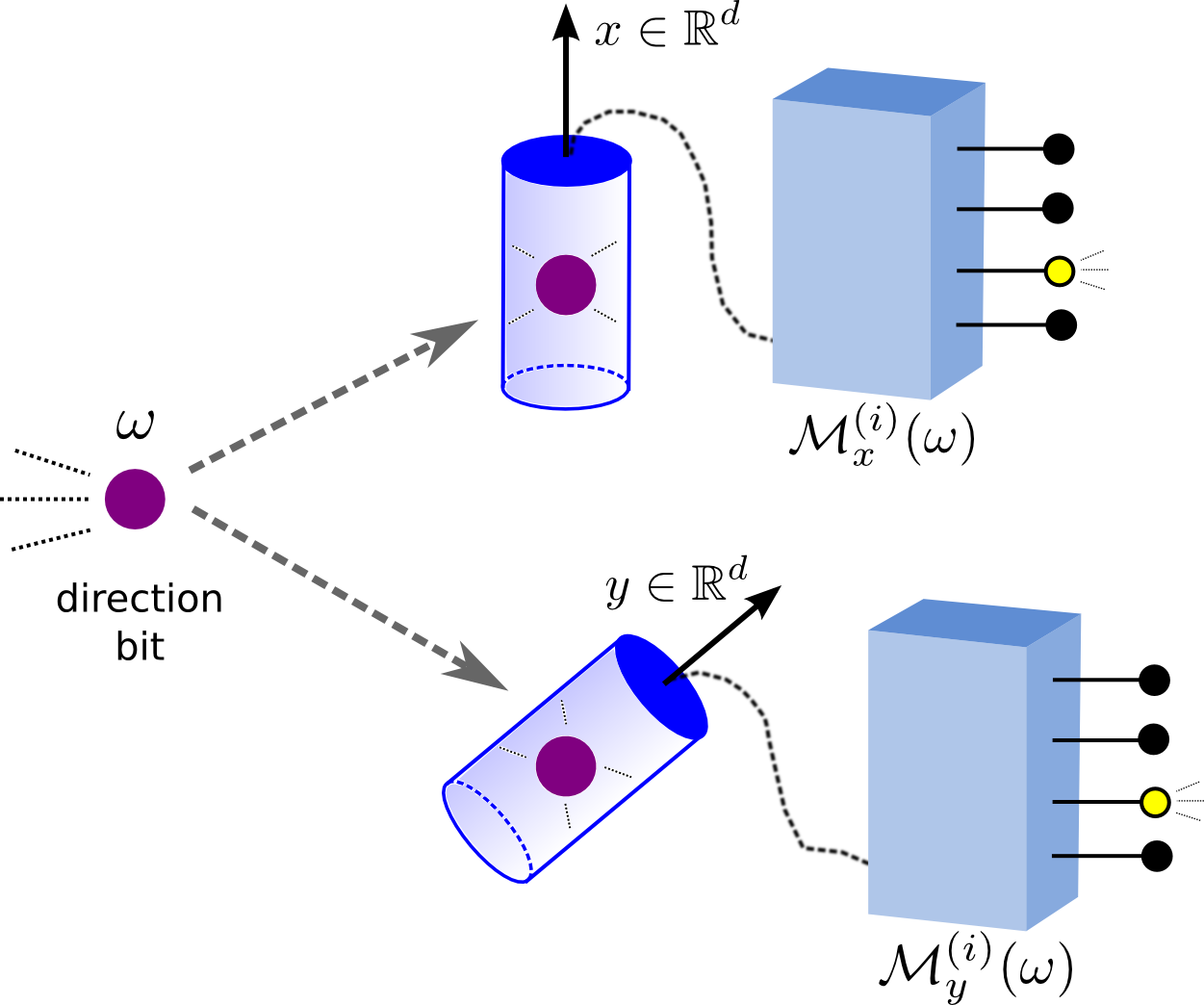}
\caption{We assume that direction bits can be measured by some macroscopic measurement device, which yields
one of several outcomes $i\in\{1,\ldots,k\}$ probabilistically. Due to symmetry,
its modus operandi depends only on a vector $y\in\mathbb{R}^d$, $|y|=1$ specifying its ``direction''
in the local laboratory frame.
The probability $\mathcal{M}_y^{(i)}(\omega)$
to obtain the $i$-th outcome depends only on the direction bit state $\omega$ and continuously on the direction $y$.
The device can be rotated in space according
to any rotation $R\in SO(d)$. In the rotated reference frame of the device, this corresponds to a reversible transformation on
the direction bit.}
\label{fig_directionbit}
\end{center}
\end{figure}

We assume that direction bits can be measured by a certain type of measurement device with a finite number of outcomes.
As shown in Fig.~\ref{fig_directionbit}, we imagine that the device is implemented as a macroscopic, massive object which can be rotated
arbitrarily, i.e.\ can be subjected to any $SO(d)$ rotation. Due to some symmetry of the device, its orientation in space (locally in the lab)
may be described by a unit vector $y\in\R^d$, $|y|=1$, choosing some arbitrary but fixed coordinate system in the local
laboratory. Instead of naively thinking of the whole device as ``pointing in direction $y$'', we may also think that one of the device's components
is a vectorial physical quantity which determines the type of measurement that is performed. A standard example in three dimensions is given
by a Stern-Gerlach device, where $y$ is the direction of inhomogeneity of a magnetic field.

The case $d=1$ is special, because $SO(1)=\{\mathbf{1}\}$ is trivial, and thus no one-dimensional rotation can map the unit vector
$+1\in\R^1$ to the unit vector $-1\in\R^1$. In order to allow Bob to collimate his device in all directions also in $d=1$,
we will thus silently replace $SO(1)$ by $O(1)=\{\mathbf{1},-\mathbf{1}\}$ in all of the following.

Since the measurement which is performed by the device may depend on its direction $y$ in space,
it is denoted $\m_y$. In the following, by a ``direction'', we shall always mean a unit vector in $\R^d$.
For obvious physical reasons, we assume that the outcome probabilities $\m_y^{(i)}(\omega)$ are continuous in the direction $y$.

Physically, we assume that we can perform a rotation $R\in SO(d)$ to the measurement device without touching the
direction bit; this transforms $\m_y$ to $\m_{Ry}$, but leaves the bit's state $\omega$ invariant. The fact that the outcome probabilities
are altered, from $\m_y^{(i)}(\omega)$ to $\m_{Ry}^{(i)}(\omega)$, should be understood as a result of the change in the \emph{relative} orientation
of the bit and the device. Thus, even though direction bits are considered as informational ``black boxes'' with arbitrary physical realization,
we are forced to adopt the interpretation that direction bits carry actual physical geometrical orientation.

This enforces a certain duality that is familiar from quantum mechanics. Suppose that, after rotating the measurement device by $R$,
we do not perform the measurement, but instead rotate the \emph{joint system of direction bit and measurement device back} by $R^{-1}$.
If it is physically unclear how to do this in practice, we can just imagine performing a passive coordinate transformation.

Since this transformation does not change the relative direction of the system and measurement apparatus,
it does not alter the outcome probabilities. However, by changing to the new coordinate system, $\m_{Ry}$ has been transformed back to $\m_y$,
hence the direction bit state must have changed from $\omega$ to some other state $\omega'$ such that $\m_y^{(i)}(\omega')=\m_{Ry}^{(i)}(\omega)$.
The state transformation $\omega\mapsto\omega'$ can be physically undone (by rotating the joint system again by $R$), hence
it must be an element of the group of \emph{reversible transformations} on $\Omega_d$. We call it $G_{R^{-1}}$, such that
we can switch from the ``Heisenberg'' to the ``Schr\"odinger'' picture via
\[
   \m^{(i)}_{Ry}(\omega)=\m^{(i)}_y(G_{R^{-1}} \omega).
\]
Clearly $G_R\circ G_S=G_{RS}$; in other words, the map $R\mapsto G_R$ is a group representation of $SO(d)$ on the direction bit state space.

Now suppose we have a situation where two agents (Alice and Bob) reside in distant laboratories as depicted in Fig.~\ref{fig_alicebob}.
Imagine that Alice holds an actual physical vector $x\in\R^d$ \textit{(all vectors and rotations will be denoted with respect to Alice's local coordinate system in the following)},
and she would like to send this geometric information to Bob. Since Alice
and Bob have never met, they have never agreed on a common coordinate system. Thus, it is useless for Bob if Alice sends him a
classical description of $x$, because he does not know what coordinate system the description is referring to.

However, if Bob holds a measurement device as in Fig.~\ref{fig_directionbit}, Alice can send him a direction bit in some state
$\omega$. As usual in information theory (taking into account the statistical definition of states), we analyze the properties of a single state $\omega$ by considering
many identical copies of that state. So suppose Alice sends many independent copies of $\omega$ to Bob. On every copy, Bob
can measure in a different direction, and he may find that some outcome probabilities are varying over the different directions $y\in\R^d$, $|y|=1$.
This breaks rotational symmetry, and so may be used by Alice to send physical direction information to Bob.

However, Alice cannot send information about the \emph{length} of the vector $x$ to Bob, if we assume that Bob can only rotate
his device (as in Fig.~\ref{fig_directionbit}) and not more.
Thus, restricting to the situation in Fig.~\ref{fig_alicebob}, we state that \emph{Alice's task is to send a direction vector
$x\in\R^d$, $|x|=1$, to Bob, by encoding it into some state}.

\begin{shaded}
\textbf{Postulate 1 (Encoding).} There is a protocol (as in Fig.~\ref{fig_alicebob}) which allows Alice to encode all spatial
directions $x\in\R^d$, $|x|=1$, into states $\omega(x)\in\Omega_d$, such that Bob is able to retrieve $x$
in the limit of many copies.
\end{shaded}

Denote the coordinates of some vector $x\in\R^d$ in Bob's local coordinate system by $x_B$. Then
we stipulate that after obtaining $n$ copies of $\omega(x)$, Bob makes a guess $x_B^{(n)}$ of $x_B$ (based on his
previous measurement outcomes) such that $x_B^{(n)}\to x_B$ for $n\to\infty$ with probability one.
For obvious physical reasons, we assume that Alice's encoding $x\mapsto \omega(x)$
is continuous~\footnote{We are only assuming that
there exists at least one choice of encoding as a continuous map $x\mapsto\omega(x)$ which works.
}. Moreover, Bob measures each direction bit individually and only once (we may imagine that direction bits get destroyed upon
measurement~\footnote{This is by no means a crucial assumption -- in general, we would have to model the measurement's effect on the
state by an outcome-dependent state transformation, i.e.\ an \emph{instrument}~\cite{DaviesLewis}. This is analogous to the construction in
quantum information theory.}).

In principle, direction bits might carry further additional information that can be read out in measurements. As a naive example,
the physical system that Alice transmits could be a simple wristwatch, with the watch hand pointing in the direction that Alice is
intending to send. However, a wristwatch is hardly ``economical'' for this task: it carries a large amount of
additional information, like the details of its head shape etc. Our second postulate says that direction bits should be ``minimal''
in their ability to carry directional information: any attempt to encode further information can only succeed at the
expense of loosing some of the directional information.

\begin{shaded}
\textbf{Postulate 2 (Minimality).}
No protocol allows Alice to encode any further information into the state without adding noise to the directional information.
\end{shaded}

To spell out the mathematical details, we need to define what it means that one state $\varphi$ is noisier in its directional information
than another state $\omega$. First, by directional information of $\varphi$, we mean the probability functions $\m_z^{(i)}(\varphi)$ as seen by
direction bit measurement device.
If we have two states $\varphi,\omega$ with directional probabilities related by a rotation, i.e.\ $\m_z^{(i)}(\varphi)=\m_{Rz}^{(i)}(\omega)$
for some rotation $R\in SO(d)$ and all $i$,
we argue that both states are \emph{equally noisy} in this respect -- they both contain the same ``amount of asymmetry'',
just pointing in different directions.

We could additionally say that $\varphi$ and $\omega$ are equally noisy if $H(\varphi)=H(\omega)$ for some entropy measure $H$;
however, there is no unique entropy definition for arbitrary state spaces~\cite{ShortEntropy,BarnumEntropy,KimuraEntropy}, and entropy measures
acquire meaning only relative to certain operationally defined tasks which is a complication we want to avoid. Therefore, we define
$\varphi$ to be \emph{at least as noisy in its directional information as $\omega$} if its directional probabilities are statistical mixtures
of those of $\omega$ and other states that are equally noisy as $\omega$; that is, if there are statistical weights $\lambda_j>0$, $\sum_j\lambda_j=1$,
and rotations $R_j\in SO(d)$ such that for all outcomes $i$,
\begin{equation}
   \m_z^{(i)}(\varphi)=\sum_j \lambda_j \m_{R_j z}^{(i)}(\omega)\qquad\mbox{for all }z.
   \label{eqNoisiness}
\end{equation}
Clearly, $\varphi$ is \emph{noisier than $\omega$ in its directional information} if it is at least as noisy, and at the same time not equally noisy as $\omega$.
In Definition~\ref{DefNoisiness} and following in Appendix~\ref{AppDirBit}, we show that this notion is a natural generalization of the \emph{majorization} relation~\cite{NielsenChuang}
from classical probability theory and quantum theory. It also has a natural interpretation in terms of \emph{resource theories}~\cite{HorodeckiOppenheim,GourSpekkens}: for any
given $\omega$, the probability functions $z\mapsto \m_z^{(i)}(\omega)$ -- or rather their directional asymmetry -- constitute a resource for Bob.
One resource is less useful -- that is, more noisy -- than the other if it can be obtained from the other by ``free'' operations; in this case, by tossing a coin and performing a
random rotation.

Suppose we had a protocol that satisfied Postulate 1, and two states $\varphi\neq \omega$ that would work as a possible encoding
of some direction $x$, in the sense that Bob would in both cases decode direction $x$ in the limit of obtaining infinitely many copies.
Then, by choosing to send either $\varphi$ or $\omega$, Alice could send an additional
classical bit to Bob. Postulate 2 says that this is only possible at the expense of adding noise -- that is, one of the two states must be noisier in its
directional information than the other.

Our goal is to determine the shape of the convex state space of a direction bit, using only Postulates 1 and 2 and the physical background
assumptions (Postulates 3 and 4 will be considered later). To this end, suppose Alice encodes some direction $x$ according to some protocol
into a state $\omega(x)$ and sends many copies of it to Bob. If the protocol satisfies Postulates 1 and 2, Bob will be able to decode $x$.

Now suppose that Bob secretly performed a rotation $R\in SO(d)$ on his laboratory before the protocol started.
Since the protocol must work regardless of the relative orientation of Alice and Bob, Bob will still succeed to obtain an accurate estimate of $x$ as before.

As we have seen, applying $R$ to a measurement device can be replaced by applying $G_{R^{-1}}$ to the direction bit state.
Therefore, the following implementation will also allow Bob to guess $x$:
\begin{itemize}
\item Apply $G_{R^{-1}}$ to every incoming direction bit; measure as in the protocol above.
\item After obtaining $n$ copies, determine the guess $x^{(n)}$ given by the protocol above.
\item To compensate for the missing lab rotation, output the guess $R x^{(n)}$.
\end{itemize}
Suppose that $R$ is in the stabilizer subgroup of $x$, i.e.\ $Rx=x$. Then the lines above prove that the original protocol also
works if Alice sends the state $G_{R^{-1}}(\omega(x))$ to Bob instead of $\omega(x)$. But these states are equally noisy in
their directional information, hence Postulate 2 implies that they are equal. In other words, we have shown the following:

\emph{For any encoding $x\mapsto \omega(x)$, the state $\omega(x)$ is invariant with respect to all rotations
that leave $x$ invariant.}

For what follows, fix $x:=(1,0,\ldots,0)^T$ and an arbitrary protocol that satisfies Postulates 1 and 2, yielding a state $\omega(x)$
that encodes direction $x$. If $Rx=x$, then
\[
   \m_y^{(i)}(\omega(x))=\m_y^{(i)}(G_{R^{-1}}\omega(x)) = \m_{Ry}^{(i)}(\omega(x)) \mbox{ for all }y.
\]
Thus, for every $i$, the probability $\m_y^{(i)}(\omega(x))$ is a function of the first component of $y$.
As we show in Lemma~\ref{LemNotTooSymmetric} in Appendix~\ref{AppDirBit}, this has the following consequence: there is at least one measurement
outcome (call it $i_0$) and one direction $y$ such that $\m_y^{(i_0)}(\omega(x))>\m_{-y}^{(i_0)}(\omega(x))$ -- otherwise,
the state $\omega(x)$ would be ``too symmetric'' to allow the transmission of direction information, and Postulate 1 would be violated.

Fix this $i_0$. For every direction $y$ that satisfies the inequality above (there might be more than one), we define a new state
$\omega'(y)$ by averaging over the stabilizer group~\cite{Simon} of $y$:
\begin{equation}
   \omega'(y):=\int_{R\in SO(d):Ry=y} G_R \omega(x)\, dR.
   \label{eqomegaprimey}
\end{equation}
From now on, we are only interested in the outcome $i_0$, and use the abbreviation $\m_z:=\m_z^{(i_0)}$.
Furthermore, for all states $\omega$ and directions $z$, we set
\[
   L_z(\omega):=\m_z(\omega)-\m_{-z}(\omega).
\]
In particular, we obtain $L_y(\omega'(y))=L_y(\omega(x))>0$. As we prove in Lemma~\ref{LemUniqueMaximum} in Appendix~\ref{AppDirBit},
if $y$ is chosen in a clever way, then this is the maximal possible value:
\emph{there is a particular choice of $y$ such that the map $z\mapsto L_z(\omega'(y))$ attains its unique global maximum at $z=y$.}

This property allows us to construct a new protocol for Alice and Bob to transmit direction information. Fix this particular choice
of $y$, and $\omega'(y)$. For all other directions $z\neq y$,
define $\omega'(z)$ by rotating $\omega'(y)$ appropriately, i.e.\ $\omega'(z):=G_S \omega'(y)$, where $S\in SO(d)$
is any rotation with $Sy=z$. The new protocol works as follows:
\begin{itemize}
\item Alice encodes some direction $x$ into the state $\omega'(x)$ and sends many copies of it to Bob.
\item By measuring the received copies,
Bob determines a good estimate of the function $f(z):= \m_z(\omega'(x))$.
Bob's guess is the vector $z$ for which $L_z(\omega'(x))\equiv f(z)-f(-z)$ is maximal.
\end{itemize}

Remarkably, from an \emph{arbitrary} protocol to transmit direction information, we have constructed a \emph{standard protocol}.
This involves a difference $L_z(\omega)=\m_z(\omega)-\m_{-z}(\omega)$, which has striking similarity to the spin-$1/2$
angular momentum expectation value in quantum mechanics~\footnote{A natural first attempt would be to construct a protocol
which simply looks for the global maximum of $\m_z(\omega)$ instead of $L_z(\omega)$. However, it is not clear how the existence
of any state $\omega$ with a unique maximum of $z\mapsto\m_z(\omega)$ can be ensured in this case, at this stage of the proof.}:
this expression is the expectation value of a random variable which assigns ``$+1$'' to
direction $z$, and ``$-1$'' to direction $(-z)$.

Since the new protocol allows Alice to transmit arbitrary spatial directions to Bob, it must satisfy Postulate 2.
Thus, if we have two states $\omega$ and $\varphi$ that satisfy $L_y(\omega)>L_z(\omega)$ \emph{and} $L_y(\varphi)>L_z(\varphi)$ for all $z\neq y$,
they must either be equal, or one must be strictly noisier than the other, as in eq.~(\ref{eqNoisiness}) (exchanging the names $\omega\leftrightarrow\varphi$ if
it is the other way round). As we prove in Lemma~\ref{LemNoisySOd} in Appendix~\ref{AppDirBit}, this equation implies for the states themselves that
$\varphi=\sum_j \lambda_j G_{R_j^{-1}}\omega$, as states turn out to be uniquely determined by their directional information.
Since both $\omega$ and $\varphi$ can be used as codewords for direction $y$ in our standard protocol, our intermediate result one page above
implies that $G_R\omega=\omega$ and $G_R\varphi=\varphi$ for all $R\in SO(d)$ with $Ry=y$.

Suppose that furthermore the maxima agree, i.e.\ $L_y(\omega)=L_y(\varphi)=:M$ holds. Then
\[
   M=L_y(\varphi)=\sum_j \lambda_j L_y(G_{R_j^{-1}}\omega)=\sum_j \lambda_j \underbrace{L_{R_j y}\omega}_{\leq M}.
\]
This is only possible if $L_{R_j y}\omega=M$ for all $j$, and thus $R_j y=y$ by construction. But then $G_{R_j}\omega=\omega$ for all $j$ as just mentioned,
and applying $G_{R_j^{-1}}$ to both sides and substituting into the relation between $\varphi$ and $\omega$ proves that $\varphi=\omega$. Thus, \emph{if
two states encode the same direction in our standard protocol, with the same maximal value of $L$, they must agree.}
This property will now be used to determine the state space of a direction bit.

From now on, $x$ and $y$ will denote arbitrary directions, disregarding the special choices above.
Call any state $\omega$ with the property that $L_x(\omega)>L_y(\omega)$ for all $y\neq x$ a codeword for $x$.
The codewords $\omega'(x)$ constructed above are in general not the ``optimal'' ones for the standard protocol -- we might be able to find
``better'' ones, $\omega''(x)$, with $L_x(\omega''(x))>L_x(\omega'(x))$.
The previous inequality can be interpreted as saying that
the $\omega''(x)$ let Bob determine $x$ more quickly than the codewords $\omega'(x)$ in the standard protocol above, because the difference
in probabilities is larger and statistically visible after transmission of fewer direction bits.

As we show in Lemma~\ref{LemHomeo} in Appendix~\ref{AppDirBit}, there is an ``optimal'' set of codewords which we
call $\omega_x$, with the property that $L_x(\omega_x)\geq L_x(\omega'(x))$ for all other codewords $\omega'(x)$.
The codewords for different directions are related by rotations: if $y=Rx$ for $R\in SO(d)$, then $\omega_y=G_R \omega_x$.
Furthermore, there is a constant $0<a\leq 1$ such that $L_y(\omega_y)=a$ for all $y$; we call $a$ the direction bit's \emph{visibility parameter}.

Given $\omega_x$, we can define a ``uniform noise'' state which we call the \emph{maximally mixed state} $\mu$:
\[
   \mu:=\int_{R\in SO(d)} G_R\omega_x\, dR = \int_{R\in SO(d)} \omega_{Rx}\, dR.
\]
Since all $\omega_y$ are related by rotations, $\mu$ is independent of the initial choice of $x$. As this is an integral
over the invariant Haar measure, there is constant $c\in(0,1)$ such that $\m_y(\mu)=c$ for all $y$. We call $c$ the
direction bit's \emph{noise parameter}.

Now suppose $\omega$ is any state which is a codeword for some direction $x$. Then $\lambda:=L_x(\omega)/a$ is in the
interval $(0,1]$. Thus, $\omega':=\lambda\omega_x+(1-\lambda)\mu$ is a valid state, and it is easy to see that it is also a codeword for $x$.
But $L_x(\omega')=L_x(\omega)$, and so the intermediate result above implies that $\omega=\omega'$. Since every state can be approximated arbitrarily
well by some codeword, we have proven that \emph{every state $\omega$ can be written in the form $\omega=\lambda\omega_x+(1-\lambda)\mu$
for some direction $x$, where $0\leq\lambda\leq 1$.}

We are free to reparametrize the state space $\Omega_d$ via some affine map $\phi:\R^D\to\R^D$, where $D$ is the dimension
of $\Omega_d$: replacing states via $\omega\mapsto \hat\omega:=\phi(\omega)$, effects via $\m\mapsto\hat\m:=\m\circ\phi^{-1}$ and
transformations via $G\mapsto  \hat G:=\phi\circ G \circ\phi^{-1}$ does not change any probabilities or physical predictions.
Basic group representation theory~\cite{Simon} tells us that we can choose $\phi$ such that the transformed group $\hat\G$
acts linearly and contains only orthogonal matrices, and the transformed states $\hat\omega_x$ (for different $x$) -- being connected by reversible
transformations -- have all the same Euclidean norm $1$. Moreover, the maximally mixed state $\hat\mu$, being invariant with respect to
all transformations, becomes the zero vector.

\begin{figure}[!hbt]
\begin{center}
\includegraphics[angle=0, width=4cm]{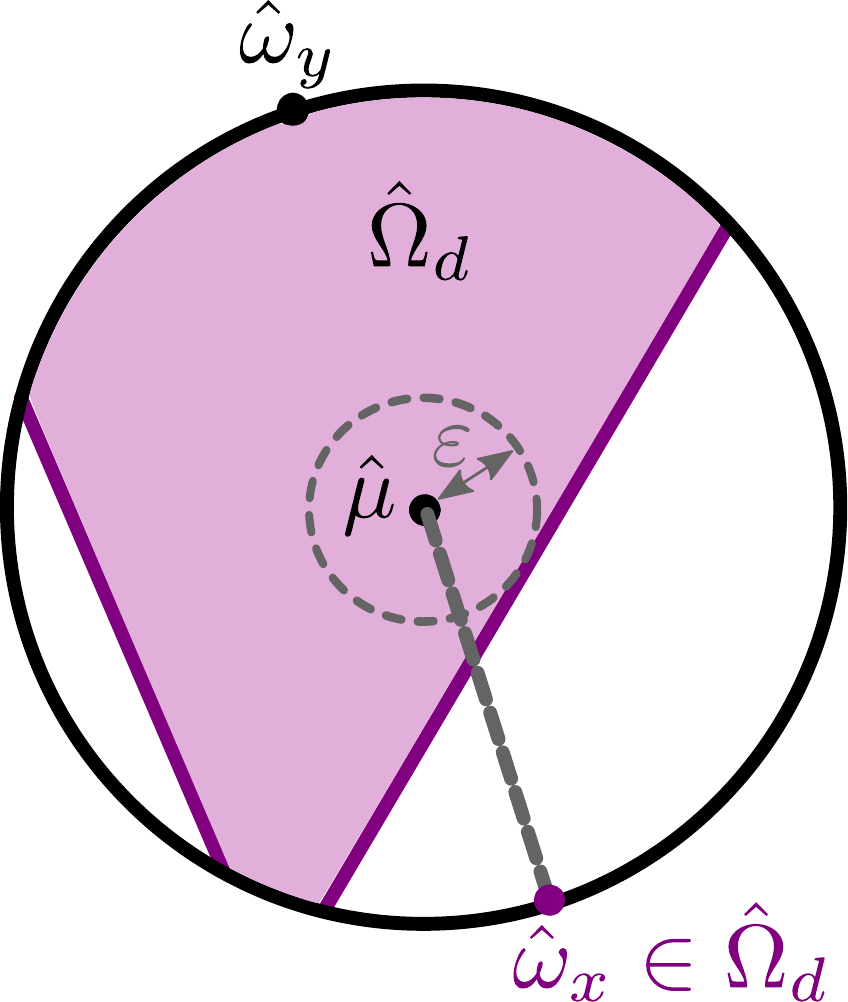}
\caption{After a reparametrization, we obtain that the direction bit state space $\hat\Omega_d$ is a compact convex subset
of a unit ball. Since the maximally mixed state $\hat\mu$ is in the interior, there is some $\varepsilon>0$ such
that the state space contains a full $\varepsilon$-ball around the origin $\hat\mu=0$. But we have proven that all states $\hat\omega$ are convex
combinations of $\hat\mu$ and some state $\hat\omega_x$ with $|\hat\omega_x|=1$, thus $\hat\omega_x$ must thus lie on the line starting at $\hat\mu=0$
and crossing $\hat\omega$. Consequently, all points on the sphere must be contained
in the state space -- we obtain the full unit ball. By dimension counting, it is $d$-dimensional.}
\label{fig_fullball}
\end{center}
\end{figure}

Since all states $\hat\omega$ are convex mixtures of some $\hat\omega_x$ and $\hat\mu$, we obtain the situation depicted in Fig.~\ref{fig_fullball}:
the transformed state space $\hat\Omega_d$ is compact convex subset of the $D$-dimensional unit ball, with all $\hat\omega_x$ on the
surface and $\hat\mu=0$ in the center.

It is easy to see that the maximally mixed state $\hat\mu$ is in the interior of $\hat\Omega_d$, since it is a mixture of all pure states. Hence
there is some ball of radius $\varepsilon>0$ around $\hat\mu=0$ which is fully contained in $\hat\Omega_d$.
Thus, if $v\in\R^D$ is any unit vector, then $\varepsilon  v/2$ must be a valid state in $\hat\Omega_d$. As we have proven above, there is some $0\leq\lambda\leq 1$
and some direction $x\in\R^d$
such that $\varepsilon v/2 =\lambda \hat\omega_x+(1-\lambda)\hat\mu$. This is only possible if $\varepsilon=2\lambda$ and $v=\hat\omega_x$ -- in
other words, $v\in\hat\Omega_d$. This proves that $\hat\Omega_d$ is the full $D$-dimensional unit ball.
By construction, the map $x\mapsto\hat\omega_x$ is a homeomorphism from the unit sphere in $\R^d$ to the unit sphere in $\R^D$. This proves that $D=d$.

\begin{framed}
\textbf{Theorem 1.} The state space of a direction bit is a $d$-dimensional unit ball.
\end{framed}

This shows that a direction bit cannot be described by a classical probability distribution: it must carry a non-classical state space, exhibiting uncertainty relations
among $d$ independent, mutually complementary measurements. Probabilistic systems of this type, i.e.\ ball state spaces, have been studied
before~\cite{hyperbits,limited,Mielnik1968}. In quantum physics as we know it, there is only one
kind of system with a ball state space: it is the qubit, a quantum $2$-level state space. It is three-dimensional,
which coincides with the spatial dimension, confirming the result we just proved. 
By classifying the affine maps from the ball to $[0,1]$, it is easy to check that we must have
\begin{equation}
   \m_x(\omega)=c+(a/2)\langle\hat\omega_x,\hat\omega\rangle.
   \label{eqSpin}
\end{equation}
In the familiar three-dimensional case, if $c=1/2$ and $a=1$, this describes a quantum spin measurement in direction $x$; if $c\neq 1/2$ or $a<1$, it is a noisy spin measurement.

To see why ball state spaces satisfy Postulate 2, note first that two states $\varphi,\omega$, with corresponding ``Bloch vectors'' $\hat\varphi,\hat\omega$
in the $d$-dimensional Euclidean unit ball, are equally noisy in their directional information if and only if $|\hat\varphi|=|\hat\omega|$; similarly, $\varphi$ is noisier than
$\omega$ if and only if $|\hat\varphi|<|\hat\omega|$ (in the case $d=3$, where the state space is a qubit, this condition becomes ${\rm tr}(\varphi^2)<{\rm tr}(\omega^2)$). For any
protocol, and any
spherical shell of fixed Bloch vector norm, there is a one-to-one correspondence of states in that shell and spatial directions that these states encode.
Thus, if two different states encode the
same direction, they must have different norms, and so one is noisier than the other. We say more about the different possible protocols in Lemma~\ref{LemFormProtocol}
in the appendix.

\section{Frame bits instead of direction bits?}
\label{SecFrameBits}
\begin{figure*}[!hbt]
\begin{center}
\includegraphics[angle=0, width=16cm]{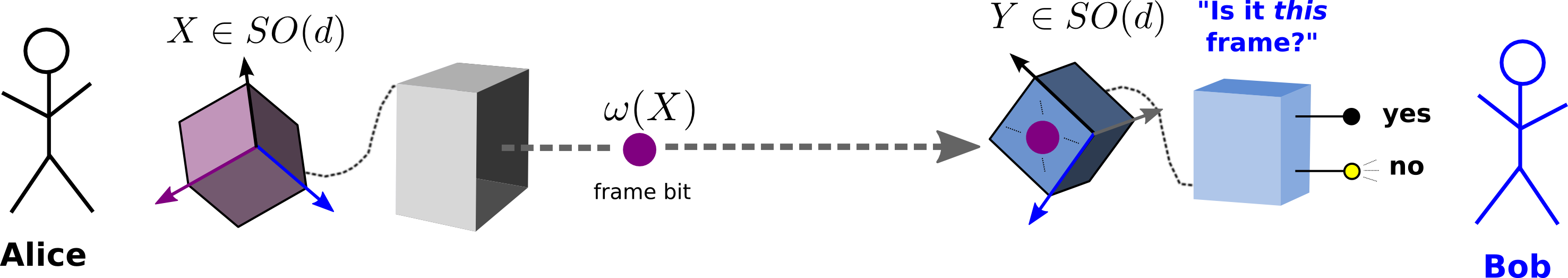}
\caption{The ``frame bit'' setup for the special case discussed in Section~\ref{SecFrameBits}.
Instead of a spatial direction, Alice's goal is now to send an orthonormal frame $X\in SO(d)$
to Bob, by transmitting some state $\omega(X)$ of some arbitrary convex state space.
Bob holds a macroscopic \emph{binary} measurement device that can be rotated arbitrarily in space. In contrast to the situation in Fig.~\ref{fig_alicebob}
for direction bits, Bob's measurement device does not possess any rotational symmetry, such that its working is specified by the orthonormal basis that defines
its spatial orientation, $Y\in SO(d)$.
Alice and Bob have agreed on the protocol that Bob detects the frame $Y\in SO(d)$ in which the ``yes''-probability is maximal. This will be his guess
for Alice's frame $X\in SO(d)$. In contrast to the  ``direction bit'' situation, we prove that there \emph{is no convex state space} that allows this protocol while
at the same time satisfying the analogue of Postulate 2 -- except for spatial dimension $d=2$, where frames and directions coincide.
}
\label{fig_alicebobX}
\end{center}
\end{figure*}
Before we formulate Postulates 3 and 4 and prove more properties of direction bits, let us reconsider one basic assumption.
As depicted in Fig.~\ref{fig_directionbit}, we
have assumed that the orientation of a measurement device in space is given by a direction vector, implicitly
assuming some internal rotational symmetry of the device. What if we drop this assumption? In general,
the orientation of a massive body in $\R^d$ is given by an orthonormal frame, that is, by some oriented orthonormal basis that can
be written in the form of an orthogonal matrix $X\in SO(d)$, instead of a unit vector $x\in\R^d$.
Thus, an interesting question is what happens if we repeat the calculations above, formulating analogues of the postulates for ``frame bits''
instead of direction bits.

While we have to leave the general answer open, we can give the answer in a particular special case.
First, note that for \emph{direction bits} as considered above, our calculations show that Alice and Bob can also apply the following protocol:
\begin{itemize}
\item Alice encodes spatial directions $x\in\R^d$ into the particular states $\omega_x$.
\item Bob holds a two-outcome measurement device, where the first outcome is described by the effect $\m_y$, with $y$ the
direction in which the device is pointing.
Upon receiving (many copies of) some state $\omega$, Bob looks for the spatial direction $y$ in which $\m_y(\omega)$ is maximal, which will be
his guess of Alice's direction $x$.
\end{itemize}
Effectively, the device that Bob holds asks the yes-no-question ``Is it \emph{this} direction that Alice encoded?'' The actual direction is the one
in which the probability to obtain ``yes'' is maximal.

Let us now ask whether we can implement the analogous protocol for the case that Alice wants to send a frame $X\in SO(d)$.
The main idea is that $X=\left(\begin{array}{ccc} | & & | \\ x_1 & \dots & x_d \\ | & & | \end{array}\right)$ is used to denote the
spatial orientation of an orthonormal frame attached to an object, with the $i$-th orthonormal vector pointing in direction $x_i$;
rotations $R\in SO(d)$ will thus map this to $RX=\left(\begin{array}{ccc} | & & | \\ Rx_1 & \dots & Rx_d \\ | & & | \end{array}\right)$.
The protocol is depicted in Fig.~\ref{fig_alicebobX}.
We formulate analogues of Postulates 1 and 2 (encoding and minimality) for this setup, and consider them only in the special case of this protocol.

As we show in Appendix~\ref{appendixX}, a calculation very similar to the one above proves that the ``frame bit'' state space must be generated
by pure states $\omega_Y$, labelled now by orthogonal matrices $Y\in SO(d)$ that are connected by rotations. In complete analogy to above,
every state $\omega$ can be written in the form $\omega=\lambda\omega_Y+(1-\lambda)\mu$, where $0\leq\lambda\leq 1$, $\mu$ is a
maximally mixed state, and $Y\in SO(d)$ some frame. Thus, the frame bit state space must also be a Euclidean ball of some dimension $D$.

At this point, we run into a topological problem: similarly as for direction bits, the map $X\mapsto \omega_X$ turns out to be a homeomorphism,
this time from $SO(d)$ to the unit sphere on $\R^D$. Since $SO(d)$ is not simply connected for $d\geq 2$, but the unit sphere in $\R^D$
is simply connected for $D\geq 3$, this is only possible if $D=2$ and thus (from dimension counting) $d=2$. Thus, we have proven that \emph{there
is no convex state space that allows implementation of this protocol, while satisfying analogues of Postulates 1 and 2} -- unless $d=2$,
where a frame is the same as a spatial direction, and the setup reduces to the concept of the direction bit. (We will rule out $d=2$
in Section~\ref{SecPairs}, using two further postulates.)

\section{Spatial geometry from probability measurements}
\label{SecSpatialGeometry}
Before continuing our derivation, we take another slight detour by asking for the relationship between the geometry
of physical space and state space.

As indicated in Fig.~\ref{fig_directionbit}, our setting assumes that macroscopic objects can
be physically rotated. The implicit assumption behind this is that local physical space in the laboratory
is a vector space with a Euclidean structure, that is, with an inner product, that determines the notion of a rotation
as a linear orthogonal map and, at the same time, allows to compute angles between vectors.

We assume that physical rotations $R\in SO(d)$ have representations $G_R\in SO(d)\subseteq \mathcal{G}_A$ on the direction
bit state space $A$. As we show in Lemma~\ref{eqR} in Appendix~\ref{AppDirBit}, group theory dictates that the map $R\mapsto G_R$ is
linear and, moreover, that it is of the form $G_R=O R O^{-1}$
for some orthogonal matrix $O$. Thus, there is automatically a correspondence between the vector space and Euclidean structures
of state space and physical space.
This has an interesting consequence: it implies that \emph{observers can measure physical
angles by measuring probabilities}. In other words: even if an observer has no meter stick to measure physical angles, she may
infer physical angles from probabilities.

In Appendix~\ref{AppSpatialGeometry}, we give a boot-strapped protocol that allows observers to determine angles from probability
measurements on direction bits. This method generalizes the simple quantum-mechanical insight that polarized electrons with spin-up
in a fixed direction give probability of spin-up in another direction (of relative angle $\theta$) with probability $\cos^2(\theta/2)$.

This structural coincidence (which is in particular true for quantum theory) seems remarkable beyond the specific derivation
in this paper. Clearly, in this work, we start with postulates that \emph{assume} a Euclidean structure in physical space, and obtain the ball state
space with its reversible rotation transformations \emph{as a consequence}. It is then not very surprising, though mathematically not trivial, that observers
can use this state space structure to obtain information on spatial angles.

However, irrespective of the specific construction in this paper, it is interesting to speculate whether the physically fundamental order
of logic (if there is any) might actually be reversed.
In Example~\ref{ExTopMf} in Appendix~\ref{AppSpatialGeometry}, we give a modification of the direction
bit setup where this speculation can be shown to make sense.

In this example, space is described by a topological manifold, and Bob's local laboratory space does not have a vector space structure or inner product to begin with.
We then assume that there are physical processes that can in a certain sense be interpreted as generalized rotations of a device,
yielding reversible transformations on some convex state space. Under specific conditions, we show that Bob can
use the coefficients of the measurement outcome probabilities in the space of effects to define natural coordinates on his local physical space.

In these new coordinates, the generalized rotations act linearly and orthogonally on the devices, establishing spatial Euclidean
structure that was not assumed to be there in the first place. Even though our particular example is not meant to describe an actual
physical mechanism, it is tempting to speculate whether the Euclidean structure of tangent space might be fundamentally inherited
from the convexity of probabilities.

\section{Pairs of systems: Postulates 3 \& 4}
\label{SecPairs}
Consider two (distinguishable) direction bits $A$ and $B$; taken together, they form a joint system $AB$, described by some convex state space $\Omega_{AB}$.
In the usual formulation of quantum theory, the joint state space $\Omega_{AB}$ would be given by the density matrices on the tensor product Hilbert space -- however,
in this paper, we do not assume quantum theory.

In full generality, for two state spaces $\Omega_A$ and $\Omega_B$, the framework of convex state spaces allows infinitely many possible
ways to combine them into some $\Omega_{AB}$, restricted only by a few physically obvious constraints. One of them says that if $\omega^A\in\Omega_A$
and $\omega^B\in\Omega_B$ are two local states, then there is a joint state $\omega^A\omega^B\in\Omega_{AB}$ which describes the independent
local preparation of both states on the subsystems, analogous to product states in quantum theory.
Similarly, if $\m^A$ and $\m^B$ are measurement outcomes (i.e.\ effects) on $A$ and $B$,
then by assumption there is a global effect $\m^A\m^B$ which asks whether both measurement
outcomes have occurred jointly. It satisfies in particular
\[
   (\m^A \m^B)(\omega^A \omega^B)= \m^A(\omega^A)\cdot \m^B(\omega^B),
\]
and can be shown to respect the \emph{no-signalling conditions}~\cite{Barrett}. Furthermore,
we assume that the local state space $A$ ($B$) is closed with respect to postselection according to measurement
outcomes on $B$ ($A$), for details see Appendix~\ref{AppDirBit}.

In physics, we are often interested in expectation values of observables such as energy or angular momentum.
Classical as well as quantum physics have an important structural property regarding composite systems: suppose we have two systems
$A$ and $B$ of the same type, and $h$ is a single-system observable (in quantum theory, where states $\omega^A$ are density matrices,
this would be a map $h(\omega^A)=\tr(\omega^A H)$, where $H=H^\dagger$). Suppose we are interested in the \emph{sum} of this
observable on systems $A$ and $B$ -- this defines a new observable $h^{(2)}$ on pairs of systems, where
\begin{equation}
   h^{(2)}(\omega^A \omega^B)=h(\omega^A)+h(\omega^B)
   \label{eqSumObservables}
\end{equation}
for uncorrelated states. What if we want to evaluate $h^{(2)}(\omega^{AB})$ for \emph{correlated} (possibly entangled) states $\omega^{AB}$?
In quantum theory, there is a \emph{unique} way to do this, because eq.~(\ref{eqSumObservables}) uniquely determines $h^{(2)}$ and its action on all states.
We necessarily have
\[
   h^{(2)}(\omega^{AB})=\tr\left[
      \omega^{AB}\left( H\otimes\mathbf{1} + \mathbf{1}\otimes H\right)
   \right].
\]
Arguably, this uniqueness is an important property of composite systems -- if it failed, it would not be clear how to add up observables on composite systems
(for example, there would be no unique notion of a ``non-interacting sum of Hamiltonians'', and thus no unique physical way to combine systems
in trivial non-interacting ways). We promote this property to a postulate.

\begin{shaded}
\textbf{Postulate 3 (Sums of observables).} If $h$ is any direction bit observable, then there is a unique
observable $h^{(2)}$ on \emph{pairs} of direction bits such that
\[
   h^{(2)}(\omega^A \omega^B)=h(\omega^A)+h(\omega^B).
\]
\end{shaded}

It is easy to see that Postulate 3 holds true if and only if the uncorrelated states $\omega^A \omega^B$ linearly span
the global state space $\Omega_{AB}$.
Thus, Postulate 3 is equivalent to a condition that is usually called ``local tomography'' in the literature~\cite{Hardy2001}:
it says that \emph{joint states on $AB$ are uniquely characterized by the local measurement outcome probabilities on $A$ and $B$ and their
correlations.}
Denoting the dimension of $\Omega_A$ by $d_A$, this is also equivalent to
\begin{equation}
   d_{AB}+1 = (d_A+1)(d_B+1).
   \label{eqLocalTomo}
\end{equation}
The global state space $\Omega_{AB}$ carries its own group of reversible transformations $\G_{AB}$. We assume
that Alice and Bob may still apply their local reversible transformations, that is, if $G^A\in\G_A$ and $G^B\in \G_B$, then
$G^A G^B\in\G_{AB}$. Due to Postulate 3, this transformation is uniquely determined by its action on uncorrelated states:
$(G^A G^B)(\omega^A \omega^B)=(G^A \omega^A)(G^B \omega^B)$.

This shows that Postulate 3 also has geometric significance: suppose we decide to carry out a local coordinate transformation;
in our case, this is a rotation $R\in SO(d)$. This transformation acts on states of direction bits via $\omega^A\mapsto G_R \omega^A$.
The third postulate now tells us that this \emph{uniquely determines the coordinate transformation map on (correlated) pairs of systems:}
they are transformed via $\omega^{AB}\mapsto (G_R G_R)\omega^{AB}$, which is the only possible linear map that transforms
$\omega^A\omega^B$ into $(G_R\omega^A)(G_R\omega^B)$.

Every pair of state spaces $\Omega_A$ and $\Omega_B$ can be combined into a joint state space $\Omega_{AB}$ in accordance
with Postulate 3: the ``smallest'' possible choice (denoted $\Omega^{\min}_{AB}$) is to define it as the convex
hull of all product states $\omega^A\omega^B$. On the other hand, the ``largest'' possible choice (denoted $\Omega^{\max}_{AB}$) is to allow all vectors $\omega^{AB}$
such that all local measurements yield valid probabilities, even after postselection~\cite{Christandl,Broadcasting,BFRW}.
Every compact convex set $\Omega_{AB}$ which satisfies
\[
   \Omega^{\min}_{AB}\subseteq \Omega_{AB}\subseteq \Omega^{\max}_{AB}
\]
is then a possible choice of the global state space, as long as local reversible transformations map $\Omega_{AB}$ into itself.
In quantum theory, $\Omega^{\min}_{AB}$ turns out to be the set of
unentangled states, while the actual global quantum state space $\Omega_{AB}$ lies strictly in between $\Omega^{\min}_{AB}$
and $\Omega^{\max}_{AB}$.

Composites of convex state spaces have been extensively studied in the quantum information literature. Some of this interest is due to the fact
that many of these state spaces contain states with non-local correlations that are stronger than those allowed by quantum theory.
For example, if $\Omega_A=\Omega_B$ is the square state space as in Fig.~\ref{Fig2}f), then the composite $\Omega^{\max}_{AB}$
is the \emph{no-signalling polytope} for two binary measurements on two parties, containing \emph{PR box states} which violate
the Bell-CHSH inequality stronger than any quantum state~\cite{Khalfin,Tsirelson,Popescu,Barrett}. This example also illustrates that the convex state spaces formalism
describes a vast landscape of theories with physical properties that can be very different from those of quantum theory.

In the case of two direction bits $A$ and $B$, where the local state spaces are $d$-balls, there are also many possible choices of the global
state space $\Omega_{AB}$ in accordance with Postulate 3, including $\Omega^{\min}_{AB}$ and $\Omega^{\max}_{AB}$.
Our fourth and final postulate now states that this global state space allows for continuous reversible interaction.

\begin{shaded}
\textbf{Postulate 4 (Interaction).} For two direction bits $A$ and $B$,
there is a continuous one-parameter group of transformations $\{T^{AB}_t\}_{t\in\R}$
which is not a product of local transformations, $T^{AB}_t \neq T^A_t T^B_t$.
\end{shaded}

The group $T^{AB}_t$ describes continuous reversible time evolution in a closed system of two direction bits:
if we start at time $t=0$ with a product state $\omega^A\omega^B$, then the state at time $t$ will be $\omega^{AB}(t):=T^{AB}_t(\omega^A\omega^B)$.
If $T^{AB}_t$ was a product transformation $T^A_t T^B_t$ for all times $t$, then the global state would remain
a product state forever: $\omega^{AB}(t)=(T^A_t\omega^A)(T^B_t\omega^B)$.
In this case, the two direction bits
could never become correlated; there would be no interaction. Postulate 4 excludes this: it states that there is at least one time $t\in\R$ such
that $T^{AB}_t$ is not of this product form.

\begin{figure}[!hbt]
\begin{center}
\includegraphics[angle=0, width=6cm]{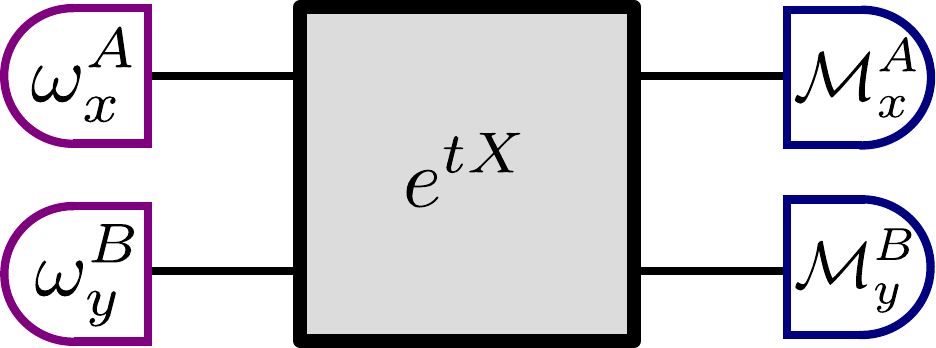}
\caption{Circuit model which yields constraints for the global Lie algebra elements $X$. We prepare a pure product state
$\omega^A_x \omega^B_y$, apply the transformation $\exp(t X)$, and perform a product measurement $\m^A_x\m^B_y$.
Since this gives probability $1$ for $t=0$, and probabilities cannot be larger than $1$ for other (small) $t$, this implies
that the derivative at $t=0$ must vanish, and the second derivative must be non-positive.}
\label{fig_circuit}
\end{center}
\end{figure}

The global transformations $T^{AB}_t$ and the local transformations $G^A G^B$ with $G^A,G^B\in SO(d)$ generate a Lie subgroup
of $\G_{AB}$; we call it $\H_{AB}$. Due to~(\ref{eqLocalTomo}), it is a matrix Lie group acting on $\R^{(d+1)(d+1)-1}$.
The corresponding Lie algebra is called $\h_{AB}$. Let $X$ be some element of $\h_{AB}$, and
consider the circuit in Fig.~\ref{fig_circuit}. It depicts the outcome probability of a product measurement on an evolved product state,
\[
   f(t):=\m^A_x \m^B_y\left( e^{tX} \omega^A_x \omega^B_y\right)\in[0,1].
\]
As we show in Lemma~\ref{LemNoisyToNoiseless} in Appendix~\ref{AppDirBit}, we may assume without loss of generality that the direction
bit state space has noise parameter $c=1/2$ and visibility parameter $a=1$. This is the ``noiseless case'', where spin measurements
give probabilities $\m_x(\omega_{-x})=0$ and $\m_x(\omega_x)=1$, implying in particular that $f(0)=1$ for the circuit in Fig.~\ref{fig_circuit}.
Since this is the maximal possible value, we must have $f'(0)=0$ and also $f''(0)\leq 0$. Thus
\begin{eqnarray*}
   \m_x^A \m_y^B X \omega_x^A \omega_y^B &=& 0 \\
   \m_x^A \m_y^B X^2 \omega_x^A \omega_y^B &\leq& 0
\end{eqnarray*}
for all $x,y\in\R^d$ with $|x|=|y|=1$. By considering other circuits of this kind, we obtain a long list of constraint equalities
and inequalities which must be satisfied by all global Lie algebra elements $X$.

Surprisingly, as shown in Appendix~\ref{AppDirBit} and in~\cite{EntDynBeyondQT}, \emph{if $d\neq 3$, then the only matrices $X$ which satisfy all constraints
are those of the form $X=X^A+X^B$}. And these elements generate non-interacting time evolution of product form:
$\exp(tX)=\exp(tX^A)\exp(tX^B)$. Thus, if $d\neq 3$, $\H_{AB}$ contains only product transformations, and
Postulate 4 cannot be satisfied.

\begin{framed}
\textbf{Theorem 2.} From Postulates 1--4 it follows that the spatial dimension must be $d=3$.
\end{framed}

The main reason why $d=3$ is special becomes visible by inspection of the proof in~\cite{EntDynBeyondQT}.
It boils down to the group-theoretic fact that (at least for $d\geq 3$) the subgroup of $SO(d)$ which fixes
a given vector (that is, $SO(d-1)$) is Abelian only if $d=3$. In other words,
\emph{the fact that rotations commute in two dimensions, but not in higher dimensions} is the main reason
why $d=3$ survives. The cases $d=1$ and $d=2$ are special as well, but are ruled out in the proof for other reasons.

It remains to show that we actually get quantum theory for two direction bits if $d=3$. We already know that the dimension of the
global state space is $\dim\Omega_{AB}=(d+1)(d+1)-1=15$, which agrees with the number of real parameters in a complex $4\times 4$ density
matrix. Thus, we can embed $\Omega_{AB}$ in the real space of Hermitian $4\times 4$-matrices of unit trace.
Now we have global Lie algebra elements $X\in\h_{AB}$ that are not just sums of local generators, i.e.\ $X\neq X^A+X^B$.
However, as shown in~\cite{LocalToGlobalQT}, these elements are still highly restricted: they generate unitary conjugations, i.e.\
transformations of the form $\rho\mapsto U\rho U^\dagger$.

By Postulate 4, at least one of these unitaries must be entangling. Moreover, all local unitary transformations are possible
(in the ball representation, these are the rotations in $SO(3)$). It
is a well-known fact from quantum computation~\cite{Bremner} that a set of unitaries of this kind generates the set of all unitaries -- that is,
\emph{every} map of the form $\rho\mapsto U\rho U^\dagger$ must be contained in the global transformation group $\H_{AB}\subseteq\G_{AB}$.

The orbit of this group on pure product states generates all pure quantum states, and one can show~\cite{LocalToGlobalQT} that there
cannot be any additional non-quantum states. Thus, we have recovered the state space of quantum theory on two qubits. Due to positivity, all effects
must be quantum effects; in the noisy case (i.e.\ $c\neq 1/2$ or $a<1$), not all quantum effects may actually be implementable -- that is, we might
have a restricted set of measurements. We have thus proven:
\begin{framed}
\textbf{Theorem 3.} From Postulates 1--4, it follows that the state space of two direction bits is two-qubit quantum state space (i.e.\
the set of $4\times 4$ density matrices), and time evolution is given by a one-parameter group of unitaries,
$\rho\mapsto U(t) \rho U(t)^\dagger$.
\end{framed}

As a simple consequence, there exists some $4\times 4$ Hermitian matrix $H$ such that $U(t)=\exp(-i H t)$,
i.e.\ a Hamiltonian which generates time evolution according to the Schr\"odinger equation.
\vskip 0.5cm

$\strut$

\section{Conclusions}
We have derived two facts about physics from information-theoretic postulates: the three-dimensionality of space~\cite{Bengtsson},
and the fact that probabilities of measurement outcomes for some systems are described by quantum theory. In order to do this, we
assumed that there exist ``reasonable'' physical systems which, in a certain sense, carry minimal amounts of directional information.

Our result supports and clarifies the point of view that the geometric structure of spacetime and the probabilistic structure of quantum theory
are closely intertwined, similar in spirit to~\cite{Weizsaecker,Lyre,Penrose,WoottersThesis,dAriano2,Fotini,Thiemann}. As one can see in Fig.~\ref{Fig2},
this conclusion becomes particularly obvious in the context of convex state spaces. This interrelation is not only axiomatic, but also operational:
as we have shown in Sec.~\ref{SecSpatialGeometry}, observers can measure -- or even define -- physical angles by measuring probabilities.

Furthermore, these findings suggest exploring possible generalizations: the approach to construct state spaces from physical symmetry
properties~\cite{Sanyal}, together with minimality assumptions, might reproduce quantum systems of higher spin, or even physically interesting non-quantum
state spaces that have so far remained unexplored.

In summary, there seem to be two possible interpretations of the results in this paper. First, the results might simply be mathematical coincidence, without
any deep physical reason underlying them. This is perfectly conceivable; in this case, the main contribution of this paper is a detailed analysis of the
structural fit between quantum theory and spacetime. Second, the results might point to an actual logical relation between geometry
and probability that arises from some unknown fundamental physics, such as quantum gravity.

If the second possibility turned out to be true, this would suggest an exciting speculation, stated also in~\cite{DiscreteHilbertSpace,Kleinmann}:
In many approaches to quantum gravity, the smoothness and/or three-dimensionality of space is considered to be only an approximation.
But then, given the close relation between smooth Euclidean space and the qubit, maybe the universe's probabilistic theory is only approximately quantum?
Taking this idea seriously would suggest to go beyond the usual ``quantization of geometry'' paradigm.

\acknowledgments
We thank Lucien Hardy and Lee Smolin for discussions and continuous encouragement, and
we are grateful to Cozmin Ududec, FJ Schmitt, Freddy Cachazo,
Giulio Chiribella, Raymond Lal, Rob Spekkens, Robert H\"ubener, Tobias Fritz and Wolfgang Binder for helpful comments and suggestions.
Research at Perimeter Institute for Theoretical Physics is supported in part by the Government of Canada through NSERC and by the Province of Ontario through MRI.
LM acknowledges CatalunyaCaixa, the EU ERC Advanced Grant NLST, the EU Qessence project and the Templeton Foundation.

\vskip 3cm

\onecolumngrid

\appendix
\section{Characterization of all direction bit state spaces}
\label{AppDirBit}
The proof consists of four steps: first, we prove that the direction bit state space is a Euclidean ball (possibly noisy, that is,
with a restricted set of measurements). Then we show that that the noisy case can always be reduced to the noiseless case.
Given this, the results from Ref.~\cite{EntDynBeyondQT} do most of the work: they show that only $d=3$ is possible. As a last step,
in order to obtain quantum theory for $d=3$, we refer to the results in Ref.~\cite{LocalToGlobalQT}.

We start with a formal definition of state spaces. As we have motivated in the main text, the set of normalized states on any system is a compact
convex set. To simplify the calculations, it makes sense to start right away with the full set of
\emph{unnormalized} states, which will be all vectors of the form $\lambda\cdot\omega$, where $\omega$ is a normalized state,
and $\lambda\geq 0$. This yields a \emph{cone} in the sense of convex geometry -- that is, a subset $C$ of a vector space
with the property that $x\in C$ implies $\lambda x\in C$ for all $\lambda\geq 0$.

For reasons of brevity, we will not give a detailed explanation and motivation of all definitions. For more discussion, we refer
the reader to the references mentioned in the main text, in particular to Chapter 3 in~\cite{Pfister}.

\setcounter{theorem}{3}
\begin{definition}[State space]
A \emph{state space} is a tuple $(A,A_+,\u^A,E_A)$, where
\begin{itemize}
\item $A$ is a real finite-dimensional vector space,
\item $A_+\subset A$ is a proper cone (that is, a closed, convex cone of full dimension with $A_+\cap (-A_+)=\{0\}$), called the
\emph{cone of unnormalized states},
\item $\u^A$ is a linear functional which is strictly positive on $A_+\setminus\{0\}$, called the \emph{order unit} of $A$,
\item $E_A$ is a closed convex set of functionals which are non-negative on all of $A_+$, containing $\u^A$, having full dimension $\dim(E_A)=\dim(A)$, and
satisfying $\m(x)\leq\u^A(x)$ for all $x\in A_+$ and $\m\in E_A$.
It is called the \emph{set of allowed effects}.
\end{itemize}
Furthermore, we define the \emph{dual cone} $A_+^*$ as the set of all linear functionals which are non-negative
on $A_+$, which implies $E_A\subseteq A_+^*$.
The set of all $\omega\in A_+$ with $\u^A(\omega)=1$ will be denoted $\Omega_A$ and is called the \emph{set of
normalized states}.
\end{definition}
The requirement $\dim(E_A)=\dim(A)$ has a simple physical motivation: if $\dim(E_A)<\dim(A)$, then we would have
states $\omega\neq\varphi$ that would yield the same outcome probabilities for all possible measurements, invalidating to call them ``different states''
in the first place.

To save some ink, we will usually just write $A$ for the state space, instead of writing the full tuple. However, keep in mind
that the choice of a state space comes also with a choice of $A_+$, $\u^A$ and $E_A$.

Given any measurement with an arbitrary number of outcomes, the probability of one of the outcomes -- if measured on some
state $\omega\in\Omega_A$ -- will be a real number in $[0,1]$. The map $\m$ which takes the state $\omega$ to the corresponding probability
$\m(\omega)$ must be linear, since statistical mixtures of states must yield mixtures of probabilities. In principle, every linear functional
$\m\in E_A^{\max}$ may describe a measurement outcome probability, where
\begin{equation}
   E_A^{\max}:=\{\m\in A_+^*\,\,|\,\, 0\leq \m(\omega)\leq 1 \mbox{ for all }\omega\in\Omega_A\}.
   \label{eqEAmax}
\end{equation}
However, one may imagine that it might be physically impossible to implement measurement devices for all these linear functionals. This is why
the set $E_A$ is introduced in the definition above: it is meant to describe the collection of all possible effects that may actually be implemented in
measurements. Clearly, we have $E_A\subseteq E_A^{\max}$. In some publications (e.g.~\cite{MuellerUdudec}), it is assumed that $E_A=E_A^{\max}$,
but not in this paper. In other words, we are \emph{not} assuming the ``no-restriction hypothesis'' here~\cite{JanottaLal}.
The possibility to have $E_A\neq E_A^{\max}$ describes situations, as we will
see below, where all measurements on a direction bit are by necessity intrinsically noisy.

As an example, in finite-dimensional $n$-level quantum theory,
\begin{itemize}
\item $A$ is the real vector space of Hermitian matrices on $\C^n$,
\item $A_+$ is the set of positive semi-definite matrices on $\C^n$,
\item $\u^A(\rho)=\tr(\rho)$ is the trace functional,
\item $E_A$ is the set of all maps of the form $\rho\mapsto \tr(\rho M)$, with
$M\geq 0$ a positive semi-definite matrix,
\item $\Omega_A$ is the set of density matrices on $\C^n$.
\end{itemize}
Similarly, the state space of classical $n$-level probability theory is $(B,B_+,\u^B,B^+)$, where
\begin{itemize}
\item $B=\R^n$,
\item $B_+=\{p=(p_1,\ldots, p_n)\,\,|\,\, \mbox{all }p_i\geq 0\}$,
\item $\u^B(p)=p_1+p_2+\ldots +p_n$,
\item $E_B$ is the set of all maps $p\mapsto p\cdot q$ with $q=(q_1,\ldots, q_n)$ such that all $q_i\geq 0$,
where $\cdot$ denotes the Euclidean inner product,
\item $\Omega_B$ is the set of all probability distributions:
\[
   \Omega_B=\left\{p=(p_1,\ldots,p_n)\,\,\left|\,\,\mbox{all }p_i\geq 0,\sum_i p_i=1\right.\right\}.
\]
\end{itemize}
In both classical probability theory and quantum theory, all effects are allowed.

We would like to talk about reversible transformations on state spaces. To this end, we define
\begin{definition}[Dynamical state space]
A tuple $(A,A_+,\u^A,E_A,\G_A)$, where $(A,A_+,\u^A,E_A)$ is a state space, and $\G_A$ is a compact (possibly finite) group
of linear maps on $A$, is called a \emph{dynamical state space}, if every $G\in\G_A$ satisfies
\begin{itemize}
\item $G \Omega_A=\Omega_A$ (or, equivalently, $\u^A\circ G=\u^A$ and $G A_+=A_+)$, and
\item $E_A\circ G=E_A$.
\end{itemize}
\end{definition}
These two conditions say that reversible transformations must respect the set of normalized states and the set of allowed effects.
It is easy to see that the first condition implies that $A_+^*\circ G=A_+^*$ for all $G\in\G_A$.

In quantum theory, $\G_A$ is the group of all maps of the form $\rho\mapsto U \rho U^\dagger$, with $U$ unitary. In classical
probability theory, $\G_B$ is a representation of the permutation group. Specifically, for every permutation $\pi$, there is a reversible transformation
$G_\pi$ with $G_\pi(p_1,\ldots, p_n)=(p_{\pi(1)},\ldots, p_{\pi(n)})$.

Here is a rigorous definition of equivalence of state spaces:
\begin{definition}[Equivalent state spaces]
\label{DefEquiv}
Two state spaces $A$ and $B$ are \emph{equivalent} if there exists a bijective linear map $L:A\to B$ such that
the following conditions are satisfied:
\begin{itemize}
\item $L A_+=B_+$,
\item $\u^B\circ L=\u^A$,
\item $E_B\circ L=E_A$.
\end{itemize}
Two \emph{dynamical} state spaces $A$ and $B$ are equivalent, if they are equivalent as state spaces and additionally satisfy
$\G_B=L \circ \G_A \circ L^{-1}$.
\end{definition}
This is clearly an equivalence relation. If two (dynamical) state spaces are equivalent, they are indistinguishable in all their physical properties.

Now we show how the notion of noisiness in Postulate 2 can be seen as a special case of ``group majorization'',
a natural definition of noisiness with respect to a group that encompasses the classical and quantum cases in the obvious way.
This definition is well-known in the mathematics literature~\cite{Marshall}; we rephrase it in Definition~\ref{DefNoisiness} below in the context of convex state spaces.
We start by showing a simple consequence of Postulate 2.
\begin{lemma}
\label{LemNoisySOd}
Suppose that $\omega$ and $\varphi$ are both possible encodings of the same direction $x\in\R^d$, $|x|=1$, in some protocol that
satisfies Postulate 1. From Postulate 2, it follows that there exist $0<\lambda_j\leq 1$, $\sum_j\lambda_j=1$, and rotations $R_j\in SO(d)$ (resp.\ $O(d)$ if $d=1$)
such that
\[
   \varphi=\sum_j \lambda_j G_{R_j^{-1}} \omega
\]
or with $\varphi$ and $\omega$ interchanged. If $\varphi\neq \omega$ then this is a proper convex combination.
\end{lemma}
\proof
According to Postulate 2, the assumptions of this lemma imply
\[
   \m_z^{(i)}(\varphi)=\sum_j \lambda_j \m_{R_j z}^{(i)}(\omega)\qquad\mbox{for all directions }z\in\R^d, |z|=1,
\]
or vice versa (in the latter case, rename $\varphi$ and $\omega$ to fit this formula). Set $\omega':=\sum_j \lambda_j G_{R_j^{-1}} \omega$,
then $\m_z^{(i)}(\omega')=\m_z^{(i)}(\varphi)$ for all $z$. But then $\omega'$ could be used as a replacement for $\varphi$ in the protocol,
namely, as yet another codeword for direction $x$. Moreover, $\omega'$ and $\varphi$ are by construction equally noisy in their directional
information, so Postulate 2 implies that they must be equal.
\qed

Now we show how this fits into a majorization framework.
\begin{definition}[Group majorization]
\label{DefNoisiness}
Let $A$ be any dynamical state space, and $\mathcal{H}$ a compact subgroup of $\G_A$. Then we define a relation $\preceq_\mathcal{H}$
on $\Omega_A$ in the following way: for $\omega,\varphi\in\Omega_A$, it holds $\omega\preceq_{\mathcal{H}}\varphi$ if and only if
there are $\lambda_i\geq 0$, $i=1,\ldots,n$, $\sum_{i=1}^n \lambda_i=1$, and $T_i\in\mathcal{H}$ such that
\begin{equation}
   \omega=\sum_i\lambda_i T_i \varphi.
   \label{eqNoisiness2}
\end{equation}
We write $\omega\preceq\varphi$ if any only if $\omega\preceq_{\G_A} \varphi$.
\end{definition}
\begin{lemma}
The noisiness relation is a partial order on the orbits. That is, for any dynamical state space $A$ and any compact subgroup $\mathcal{H}\subseteq\G_A$, we have
\begin{itemize}
\item[(i)] if $\omega\preceq_\mathcal{H} \varphi$ and $\varphi\preceq_\mathcal{H} \rho$, then $\omega\preceq_\mathcal{H}\rho$.
\item[(ii)] $\omega\preceq_\mathcal{H}\omega$ for all $\omega\in\Omega_A$, and
\item[(iii)] if $\omega\preceq_\mathcal{H}\varphi$ and $\varphi\preceq_\mathcal{H}\omega$, then there exists $T\in\mathcal{H}$ such that $\omega=T\varphi$.
\end{itemize}
Moreover, we have
\begin{itemize}
\item[(iv)] if $\omega\preceq_\mathcal{H}\varphi$ then $T_1 \omega\preceq_\mathcal{H}T_2\varphi$ for all $T_1,T_2\in\mathcal{H}$.
\end{itemize}
\end{lemma}
\proof
Property (ii) is trivial, by setting $\lambda_1=1$ and $T_1=\mathbf{1}$ in~(\ref{eqNoisiness2}). If $\omega\preceq_\mathcal{H}\varphi$, then
\[
   T_1 \omega = \sum_i \lambda_i (T_1 T_i T_2^{-1}) T_2\varphi\preceq_\mathcal{H} T_2\varphi
\]
if $T_1,T_2\in\mathcal{H}$, proving (iv). If additionally $\varphi\preceq_\mathcal{H}\rho$ such that $\varphi=\sum_j \lambda'_j T'_j \rho$,
then $\omega=\sum_{ij} \lambda_i\lambda'_j T_i T'_j \rho\preceq_\mathcal{H}\rho$. This proves (i). It remains to prove (iii). To this end,
introduce an inner product $\langle\cdot,\cdot\rangle$ on $A$ which is invariant with respect to $\G_A$ (and thus $\mathcal{H}$), i.e.
\[
   \langle x,y\rangle = \langle Tx,Ty\rangle \qquad\mbox{for all }x,y\in A, T\in\G_A.
\]
Moreover, let $\|\cdot\|$ be the corresponding norm. Then~(\ref{eqNoisiness2}) and the triangle inequality yield
\[
   \|\omega\|=\left\|\sum_i\lambda_i T_i\varphi\right\|\leq \sum_i \lambda_i \|T_i\varphi\| = \sum_i \lambda_i \|\varphi\|=\|\varphi\|.
\]
Thus, if both $\omega\preceq_\mathcal{H} \varphi$ and $\varphi\preceq_\mathcal{H}\omega$, then $\|\omega\|=\|\varphi\|=\|T_i\varphi\|=:r$.
Let $S_r$ be the unit sphere of radius $r$, then~(\ref{eqNoisiness2}) says that $\omega\in S_r$ is a convex combination of the $T_i\varphi\in S_r$.
Geometrically, it is clear that this is only possible if $T_i\varphi=\omega$ whenever $\lambda_i\neq 0$ (formally, it follows from the fact
that all boundary points of the ball are exposed points). Setting $T:=T_i$ for any of these $i$
proves (iii).
\qed

Now we see how our definition of noisiness from Postulate 2 fits naturally into the well-known notion of majorization. In the case of quantum theory (with the full
unitary group), it follows from~\cite[Thm.\ 12.13]{NielsenChuang} that our relation $\preceq$ is identical to Nielsen's majorization relation on density matrices.
From Lemma~\ref{LemNoisySOd}, we obtain the following:
\begin{theorem}[Noisiness and group majorization]
A state $\varphi$ is at least as noisy in its directional information as another state $\omega$ if and only if
\[
   \varphi\preceq_{SO(d)}\omega,
\]
where $SO(d)$ denotes the representation of the rotation group within the group of reversible transformations of a direction bit. (If $d=1$, then $SO(d)$
has to be replaced by $O(1)$.)
\end{theorem}

Given two state spaces $A$ and $B$, we would like to define a composite state space $AB$ which, according to Postulate 3, satisfies the \emph{local tomography
property}~\cite{MasanesMueller}: states on $AB$ are uniquely characterized by the statistics of local measurements. Eq.~(\ref{eqLocalTomo}) in the
main text translates into $\dim(AB)=\dim A \dim B$; thus, we may choose the vector space $AB$
to be the tensor product $A\otimes B$. This will turn out to be a handy choice: we can represent independent preparations $\omega^A\omega^B$
by products $\omega^A\otimes \omega^B$. We get the following definition:
\begin{definition}[Locally tomographic composite]
\label{DefTomoComposite}
Given two dynamical state spaces $A$ and $B$, a dynamical state space $(AB,(AB)_+,\u^{AB},E_{AB},\G_{AB})$ will be called a \emph{composite} of $A$ and $B$,
if the following conditions are satisfied:
\begin{itemize}
\item the linear space which carries the state space is $AB=A\otimes B$,
\item $\u^{AB}=\u^A \otimes \u^B$,
\item if $\varphi^A\in A_+$ and $\omega^B\in B_+$, then $\varphi^A\otimes \omega^B\in (AB)_+$,
\item if $\m^A\in E_A$ and $\n^B\in E_B$, then $\m^A\otimes \n^B\in E_{AB}$,
\item if $G_A\in \G_A$ and $G_B\in\G_B$, then $G_A\otimes G_B\in \G_{AB}$,
\item for every $\n^B\in E_B$ and $\omega^{AB}\in (AB)_+$, the vector $\omega^A_{\rm cond}$ (``conditional state'') defined by
\begin{equation}
   \m^A(\omega^A_{\rm cond}) = \frac { \m^A\otimes \n^B(\omega^{AB})}{\u^A\otimes \n^B(\omega^{AB})}\quad\mbox{ for all }\m^A\in E_A
   \label{eqCondApp}
\end{equation}
is a valid state, i.e.\ $\omega^A_{\rm cond}\in A_+$, and similarly for $A$ and $B$ interchanged.
\end{itemize}
\end{definition}
Note that eq.~(\ref{eqCondApp}) is automatically satisfied if all effects on $A$ are allowed. It means that we cannot get ``new'' states outside
of $\Omega_A$ by preparing global states and postselecting on local measurement outcomes. Similarly, we might demand that
any map of the form
\[
   \omega^A \mapsto \m^{AB}(\omega^A \otimes \omega^B)
\]
for a fixed bipartite effect $\m^{AB}$ and fixed state $\omega^B$ is itself a valid effect on $A$. If this is violated, then the set of possible local measurements on $A$ is increased by
composing it with the other system $B$. However, since we do not need this condition in the following, we decided not to have it as part of the definition
in order to have a result which is as general as possible.

By setting $\n^B:=\u^B$ in eq.~(\ref{eqCondApp}), we obtain the conditional state which Alice sees if Bob does not perform any measurement. This is the reduced state $\omega^A$
of $\omega^{AB}\in\Omega_{AB}$, satisfying
\[
   \m^A(\omega^A)=\m^A\otimes \u^B(\omega^{AB})\quad\mbox{ for all }\m^A\in E_A.
\]
Thus, Definition~\ref{DefTomoComposite} ensures that global states have valid reduced states (marginals).

We continue by proving two claims in the main text in the following two lemmas:
\begin{lemma}
\label{LemNotTooSymmetric}
With the notation of the main text (in particular, $x=(1,0,\ldots,0)^T$), there is some outcome $i_0$ and some direction $y\in\R^d$, $|y|=1$,
such that $\m_y^{(i_0)}(\omega(x))>\m_{-y}^{(i_0)}(\omega(x))$.
\end{lemma}
\proof
As we have shown in the main text, the probabilities $\m_y^{(i)}(\omega(x))$ depend only on the first component $y_1$ of $y$.
Suppose that $\m_y^{(i)}(\omega(x))=\m_{-y}^{(i)}(\omega(x))$ for all $i$ and all $y$.
Let $S\in SO(d)$ be the matrix $S:={\rm diag}(-1,-1,1,\ldots,1)$ (if $d=1$ or $d=2$ take $S=-\mathbf{1}$), then it follows that
$\m_y^{(i)}(\omega(x))=\m_{Sy}^{(i)}(\omega(x))$ for all $i$ and $y$. Now consider the following two situations under which
Alice attempts to send the spatial direction $x=(1,0,\ldots,0)^T$ to Bob:
\begin{enumerate}
\item Bob's laboratory is aligned in exactly the same way as Alice's -- that is, both share the same coordinate system (maybe by chance).
In this case, Bob's coordinates $x_B$ of $x$ are the same as Alice's: $x_B=(1,0,\ldots,0)^T$.
\item Compared to Alice's laboratory, Bob's lab is oriented differently, namely it is rotated by $S$ relative to Alice. In this case, Bob's coordinates
$x_B$ of $x$ are $x_B=(-1,0,\ldots,0)^T$.
\end{enumerate}
Since Alice does not know which of the two situations (or any of the infinitely other possible ones) apply to Bob's laboratory, her encoding $x\mapsto\omega(x)$
must work in both cases. However, due to $\m_y^{(i)}(\omega(x))=\m_{Sy}^{(i)}(\omega(x))$ for all $i$ and $y$,
Bob sees exactly the same outcome probabilities in both cases, leading with probability one to the same estimate $x_B$.
This contradicts the soundness of the protocol, i.e.\ Postulate 1.
\qed

\begin{lemma}
\label{LemUniqueMaximum}
Let $i_0$ be any outcome that satisfies the statement of Lemma~\ref{LemNotTooSymmetric}. Then there is some direction $y\in\R^d$, $|y|=1$, such that the state
\begin{equation}
   \omega'(y):=\int_{R\in SO(d):Ry=y} G_R \omega(x)\, dR
   \label{eqOmegaPrime}
\end{equation}
has the property that the map $z\mapsto L_z(\omega'(y))$ attains its unique global maximum at $z=y$.
\end{lemma}
\proof
If $d=1$, then eq.~(\ref{eqOmegaPrime}) becomes $\omega'(y)=\omega(x)$, and thus $L_y(\omega'(y))=L_y(\omega(x))>0$ according to Lemma~\ref{LemNotTooSymmetric}.
Since there are only two possible directions $y=\pm 1$, and since $L_{-y}=-L_y$, this must be the global maximum.

Now consider the case $d\geq 2$. As in the main text, write $\m_z:=\m_z^{(i_0)}$. Then $L_z(\omega(x))=\m_z(\omega(x))-\m_{-z}(\omega(x))$
depends only on the first component $z_1$ of $z$;
denote this value by $\ell(z_1)$. This defines a real continuous function $\ell:[-1,1]\to[-1,1]$. By construction, it is an odd function: $\ell(-z_1)=-\ell(z_1)$,
and we know from Lemma~\ref{LemNotTooSymmetric} that there is some $z_1$ with $\ell(z_1)>0$. Since $\ell$ is continuous, it attains its global maximum $\ell_{max}>0$ somewhere.
Since the set of points where this maximum is attained is compact, and since $\ell(z_1)=\ell_{max}$ implies $\ell(-z_1)\neq\ell_{max}$, the expression
\[
   z_1^*:={\rm argmin}\{|z_1|\,\,|\,\, \ell(z_1)=\ell_{max}\}
\]
is well-defined. Let $y\in\R^d$, $|y|=1$ be any direction with first component $y_1=z_1^*$. By construction, $\ell_{max}=L_y(\omega(x))\geq L_z(\omega(x))$
for all $z\neq y$. Define $\omega'(y)$ as in~(\ref{eqOmegaPrime}). Since $L_z\circ G_R=L_{R^{-1} z}$, we obtain
\[
   L_y(\omega'(y))=L_y(\omega(x))=\ell_{max}.
\]
Consequently, $L_{-y}(\omega'(y))=-\ell_{max}$. Let $z\not\in\{y,-y\}$, then
\begin{equation}
   L_z(\omega'(y))=\int_{R\in SO(d):Ry=y} L_{R^{-1} z}(\omega(x))\, dR.
   \label{eqIntForUniqueMax}
\end{equation}
We have to show that this is strictly less than $\ell_{max}$. To this end, we define a continuous path on the surface of the $d$-dimensional unit ball.
We will assume that $z_1^*>0$; the case $z_1^*<0$ is treated analogously ($z_1^*=0$ is excluded from $\ell(0)=-\ell(0)=0<\ell_{max}$).
For $t\in[-z_1^*,z_1^*]$, let $z(t)\in\R^d$ be some vector with $|z(t)|=1$ such that $t\mapsto z(t)$ is continuous, $z(-z_1^*)=-y$, $z(z_1^*)=y$,
and such that the first component of $z(t)$ equals $t$. If $z\not\in\{-y,y\}$, then there is some $t\in(-z_1^*,z_1^*)$ such that
$|z-y|=|z(t)-y|$. Hence there is some $R\in SO(d)$ with $Ry=y$ such that $R^{-1} z = z(t)$, and since $|t|<|z_1^*|$, we have
\[
   L_{R^{-1} z}(\omega(x))=L_{z(t)}(\omega(x))=\ell(t)<\ell_{max}.
\]
But this expression appears in~(\ref{eqIntForUniqueMax}): the integrand is upper-bounded by $\ell_{max}$ for all $R$, and is strictly less than $\ell_{max}$
for the rotation $R$ that we have just found. This proves that $L_z(\omega'(y))<\ell_{max}$.
\qed

Now we are ready to give a thorough definition of a ``direction bit''. It is arguably difficult to formalize Postulates 1 and 2 from the main text into a rigorous
mathematical definition: rigorously defining what is meant by a ``protocol'' seems hardly worth the effort (the result would be long and not very illuminating);
similarly, a formalization of the physical intuition about spatial symmetries (rotating the device versus the direction bit etc.), as used in the initial stage of the proof,
seems over the top for the purpose of this paper. Instead, we use two \emph{consequences} of Postulates 1 and 2, called Assumptions 1 and 2,
as derived in the main at an intermediate
stage of the proof, to write down a definition of direction bits. This avoids talking about the physical background situation, but ensures that all the ``convex state
space'' argumentation rests on rigorous mathematical grounds.

The meaning of the assumptions is as follows. Assumption 1 states that the standard protocol which we have constructed in the main text works:
There is some state $\omega$ which may serve as a codeword for some direction $x$ in the standard protocol. This is because the quantity $L_y(\omega)$,
i.e.\ the difference of probabilities in directions $y$ and $(-y)$, has unique maximum in $y=x$. Assumption 2 formalizes the consequence of applying Postulate 2 to the
special case of the standard protocol, proven in the main text: if two states encode the same direction in the standard protocol, with the same maximal value of
$L$, they must agree. Assumption 3 subsumes Postulates 3 and 4.

\begin{definition}[Direction bit]
\label{DefDirectionBit}
For $d\in\mathbb{N}$, a dynamical state space $(A,A_+,\u^A,E_A,\G_A)$ together with a distinguished continuous representation of $SO(d)$ (resp.\ $O(1)$ if $d=1$)
as a subgroup of $\G_A$ (denoted $R\mapsto G_R$), a distinguished vector $x\in\R^d$ with $|x|=1$, and a distinguished effect $\m\in E_A$, $0\leq \m \leq \u^A$,
will be called a \emph{direction bit for spatial dimension $d$}, if the following conditions are satisfied:
\begin{itemize}
\item For $R\in SO(d)$ (resp.\ $O(1)$ if $d=1$), the effect $\m_y:=\m\circ G_{R^{-1}}$ depends only on $y:=Rx$.
\item \textbf{Assumption 1:} There exists $\omega\in \Omega_A$ with $L_x(\omega)>L_y(\omega)$ for all $y\neq x$, where $L_y:=\m_y-\m_{-y}$.
\item \textbf{Assumption 2:} Suppose that $\omega,\omega'\in\Omega_A$ are states with the property that the maps $y\mapsto L_y(\omega)$
and $y\mapsto L_y(\omega')$ both have a unique maximum in the same direction $y_0$, and the maximal value is the same: $L_{y_0}(\omega)=L_{y_0}(\omega')$.
Then $\omega=\omega'$.
\item \textbf{Assumption 3:} There exists a locally tomographic composite $AB$ of $A$ and $B:=A$ with the property that $\G_{AB}$ contains a one-parameter
subgroup $\{G^{AB}_t\}_{t\in \R}$ for which there exists $t\in\R$ such that $G^{AB}_t$ cannot be written in the form $G^A\otimes G^B$
with $G^A\in\G_A$ and $G^B\in\G_B$.
\end{itemize}
\end{definition}

Now the claims of the main text will be proven in detail.

\begin{lemma}
\label{LemHomeo}
Let $A$ be a direction bit for spatial dimension $d$, with distinguished direction $x\in\R^d$.
Then there is a constant $0<a\leq 1$ which we call \emph{visibility parameter} with the following property:
\[
   a=\max\{L_x(\omega)\,\,|\,\ \omega\in\Omega_A, L_x(\omega)>L_y(\omega)\mbox{ for all }y\neq x\}.
\]
Moreover, for every $y\in\R^d$ with $|y|=1$, there is a unique state $\omega_y\in\Omega_A$
such that $L_y(\omega_y)=a$ and $L_z(\omega_y)<a$ for all $z\neq y$. Furthermore, $G_S\omega_y = \omega_{Sy}$ and $\m_y\circ G_S=\m_{S^{-1}y}$
for all $S\in SO(d)$ (resp.\ $S\in O(1)$ if $d=1$),
and the maps $y\mapsto \omega_y$ and $R\mapsto G_R$ are both homeomorphisms into their images (in the subspace topology).
\end{lemma}
\proof
In all of this proof, if $d=1$, then all appearances of $SO(d)$ shall be replaced by $O(1)$.
Let $x\in\R^d$ be the direction bit's distinguished direction (cf.\ Definition~\ref{DefDirectionBit}), and $\m\equiv\m_x$ the distinguished effect.
Let $\omega\in\Omega_A$ be any state with $L_x(\omega)>L_y(\omega)$ for all $y\neq x$ (it follows that $L_x(\omega)>0$). Let $\omega'\in\Omega_A$ be any other state
satisfying $L_x(\omega')> L_y(\omega')$ for all $y\neq x$ and at the same time $L_x(\omega')>L_x(\omega)$ (if no such state exists, we are done:
just set $a:=L_x(\omega)$ and $\omega_x:=\omega$). Define the state
$\mu:=\int_{R\in SO(d)} G_R \omega\, dR$. By invariance of the Haar measure, there is a constant $\beta\geq 0$ such that
$\m_y(\mu)=\beta$ for all $y$, and thus $L_y(\mu)=0$. Set $\lambda:=1-L_x(\omega)/L_x(\omega')\in (0,1)$, and $\varphi:=\lambda\mu+(1-\lambda)\omega'$,
then by construction $L_x(\varphi)>L_y(\varphi)$ for all $y\neq x$, and $L_x(\varphi)=L_x(\omega)$.
Thus, Assumption 2 implies that $\omega=\varphi=\lambda\mu+(1-\lambda)\omega'$. In summary, all states that have $x\in\R^d$ as their unique
maximizing direction of $L_\bullet$ lie on the line which starts at $\mu$ and extends through $\omega$ to infinity.

Since the state space is compact, this line will hit the topological boundary of $\Omega_A$ in some state that we call $\omega_x$. By construction,
there is some $\lambda\in[0,1)$ such that $\omega=\lambda\mu+(1-\lambda)\omega_x$. But then, $L_x(\omega)>L_y(\omega)$ for all $y\neq x$
implies the analogous strict inequality for $\omega_x$. Set $a:=L_x(\omega_x)$, then it has the claimed property. For every $y\in\R^d$ with $|y|=1$, choose some $R\in SO(d)$ with
$Rx=y$, and set $\omega_y:=G_R\omega_x$. Since $L_y=L_x\circ G_{R^{-1}}$, we have
\[
   L_y(\omega_y)=L_x(G_{R^{-1}} \omega_y)=L_x(\omega_x)=a.
\]
Let $z\neq y$ be an arbitrary vector with $|z|=1$, and let $S\in SO(d)$ be any transformation with $Sx=z$. Then $z\neq Rx$, hence $R^{-1}Sx\neq x$, and so
\[
   L_z(\omega_y)=L_{Sx}(G_R \omega_x)=L_{R^{-1} S x}(\omega_x)<a.
\]
It follows directly from Assumption 2 that $\omega_y$ is the unique state with these two properties. Recalling Definition~\ref{DefDirectionBit}, we also have
\[
   \m_y\circ G_S=\m\circ G_{R^{-1}}\circ G_S=\m\circ G_{R^{-1}S}=\m\circ G_{(S^{-1}R)^{-1}} = \m_{S^{-1} R x}=\m_{S^{-1}y}.
\]
A simple calculation also shows that $\omega:=G_S \omega_y$ has the properties $L_{Sy}(\omega)=a$ and $L_z(\omega)<a$ for all $z\neq Sy$,
which shows that $\omega=\omega_{S y}$. Next we show that the map $y\mapsto \omega_y$ is continuous.
To this end, let $\{y_n\}_{n\in\mathbb{N}}$ be a sequence of vectors in $\R^d$ with $|y_n|=1$ which converges to some vector $y$.
Clearly, we can find a sequence of orthogonal linear maps $\{R_n\}_{n\in\mathbb{N}}$ with $R_n y_n=y$ and $R_n\stackrel{n\to\infty}\longrightarrow \mathbf{1}$.
By continuity of the group representation, we have $G_{R_n}\stackrel{n\to\infty}\longrightarrow \mathbf{1}$, and thus
\[
   |\omega_y - \omega_{y_n}|=|\omega_{R_n y_n} - \omega_{y_n}| = |G_{R_n} \omega_{y_n} - \omega_{y_n}|
   \leq |G_{R_n} - \mathbf{1}|_\infty \cdot |\omega_{y_n}|\stackrel{n\to\infty}\longrightarrow 0
\]
since the state space is compact. Since $\omega_y\neq \omega_z$ for $y\neq z$, the map $y\mapsto \omega_y$ is a continuous injective map from
the compact unit sphere in $\R^d$ to its image. Thus~\cite{Jaenich,Soergel}, it is a homeomorphism into its image.

Similarly, the calculations above show that $R\neq S$ implies that $G_R\neq G_S$. Since the map $R\mapsto G_R$ is continuous,
it is a homeomorphism into its image.
\qed

The next lemma also serves as a definition of the maximally mixed state.
\begin{lemma}
Let $A$ be a direction bit for $d$ spatial dimensions. Fix any $x\in\R^d$ with $|x|=1$, and
define the \emph{maximally mixed state} by integration over the Haar measure of $SO(d)$ (resp.\ $O(1)$ if $d=1$),
\[
   \mu:=\int_{R\in SO(d)} G_R \omega_x dR= \int_{R\in SO(d)} \omega_{R x} dR.
\]
The resulting state $\mu$ does not depend on the choice of $x$. Moreover, there is a constant $0<c<1$ such that $\m_x(\mu)=c$ for all
$x\in\R^d$ with $|x|=1$, and $G_R\mu=\mu$ for all $R\in SO(d)$ (resp.\ $R\in O(1)$ if $d=1$). We call $c$ the \emph{noise parameter} of the direction bit $A$.
\end{lemma}
\proof
If $d\geq 2$, it follows from $G_R \omega_y=\omega_{Ry}$ that the definition of $\mu$ does not depend on the choice of $x$. The identity $G_R\mu=\mu$
follows from the invariance of the Haar measure. Set $c:=\m_x(\mu)$, and let $y\in\R^d$ be any vector with $|y|=1$. Then there is $S\in SO(d)$
with $y=Sx$, thus $\m_y=\m_{Sx}=\m_x\circ G_{S^{-1}}$, and so
\[
   \m_y(\mu) = \int_{R\in SO(d)} \m_{x}\circ G_{S^{-1}}(\omega_{Rx}) dR= \m_x \int_{R\in SO(d)} \omega_{S^{-1} R x} dR = \m_x(\mu)=c
\]
again by invariance of the Haar measure. For the case $d=1$, replace all appearances of $SO(d)$ in the proof by $O(1)$.
\qed
\begin{lemma}
\label{LemMaxMixMixture}
Every state $\omega$ of a direction bit can be written in the form $\omega=\lambda\omega_x+(1-\lambda)\mu$, with $0\leq\lambda\leq 1$,
some direction $x$ with $|x|=1$, and $\mu$ the maximally mixed state.
\end{lemma}
\proof
Let $\omega$ be an arbitrary direction bit state.
By compactness of the unit sphere and continuity, there exists $x\in\R^d$, $|x|=1$ such that $L_x(\omega)\geq L_y(\omega)$ for
all $y\in\R^d$ with $|y|=1$ (there may be several maximizers $x$; we choose one of them arbitrarily). For $0<\varepsilon<1$, define
$\omega_\varepsilon:=(1-\varepsilon)\omega+\varepsilon\omega_x$. Clearly $L_x(\omega_\varepsilon)>L_y(\omega_\varepsilon)$ for all $y\neq x$.
According to Lemma~\ref{LemHomeo}, we have $a=L_x(\omega_x)\geq L_x(\omega_\varepsilon)$, and thus $a\geq L_x(\omega)$ by continuity.
Define $\omega'_\varepsilon:=\lambda_\varepsilon \omega_x + (1-\lambda_\varepsilon)\mu$, where
$\lambda_\varepsilon:=\varepsilon+(1-\varepsilon)L_x(\omega)/a\in(0,1]$. It follows that $L_x(\omega'_\varepsilon)>L_y(\omega'_\varepsilon)$ for all $y\neq x$.
Moreover, $L_x(\mu)=0$ implies $L_x(\omega'_\varepsilon)=L_x(\omega_\varepsilon)$,
so Assumption 2 proves that $\omega'_\varepsilon=\omega_\varepsilon$. Since $\lambda:=\lim_{\varepsilon\to 0}\lambda_\varepsilon$ exists (and
equals $L_x(\omega)/a$), we can take the limit $\varepsilon\to 0$ of this equation and obtain $\omega=\lambda\omega_x + (1-\lambda)\mu$.
\qed
\begin{lemma}
\label{LemDirBitProps}
Every direction bit state space for spatial dimension $d$ is a $d$-dimensional unit ball.

In more detail, every direction bit state space for spatial dimension $d$ is equivalent to the dynamical state space $(B,B_+,\u^B,E_B,\G_B)$, where
\begin{itemize}
\item $B=\R^{d+1}$, 
\item $\Omega_B=\{(1,r)^T\,\,|\,\, r\in\R^d, |r|\leq 1\}$, i.e.\ $\Omega_B$ is a Euclidean unit ball of dimension $d$,
\item $\u^B(x_0,x_1,\ldots,x_d)=x_0$,
\item we have $SO(d)\subseteq \G_B$ (resp.\ $O(1)\subseteq \G_B$ if $d=1$), where the inclusion means that a rotation $R\in SO(d)$
(resp.\ $R\in O(1)$ if $d=1$) acts on states as $G_R(1,r)^T=(1,\hat G_R r)^T$, where $\hat G_R=O R O^{-1}$ for some fixed orthogonal matrix
$O$ that does not depend on $R$,
\item the set of allowed effects $E_B$ contains the ``noisy spin measurements'' $\m_x(\omega)=c+(a/2) \langle \hat\omega_x,\hat\omega\rangle$,
where $x\in \R^d$ with $|x|=1$ is a fixed unit vector,
$\hat \omega$ denotes the vector corresponding to the state $\omega$ via
$\omega=(1,\hat \omega)^T$, $\hat\omega_x$ is the analogous vector corresponding to the state $\omega_x$ and satisfying $|\hat\omega_x|=1$,
while $a$ and $c$ are the visibility and noise parameters.
\end{itemize}
\end{lemma}
\proof
If $d=1$, then replace all appearances of $SO(d)$ in this proof by $O(1)$.
Let $(A,A_+,\u^A,E_A,\G_A)$ be a direction bit for spatial dimension $d$. Let $D:=\dim A -1$.
We can reparametrize the normalized state space
 $\Omega_A$ by an affine map $\phi: A^1\to\R^D$, where $A^1:=\{x\in A\,\,|\,\,\u^A(x)=1\}$ is the affine hyperplane that
 contains the normalized states.
We define $\phi$ by first setting $M(\omega):=\omega-\mu$, and $\tilde G_R := M\circ G_R\circ M^{-1}$ for $R\in SO(d)$. Both $M$
and all $\tilde G_R$ are affine maps; moreover, $\tilde G_R(0)=0$, hence $\tilde G_R$ is a linear map, or a $D\times D$-matrix.
Define the positive matrix $X>0$ by $X:=\int_{R\in SO(d)} \tilde G_R^T \tilde G_R \, dR$, then $\tilde G_S^T X \tilde G_S=X$
for all $S\in SO(d)$. Now we set $\phi(\omega):=\alpha\sqrt{X}(\omega-\mu)$, where $\alpha>0$ is a constant that we
will determine later. Let $\langle\cdot,\cdot\rangle$ denote the standard inner product on $\R^D$, and let $x$ and
$y$ be directions, and $R\in SO(d)$ a rotation with $y=Rx$. Abbreviate $\hat\omega:=\phi(\omega)$. Then
\[
   \langle\hat\omega_y,\hat\omega_y\rangle = \alpha^2 \langle \sqrt{X}(\omega_y-\mu),\sqrt{X}(\omega_y-\mu)\rangle
   = \alpha^2 \langle \omega_y-\mu,X (\omega_y-\mu)\rangle
   = \alpha^2 \langle \tilde G_R(\omega_x-\mu), X \tilde G_R(\omega_x-\mu)\rangle\\
   = \langle \hat\omega_x,\hat\omega_x\rangle.
\]
Hence, by choosing $\alpha>0$ appropriately, we achieve that the Euclidean norm satisfies $|\hat\omega_x|=1$ for all directions $x$.
Now we define a linear map $L:A\to\R^{D+1}$ by linear extension of $L(\omega):=(1,\hat\omega)$ for all $\omega\in A^1$. This is clearly
an invertible map. Set $B:=\R^{D+1}$, $\Omega_B:=L\Omega_A$, and $\u^B:=\u^A\circ L^{-1}$. Then a vector $x\in B$ satisfies
$\u^B(x)=1$ if and only if $\u^A(L^{-1}(x))=1$, that is, iff $L^{-1}(x)\in A^1$, or $x\in L(A^1)$, which is equivalent to $x=(x_0,\ldots,x_D)$ with $x_0=1$.
Thus, $\u^B(x)=x_0$ as claimed.

Due to Lemma~\ref{LemMaxMixMixture}, every $L(\omega)\in\Omega_B$ can be written in the form $L(\omega)=\lambda L(\omega_x)+(1-\lambda)L(\mu)$
with some $\lambda\in[0,1]$; that is, $(1,\hat\omega)^T=\lambda (1,\hat\omega_x)^T+(1-\lambda)(1,0)^T$. Thus $|\hat\omega|=\lambda|\hat\omega_x|\leq 1$.
That is, the set of all $\hat\omega$, i.e. $\hat\Omega_B:=\{\hat\omega\,\,|\,\,\omega\in\Omega_A\}$ is a compact subset of the Euclidean unit ball in $\R^D$ (and it has full dimension
$D$ by our background assumptions).

Suppose that $\hat\mu=0$ was a (relative) boundary point of $\hat\Omega_B$~\cite{Webster}. Then there would be a non-trivial supporting hyperplane to $\hat\Omega_B$
which contains $\hat\mu$; that is, an affine functional $f:B^1\to\R$ (where $B^1:=\{x\in B\,\,|\,\,x_0=1\}$), such that $f(\hat\mu)=0$ and $f(\hat\omega)\geq 0$ for all
$\hat\omega\in\hat\Omega_B$. If we had $f(\hat\omega)=0$ for all $\hat\omega\in\hat\Omega_B$, then $\hat\Omega_B$ would be fully contained in the corresponding
hyperplane, contradicting the fact that it has full dimension. Thus, there are states $\hat\omega$ with $f(\hat\omega)>0$. Since every $\hat\omega\neq\hat\mu$ equals
$\lambda\hat\omega_x+(1-\lambda)\hat\mu$ for some $\lambda\in (0,1]$ and some direction $x$, this proves the existence of some directions $x$ with
$f(\hat\omega_x)>0$. Thus,
\[
   \hat\mu=\int_{R\in SO(d)} \hat\omega_{Rx}\, dR\qquad\Rightarrow\qquad f(\hat\mu)=\int_{R\in SO(d)} f(\hat\omega_{Rx})\, dR >0.
\]
This is a contradiction. Thus, $\hat\mu$ must be in the relative interior of $\hat\Omega_B$. That is, there is some $\varepsilon>0$ such that the $\varepsilon$-ball
around $\hat\mu=0$ is contained in $\hat\Omega_B$. Now let $v\in\R^D$ be any vector with Euclidean norm $|v|=1$. Then $(\varepsilon/2)v\in\hat\Omega_B$,
that is, $(\varepsilon/2)v=\hat\omega$ for some $\omega\in\Omega_A$. Thus, there is some $\lambda\in[0,1]$ and some direction $x\in\R^d$ such that
$\hat\omega=\lambda\hat\omega_x$. Since $|\hat\omega_x|=1$, this is only possible if $\lambda=\varepsilon/2$ and $v=\hat\omega_x$. But this implies
that $v\in\hat\Omega_B$. We conclude that all unit vectors are contained in $\hat\Omega_B$ -- thus, by convexity, $\hat\Omega_B$ is the full unit ball.
Since all points on its surface are of the form $\hat\omega_x$ for some direction $x$, Lemma~\ref{LemHomeo} implies that the map $x\mapsto \hat\omega_x$
is a homeomorphism from the unit sphere in $\R^d$ to the unit sphere in $\R^D$. It follows that $d=D$.

Set $\G_B:=L\circ \G_A\circ L^{-1}$. Suppose $d\geq 2$. Let $R\mapsto G_R^A$ be the distinguished continuous representation of $SO(d)$ according to
Definition~\ref{DefDirectionBit}, and let $G_R:=L\circ G_R^A \circ L^{-1}\in\G_B$. Since reversible transformations preserve the normalization, this is
only possible if $G_R(1,r)^T =(1,\hat G_R r)^T$, where every $\hat G_R$ preserves the Euclidean norm. The results of Lemma~\ref{LemHomeo} imply
that the map $R\mapsto \hat G_R$ is continuous and injective, hence $\hat \G_B:=\{\hat G_R\,\,|\,\, R\in SO(d)\}$ is a compact connected subgroup
of $O(d)$ containing the identity, hence a Lie subgroup of $SO(d)$. Since $R\mapsto \hat G_R$ is in particular injective, this is only possible if $\hat G_B=SO(d)$.
In other words, the map $R\mapsto \hat G_R$ is a continuous group automorphism of $SO(d)$. According to Lemma~\ref{eqR} below, there exists an orthogonal
matrix $O\in O(d)$ such that $\hat G_R=ORO^{-1}$.

On the other hand, if $d=1$, then every $G_B\in\G_B$ preserves the set of pure states $\{(1,1)^T, (1,-1)^T\}$ as well as the normalization.
This is only possible if $G_{\mathbf{1}} (1,r)^T=(1,r)^T$ and $G_{-\mathbf{1}} (1,r)^T=(1,-r)^T$.

Let $x$ be the direction bit's distinguished direction as given in Definition~\ref{DefDirectionBit}, and $\m_x$ the corresponding allowed effect.
Let $E_B:=E_A\circ L^{-1}$, and in particular $\mathcal{E}_x:=\m_x\circ L^{-1}$. By linearity, we can write $\mathcal{E}_x$ in the form
$\m_x(\omega)=\mathcal{E}_x(1,\hat\omega)^T=\alpha+\beta\langle\hat v_x,\hat\omega\rangle$, where $\alpha\in\R$, $\beta> 0$, and $\hat v_x\in \R^d$ is some unit vector.
First, $\m_x(\mu)=c$ and $\hat \mu=0$ implies $\alpha=c$.
For every rotation $R\in SO(d)$, acting on direction bit states via $G_R$, denote by $\hat G_R$ the corresponding transformation in the ball picture, i.e.\
\[
   L\circ G_R\circ L^{-1}(1,\hat\omega)=(1,\hat G_R\hat\omega)\qquad\mbox{for all }\omega\in\Omega_A.
\]
We know that $\hat G_R\in SO(d)$, too. For arbitrary directions $y\in\R^d$, $|y|=1$, choose $R\in SO(d)$ with $Rx=y$, then
\begin{equation}
   \m_y(\omega)=\m_{Rx}(\omega)=\m_x\circ G_{R^{-1}}(\omega)=c+\beta\langle\hat v_x,\hat G_{R^{-1}} \hat\omega\rangle =
   c+\beta \langle \hat G_R \hat v_x,\hat \omega\rangle.
   \label{eqOtherM}
\end{equation}
If $d=1$, we have $\hat G_{-\mathbf{1}}=-\mathbf{1}$; if $d\geq 2$, this follows from $\hat G_{-\mathbf{1}}=O(-\mathbf{1})O^{-1}=-\mathbf{1}$. Thus
\[
   L_x(\omega_y)=\m_x(\omega_y)-\m_{-\mathbf{1}\cdot x}(\omega_y)=\m_x(\omega_y)-\m_x\circ G_{-\mathbf{1}}(\omega_y)
   =\beta\langle \hat v_x,\hat\omega_y\rangle-\beta\langle\hat v_x,\hat G_{-\mathbf{1}}\hat\omega_y\rangle = 2\beta \langle\hat v_x,\hat\omega_y\rangle.
\]
This expression attains its maximum in $y$ for $y=x$, thus $\hat v_x=\hat\omega_x$. It follows that
$a=L_x(\omega_x)=2\beta\langle\hat\omega_x,\hat\omega_x\rangle=2\beta$, hence $\beta=a/2$, and we obtain $\m_x(\omega)=c+(a/2)\langle\hat\omega_x,\hat\omega\rangle$.
Due to eq.~(\ref{eqOtherM}), the analogous equation holds true for all other directions $y\neq x$.
Thus, we know that all these $\m_y$ must be allowed effects, i.e.\ elements of $E_B$. In the following, we always assume that we have chosen the ball representation
right from the start, such that $\mathcal{E}_x=\m_x$.
\qed

Lemma~\ref{LemDirBitProps} implies that direction bits have at most two perfectly distinguishable states in their state space, and not more.
This justifies the name ``direction bits''. In more detail, if $A$ is any state space, call a set of states $\omega_1,\ldots,\omega_n\in\Omega_A$ \emph{perfectly
distinguishable} if there are effects $\mathcal{E}_1,\ldots,\mathcal{E}_n\in E_A$ with $\mathcal{E}_1+\ldots+\mathcal{E}_n=\mathcal{U}^A$ such that
$\mathcal{E}_i(\omega_j)=\delta_{i,j}$, that is $1$ if $i=j$ and $0$ otherwise. The maximal number of any set of perfectly distinguishable states will be
called the \emph{capacity} $N_A$~\cite{Wootters,Hardy2001}. In the special case of a quantum system, $N_A$ equals the system's Hilbert space dimension.
The following lemma is well-known in the context of general probabilistic theories; we give the proof for completeness.
It says that ball state spaces of any dimension $d$ are bits, i.e.\ have capacity $N=2$; this includes classical bits ($d=1$) and quantum bits ($d=3$) as special cases.
\begin{lemma}
\label{LemBits}
If $A$ is a Euclidean ball state space with all effects allowed, i.e.\
\[
   A=\R^{d+1}, \quad \Omega_A=\left\{(1,r)^T\,\,|\,\, r\in\R^d,|r|\leq 1\right\},\quad E_A=E_A^{\max}\equiv\{\m\in A_+^*\,\,|\,\, 0\leq\m(\omega)\leq 1 \mbox{ for all }\omega\in\Omega_A\},
\]
then it has capacity $N_A=2$, i.e.\ it is a generalized bit.
\end{lemma}
\proof
Let $r\in\R^d$ be any unit vector, $|r|=1$. Set $\omega_1:=(1,r)^T\in\Omega_A$ and $\omega_2:=(1,-r)^T\in\Omega_A$, then the two functionals
\[
   \mathcal{E}_1(\omega):=\frac 1 2 \langle (1,r)^T,\omega\rangle,\qquad \mathcal{E}_2(\omega):=\frac 1 2 \langle (1,-r)^T,\omega\rangle
\]
are effects in $E_A$ that perfectly distinguish $\omega_1$ and $\omega_2$ and sum up to $\mathcal{U}^A$. Thus $N_A\geq 2$.

Suppose there are $n\geq 3$ perfectly distinguishable states $\omega_1,\ldots,\omega_n\in\Omega_A$, with corresponding effects $\mathcal{E}_1,\ldots,\mathcal{E}_n$.
Consider the hyperplane $H:=\{x\in\R^{d+1}\,\,|\,\, \mathcal{E}_1(x)=0\}$; it is a support hyperplane~\cite{Webster}
of $\Omega_A$. Furthermore, since $\omega_2,\ldots,\omega_n\in H$, it contains more than one point of $\Omega_A$, so $H\cap\Omega_A$ is a face of $\Omega_A$
that contains more than one point. However, all faces of Euclidean balls contain only one point; we obtain a contradiction. Hence $N_A\leq 2$.
\qed

According to Lemma~\ref{LemDirBitProps}, direction bit state spaces are Euclidean balls. In the case that all effects are allowed (which, as we show later,
corresponds to the ``noiseless'' case with visibility and noise parameters $a=1$ and $c=1/2$), they have therefore capacity $N=2$, i.e.\ they are in fact
\emph{bits} as the name suggests. If not all effects are allowed, then direction bits are noisy versions of bits (formally they have capacity $N=1$).
Thus, in contrast to von Weizs\"acker~\cite{Weizsaecker}, we do not assume from the beginning that our physical systems under consideration are $2$-level systems,
but we prove this from the postulates.

\begin{corollary}
\label{CorNoiseParameter}
Given the noise parameter $0<c<1$ of any direction bit, the intensity parameter $a$ satisfies $0<a\leq 2\min\{c,1-c\}$, and the spin measurements
satisfy $\m_x(\omega)\in[c-a/2,c+a/2]$ for all $|x|=1$ and $\omega\in\Omega_A$.
\end{corollary}
\proof
We know that $a>0$ due to Lemma~\ref{LemHomeo}.
In $\m_x(\omega)=c+(a/2)\langle\hat\omega_x,\hat\omega\rangle$, the inner product can attain any value in the interval $[-1,1]$ by choosing $\hat\omega$
in the unit ball appropriately. But $\m_x(\omega)$ is an outcome probability, hence in the interval $[0,1]$. Working out the corresponding inequalities
proves the claimed constraint on $a$.
\qed

Now that we know that direction bit state spaces are unit balls, we can say a bit more on the set of possible protocols satisfying Postulates 1 and 2.
Surprisingly, dimension $d=2$ turns out to be special, as illustrated in Fig.~\ref{fig_d2} below.
\begin{lemma}
\label{LemFormProtocol}
Consider any protocol satisfying Postulates 1 and 2, under the additional requirement that \emph{every} state $\omega\neq\mu$ may be used to encode some direction
$x(\omega)\in\R^d$, $|x(\omega)|=1$, such that Bob's decoding $\omega\mapsto x(\omega)$ is a continuous map. If $d\neq 2$ then there is an
orthogonal matrix $O\in O(d)$ such that
\[
   x(\omega)=O\frac{\hat\omega}{|\hat\omega|};
\]
that is, up to a fixed rotation (and possibly reflection), physical directions are encoded into Bloch vectors that point into the corresponding direction in state space.
In particular, for $d\neq 2$, if $\omega$ and $\varphi$ encode the same physical direction, then there is $\lambda\in[0,1]$ such that $\omega=\lambda\varphi+(1-\lambda)\mu$
or vice versa (i.e.\ with $\omega$ and $\varphi$ exchanged) -- that is, one of the states is obtained from the other by adding uniform noise.
\end{lemma}
\proof
Suppose that $d=1$. Then every state $\omega\neq\mu$ has one-dimensional ``Bloch vector'' $\hat\omega\in[-1,1]\setminus\{0\}$. There are two possible directions, $+1$ and $-1$,
which have to be encoded in accordance with Postulate 1. If this is done in a continuous way, the only two possibilities are
\[
   x(\omega)=\left\{
      \begin{array}{cl}
         \pm 1 & \mbox{if }\hat\omega>0,\\
         \mp 1 & \mbox{if }\hat\omega<0,
      \end{array}
   \right.
\]
proving the claim. Now consider the cases $d\geq 3$.
For every $x\in S^{d-1}$, define the stabilizer subgroup $\G_x:=\{R\in SO(d)\,\,|\,\, Rx=x\}$. Let $\omega\neq\mu$ be an arbitrary state. From the main text, we know that
$R\in \G_{x(\omega)}$ implies that $G_R\omega=\omega$, hence $\hat G_R\hat\omega=\hat\omega=O^{-1}RO\hat \omega$ with some orthogonal matrix $O$,
so $R\in\G_{O\hat\omega}$.
We get $\G_{x(\omega)}\subseteq \G_{O\hat\omega}$. Since both groups are isomorphic to $SO(d-1)$ this implies equality. Since $d\geq 3$, this in turn implies
that $x(\omega)$ is parallel to $O\hat \omega$; thus, there is a sign $\sigma(\omega)\in\{-1,+1\}$ such that
 $x(\omega)=\sigma(\omega) O\hat\omega/|\hat\omega|$, and the sign (plus or minus) cannot depend on $\omega$ due to continuity of $\omega\mapsto x(\omega)$.
This proves the claim after possibly redefining $O\mapsto (-O)$.
\qed

\begin{figure*}[!hbt]
\begin{center}
\includegraphics[angle=0, width=4cm]{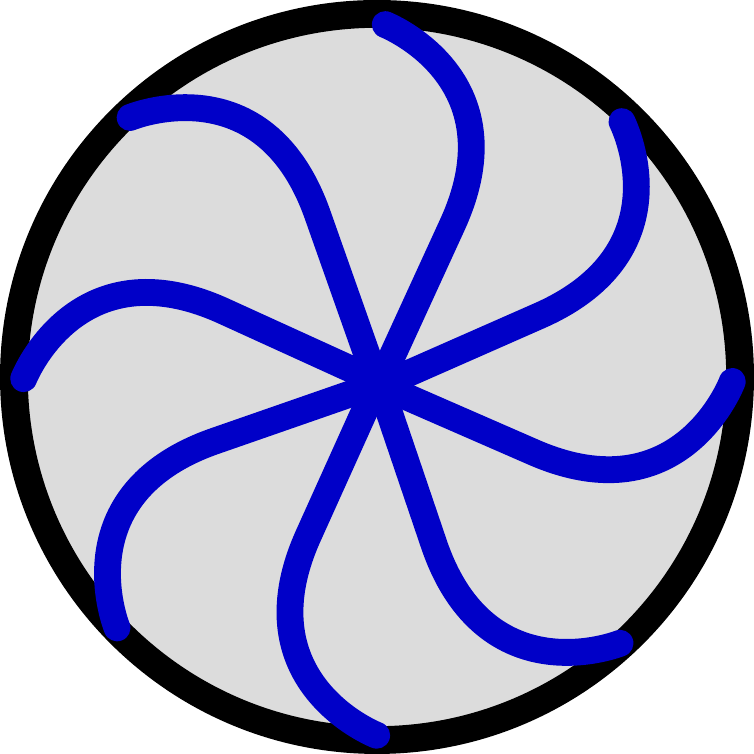}
\caption{In $d=2$ spatial dimensions, Alice and Bob may use a protocol that is unavailable in other dimensions: they may agree that
Bob decodes mixed states with a purity-dependent rotation. That is, if Bob obtains many copies of some state $\omega$ and determines $r:=|\hat\omega|$,
his output will be $R_r x$, where $x$ is the direction encoded in the pure state with Bloch vector $\hat\omega/r$, and $R_r\in SO(d)$ is a rotation depending on $r$.
The figure shows possible level sets of states in the disc state space that encode the same spatial direction. This strategy is impossible in higher dimensions, because
without any shared reference frame, Alice and Bob will not be able to agree on a $2$-dimensional reference subspace which carries the corresponding rotation.
As a result, the level sets must be straight lines for $d\geq 3$, as proven in Lemma~\ref{LemFormProtocol}.}
\label{fig_d2}
\end{center}
\end{figure*}

We now prove a technical lemma which is related to the claim in the main text that the angles inferred from state space must agree with those
in physical space (discussed in more detail in Appendix~\ref{AppSpatialGeometry} below). We show that the map $x\mapsto\hat\omega_x$ which maps direction vectors $x\in\R^d$, $|x|=1$,
to pure states' Bloch vectors $\hat\omega_x$ is linear:
there is some orthogonal matrix $O\in O(d)$ such that $\hat\omega_x=O x$. This follows from the following lemma:

\begin{lemma}
\label{LemLinearity}
Let $R\mapsto \hat G_R$ be a continuous group automorphism of $SO(d)$, and suppose $x\mapsto \hat\omega_x$ is a continuous map of the
unit sphere $S^{d-1}:=\{y\in\R^d\,\,|\,\, |y|=1\}$ to itself such that $\hat G_R \hat\omega_x=\hat\omega_{Rx}$. Then there is an orthogonal matrix
$O\in O(d)$ such that $\hat\omega_x=Ox$.
\end{lemma}
\proof
According to Lemma~\ref{eqR}, there is an orthogonal matrix $O\in O(d)$ with $\hat G_R=O R O^{-1}$ for all $R\in SO(d)$. The lemma will be proven
by distinguishing several cases.

First, consider the case that $d$ is odd. Let $x\in\R^d$, $|x|=1$ be arbitrary, then there is some $R\in SO(d)$ for which the multiples of $x$ are the only eigenvectors of eigenvalue $1$,
i.e.\ $Ry=y$ with $y\in\R^d$ is equivalent to $y=\alpha x$ with $\alpha \in\R$. But then
\[
   O R O^{-1}\hat\omega_x=\hat G_R \hat\omega_x = \hat\omega_{Rx}=\hat\omega_x\qquad\Rightarrow\qquad R(O^{-1}\hat\omega_x)=O^{-1}\hat\omega_x,
\]
and so $O^{-1}\hat\omega_x\in\{-x,x\}$. Since this is true for every direction $x$, and the map $x\mapsto \hat\omega_x$ is continuous, we either
have $\hat\omega_x=O x$ for all directions $x$ (in which case the lemma is proven), or $\hat\omega_x=-O x$ for all directions $x$, in which case
we can replace $O$ by $(-O)$ and obtain the statement of the lemma as well.

Next, consider the case $d=2$. For every $x\in\R^d$ with $|x|=1$, define
\[
   R_x:=\left(\begin{array}{cc} x_1 & -x_2 \\ x_2 & x_1 \end{array}\right) \in SO(2).
\]
We have $\hat\omega_x=\hat G_{R_x} \hat\omega_{(1,0)^T}=O R_x O^{-1} \hat\omega_{(1,0)^T}$. As a map $x\mapsto\hat\omega_x$, this is
manifestly linear. Since it preserves the Euclidean norm, it must be orthogonal.

Finally, consider the cases of even $d\geq 4$. Let $S\subset\R^d$ be any $2$-dimensional subspace. Suppose that $x\in S$. Clearly, there is $R\in SO(d)$
which acts as the identity on $S$ (and nowhere else), i.e.\ $y\in S \Leftrightarrow Ry=y$. But $x\in S$, hence $Rx=x$ which implies $\hat G_R\hat\omega_x=\hat\omega_{Rx}
=\hat\omega_x$, so $R(O^{-1}\hat\omega_x)=O^{-1}\hat\omega_x$, and thus $O^{-1}\hat\omega_x\in S$.

Now let $S,S'$ be two $2$-dimensional subspaces with $S\cap S'={\rm span}\{x\}$. Since $x\in S$ and $x\in S'$, we have $O^{-1}\hat\omega_x\in S\cap S'$,
so $O^{-1}\hat\omega_x\in\{-x,x\}$. Similarly as above, we conclude that either $O^{-1}\hat\omega_x=x$ for all directions $x$, or $O^{-1}\hat\omega_x=-x$
for all $x$, in which case we can redefine $O\mapsto (-O)$.
\qed

\begin{lemma}\label{eqR}
All continuous automorphisms of the special orthogonal groups $\phi: SO(d) \to SO(d)$ are of the form $\phi: g \mapsto PgP^{-1}$ with $P\in O(d)$.
\end{lemma}
\proof
Since every continuous homomorphism of Lie groups is analytic~\cite{Belinfante}, $\phi$ induces an automorphism on the Lie algebra $\mathfrak{so}(d)$, uniquely determined by its action on the neighborhood of the identity. But not every automorphism of a Lie algebra $\g$ necessarily induces an automorphism on the corresponding group $G$.
Ref.~\cite{FultonHarris} contains the automorphisms of the Lie algebras $\mathfrak{so}(d)$, and in what follows, we figure out which of these correspond to automorphisms of $SO(d)$.

One particular type of automorphisms for both, $G$ and $\g$, are conjugations by group elements, that is $X \mapsto gXg^{-1}$ where $g\in G$. These are called inner automorphisms. Proposition D.40 from~\cite{FultonHarris} (page 498) tells us that all automorphisms of a Lie algebra $\g$ are generated by the inner automorphisms times the symmetries of the associated Dynkin diagram. The Dynkin diagram of $\mathfrak{so}(2n+1)$ has no symmetries, hence all the corresponding automorphisms are inner. This proves the lemma for odd dimension.

Exercise 22.25 in~\cite{FultonHarris} (page 362 with answer on page 529) states that for $n \geq 5$, the symmetry of the Dynkin diagram of $\mathfrak{so} (2n)$ is implemented by a conjugation $X \mapsto PXP^{-1}$, with $P\in O(2n)$. This proves the lemma for even dimension $d \geq 10$. In what follows we consider separately the cases $d=2,4,6,8$.

{\em Case $d=2$.} The Lie algebra $\mathfrak{so}(2)$ is a one-dimensional real vector space with trivial commutator. Hence, the automorphisms are $X \mapsto \alpha X$ for any real $\alpha$. It is easy to see that among these, the only ones which induce an automorphism in $SO(2)$ are the identity and $\alpha=- 1$. The second one can be implemented as
\[
	g \mapsto 
	\left(\begin{array}{cc} 1 & 0 \\ 0 & -1 \end{array}\right)
	g
	\left(\begin{array}{cc} 1 & 0 \\ 0 & -1 \end{array}\right)^{-1}
	\ .
\]

{\em Case $d=4$.} In page 274 of Ref.~\cite{FultonHarris}, it is shown that $\mathfrak{so}(4) \cong \mathfrak{su}(2) \oplus \mathfrak{su}(2) \cong \mathfrak{so}(3) \oplus \mathfrak{so}(3)$. Hence, all the automorphisms of $\mathfrak{so}(4)$ are those of $\mathfrak{so}(3)$, which as shown above are inner, together with the exchange of the two summands in $\mathfrak{so}(3) \oplus \mathfrak{so}(3)$, which can also be implemented by conjugation.

{\em Case $d=6$.} The standard representation of $SO(6)$ is equivalent to the antisymmetric product of two copies of the standard representation of $SU(4)$ (see page 284 in~\cite{FultonHarris}), which is irreducible. This also implies that this representation of $SU(4)$ is real, and hence, equivalent to its dual (or complex conjugated) representation (see page 218 in~\cite{FultonHarris}). Exercise 22.25 in~\cite{FultonHarris} (page 362 with answer on page 529) shows that the symmetries of the Dynkin diagram of $SU(4)$ are implemented by complex conjugation $X \mapsto X^{*}$. Since the representation of $SU(4)$ equivalent to the standard representation of $SO(6)$ is real, complex conjugation leaves the algebra and the group invariant. So, the only automorphisms of $\mathfrak{so}(6)$ and $SO(6)$ are inner.

{\em Case $d=8$.} In this case the Dynkin diagram has the larger symmetry called triality. Section 20.3 in~\cite{FultonHarris} shows that this symmetry permutes the defining representation of $\mathfrak{so}(8)$ and the two fundamental spin representations $S^+, S^-$. This cannot be a symmetry of $SO(8)$, since the exponentiation of $S^+$ or $S^-$ gives the group $Spin(8)$, which is different from $SO(8)$. So the only nontrivial automorphisms of $SO(8)$ are inner.
\qed

We will now show that it is sufficient to consider ``optimal'' direction bits, i.e.\ ones with visibility parameter $a=1$ and noise parameter $c=1/2$. The idea is to take
the state space $A$ of a direction bit with $c\neq 1/2$ and/or $a<1$ and to modify it by allowing all effects $A_+^*$. The bipartite state space of two modified
direction bits is then defined as the orbit of $\G_{AB}$ on the product states and effects. However, it has to be shown that this results
in a valid state space; in particular, all probabilities must be positive. This is shown in the following lemma. It uses the definition of $E_A^{\max}$ as
given in eq.~(\ref{eqEAmax}).
\begin{lemma}
\label{LemNoisyToNoiseless}
Suppose that $A=(A,A_+,\u^A,E_A,\G_A)$ is a direction bit for spatial dimension $d$ with arbitrary visibility and noise parameters $a$ and $c$, with joint state space for
two direction bits $AB=(A\otimes B,(AB)_+,\u^A\otimes \u^B, E_{AB},\G_{AB})$. Then
\[
   A':=(A,A_+,\u^A,E_A^{\max},\G_A)
\]
is a direction bit for spatial dimension $d$ with visibility parameter $a'=1$ and noise parameter $c'=1/2$, with a possible state space of two direction bits given by
\[
   A'B'=(A\otimes B,\G_{AB}(A_+\otimes B_+),\u^A\otimes\u^B,(E_A^{\max}\otimes E_B^{\max})\circ \G_{AB},\G_{AB}),
\]
where $\G_{AB}(A_+\otimes B_+)$ is the convex hull of all unnormalized states of the form $G(\omega^A\otimes \omega^B)$ with $G\in G_{AB}$ and
$\omega^A\in A_+$, $\omega^B\in B_+$,
while $(E_A^{\max}\otimes E_B^{\max})\circ \G_{AB}$ is the convex hull of all effects of the form $(\m\otimes \n)\circ G$ with $\m\in E_A^{\max}$, $\n\in E_B^{\max}$,
and $G\in\G_{AB}$.
\end{lemma}
\proof
Throughout the proof, if $d=1$, replace $SO(d)$ by $O(1)$.
Clearly, $A'$ is a valid state space. We know that to every direction $x\in\R^d$, there is a state $\omega_x\in\Omega_A$ such that
$\m_x(\omega)=c+(a/2)\langle\hat\omega_x,\hat\omega\rangle$ for all $\omega\in\Omega_A$. The linear map
$\m'_x(\omega):=\frac 1 2 +\frac 1 2 \langle \hat\omega_x,\hat\omega\rangle$ is in $[0,1]$ for all $\omega\in\Omega_A$,
hence contained in $E_A^{\max}=E_{A'}$. It is easy to check that
\[
   \m'_x=\frac 1 a \m_x + \left(\frac 1 2 -\frac c a \right) \u^A.
\]
Let $\m$ be the effect from Definition~\ref{DefDirectionBit}, and $\m'$ the effect on $A'$ related to it by the previous equation. Using the notation from
Definition~\ref{DefDirectionBit}, where $R\in SO(d)$ is a rotation with $y=Rx$, we have
\[
   \m'\circ G_{R^{-1}} = \frac 1 a \m\circ G_{R^{-1}} + \left(\frac 1 2 -\frac c a \right)\u^A\circ G_{R^{-1}} = \frac 1 a \m_y + \left(\frac 1 2 -\frac c a\right)\u^A=\m'_y.
\]
Thus, the prerequisites and Assumptions 1 and 2 from Definition~\ref{DefDirectionBit} are satisfied for $A'$. In order to show that Assumption 3 holds true,
we have to prove that $A'B'$ is a valid composite of $A'$ and $B'$.

Clearly, $(A'B')_+=\G_{AB}(A_+\otimes B_+)$ is a closed convex cone, and $(A'B')_+\subseteq (AB)_+$ implies that $\Omega_{A'B'}=\{\omega\in (A'B')_+\,\,|\,\, \u^{AB}(\omega)=1\}
\subseteq \Omega_{AB}$ is closed. Since $\Omega_{AB}$ is compact, so must be $\Omega_{A'B'}$. Since $A_+\otimes B_+$ spans the full space $A\otimes B$, so does
$(A'B')_+$. This shows that $(A'B')_+$ is a proper cone.

According to Definition~\ref{DefTomoComposite}, all that remains to do is to show
that $(E_A^{\max}\otimes E_B^{\max})\circ\G_{AB}$ is a valid choice for $E_{A' B'}$. Clearly this is a closed convex set spanning $A\otimes B$.
It remains to show that all its elements are non-negative and no larger than one
on $\Omega_{A'B'}$; by convexity, we only have to show this for the elements $(\m\otimes \n)\circ G$, where $\m\in E_A^{\max}$, $\n\in E_B^{\max}$ and $G\in\G_{AB}$.
Finally, convexity for the state cone additionally implies that it is sufficient to prove that
\begin{equation}
   0\leq\m\otimes \n\left( G(\omega^A\otimes \omega^B)\right)\leq 1\qquad\mbox{for all }\m\in E_A^{\max}, \n\in E_B^{\max}, \omega^A\in \Omega_A, \omega^B\in\Omega_B, G\in\G_{AB}.
   \label{eqToProve}
\end{equation}
The set $E_A^{\max}$ is easy to characterize: for every effect $\m\in E_A^{\max}$, there are $\lambda,\kappa\in\R$
and $x\in\R^d$ with $|x|=1$ such that $\m(\omega)=\lambda\left(\langle\hat\omega_x,\hat\omega\rangle+1\right)+\kappa$ for all $\omega\in\Omega_A$.
A negative sign of $\lambda$ can be removed by the substitution $\hat\omega_x\mapsto -\hat\omega_x=\hat\omega_{-x}$, so we may assume $\lambda\geq 0$.
Since $\m(\omega)\in[0,1]$ for all $\omega\in\Omega_A$, we get $0\leq\kappa\leq 1-2\lambda$ and $\lambda\leq 1/2$.
It follows that
\[
   \m=\frac{2\lambda} a \m_x + \left[ \lambda\left(1-\frac{2c}a\right)+\kappa\right] \u^A.
\]
We can express $\n\in A_+^*$ in an analogous form, replacing $\lambda,\kappa,x$ by $\lambda',\kappa',y$.
Since conditional states are included in the local state spaces by definition, eq.~(\ref{eqCondApp})
and Corollary~\ref{CorNoiseParameter} imply for every $\omega^{AB}\in\Omega_{AB}$ that
\[
   \m_x\otimes \m_y(\omega^{AB})=\m_x(\omega_{\rm cond}^A)\m_y(\omega^B)\in\left[ \left(c-\frac a 2\right)^2,\left(c+\frac a 2\right)^2\right],
\]
where $\omega_{\rm cond}^A$ is the conditional state on $A$ after having obtained $\m_y$ on $B$, and $\omega^B$ is the marginal on $B$.
If $\omega^A\in\Omega_A$, $\omega^B\in\Omega_B$, and $G\in\G_{AB}$, then by definition the vector $\omega^{AB}:=G(\omega^A\otimes\omega^B)$ is
a valid state in $\Omega_{AB}$. Using that $\u^A\otimes\m_y(\omega^{AB})=\m_y(\omega^B)$ with $\omega^B$ the marginal of $\omega^{AB}$,
the expression in eq.~(\ref{eqToProve}) can be lower-bounded by
\begin{eqnarray}
   \m\otimes \n(\omega^{AB})
   &\geq& \frac{4\lambda\lambda'}{a^2}\left(c-\frac a 2 \right)^2 + \frac{2\lambda}  a\left[ \lambda'\left(1-\frac{2c} a \right)+\kappa'\right] \left(c-\frac a 2 \right)
   +\frac{2\lambda'} a \left[ \lambda\left(1-\frac{2c} a \right)+\kappa\right] \left(c-\frac a 2 \right)\\
   && + \left[\lambda\left(1-\frac{2c} a \right)+\kappa\right] \left[ \lambda'\left(1-\frac{2c} a \right)+\kappa'\right] = \kappa\kappa' \geq 0.
   \label{ineqPositiveBack}
\end{eqnarray}
An analogous calculation yields the upper bound $ \m\otimes \n(\omega^{AB})\leq(2\lambda+\kappa)(2\lambda'+\kappa')\leq 1$.
We have proven that $A'B'$ is, as given, a valid composite state space.
Finally, we have $b'=\m'_x(\omega_x)-\m'_{-x}(\omega_x)=1$ and $c'=\m'_x(\mu)=1/2$.
\qed

We obtain an immediate consequence:
\begin{corollary}
There is no direction bit for spatial dimension $d=1$.
\end{corollary}
\proof
Suppose $A=(A,A_+,\u^A,E_A,\G_A)$ is a direction bit for spatial dimension $d=1$, and $A'$ is its optimal modification from Lemma~\ref{LemNoisyToNoiseless},
with $A'B'$ the composite state space of two modified direction bits. We know that $\Omega_{A'B'}$ contains at least all combinations of product states; that is,
$\Phi_{A'B'}\subseteq \Omega_{A'B'}$, where
\[
   \Phi_{A'B'}:={\rm conv}\left\{
      \left(\begin{array}{c} 1 \\ \pm 1 \end{array}\right)\otimes\left(\begin{array}{c} 1 \\ \pm 1 \end{array}\right)
   \right\}
   ={\rm conv}\left\{
      \left(\begin{array}{c} 1 \\ 1 \\ 1 \\ 1 \end{array}\right),
      \left(\begin{array}{c} 1 \\ -1 \\ 1 \\ -1 \end{array}\right),
      \left(\begin{array}{c} 1 \\ 1 \\ -1 \\ -1 \end{array}\right),
      \left(\begin{array}{c} 1 \\ -1 \\ -1 \\ 1 \end{array}\right)
   \right\}.
\]
On the other hand, $E_{A'B'}$ contains all product effects. That is, if $\omega^{A'B'} \in\Omega_{A'B'}$, then
\[
   \left\langle \left(\begin{array}{c} 1/2 \\ \pm 1/2 \end{array}\right) \otimes \left(\begin{array}{c} 1/2 \\ \pm 1/2 \end{array}\right) , \omega^{A'B'}\right\rangle\geq 0
\]
for all possible choices of signs. It is easy to see that this is only possible of $\omega^{A'B'}\in\Phi^{A'B'}$: the four inequalities give the half-space
representation~\cite{Gruenbaum} of the tetrahedron $\Phi_{A'B'}$. It follows that $\Omega_{A'B'}\subseteq\Phi_{A'B'}$, and thus equality of these sets:
the state space for two modified direction bits is a tetrahedron, that is, a classical four-level system. It has only finitely many (four) pure states; thus, $\G_{A'B'}=\G_{AB}$ must be a
finite group. This contradicts Assumption 3 on direction bits.
\qed

All dimensions $d\geq 2$ for the ideal case $a=1$ and $c=1/2$ have been examined in~\cite{EntDynBeyondQT}: there it is shown that for $d\neq 3$, all
possible composites of $d$-dimensional ball state spaces have transformation groups that are non-interacting, therefore contradicting
Postulate 4 resp.\ Assumption 3 in the definition of direction bits. (In Appendix~\ref{AppSimpler}, we give a simplification of a key lemma in~\cite{EntDynBeyondQT} for the
special case of this paper.)
Lemma~\ref{LemNoisyToNoiseless} extends this result to noisy direction bits with $c\neq 1/2$ and/or $a<1$:
\begin{theorem}
There are no direction bits for spatial dimensions $d\neq 3$.
\end{theorem}

As mentioned in~\cite{EntDynBeyondQT}, we prove in~\cite{LocalToGlobalQT} that the only possible composite of two noiseless $3$-dimensional
ball state spaces (up to equivalence), under the assumptions of Definition~\ref{DefTomoComposite}, is quantum theory on two qubits. Now we show that this extends
to noisy $3$-dimensional balls, with the only difference that the set of effects might get reduced:
\begin{theorem}
Every direction bit for spatial dimension $d=3$ (regardless of visibility and noise parameters $a$ and $c$) can be represented as a ``noisy qubit''
$A=(A,A_+,\u^A,E_A,\G_A)$, where
\begin{itemize}
\item $A$ is the real vector space of Hermitian $2\times 2$ matrices,
\item $A_+$ is the set of positive semidefinite complex $2\times 2$ matrices,
\item the unit functional $\u^A$ is the map $\rho\mapsto \tr(\rho)$,
\item $E_A$ is a subset of the quantum effects, containing all maps of the form $\rho\mapsto\tr(\rho M)$, with $M$ a positive semidefinite $2\times 2$ matrix
satisfying $\tr(M)=2c$ and the operator inequality $M\leq (c+a/2)\cdot\mathbf{1}$,
\item $\G_A$ is the projective unitary group, $\rho\mapsto U\rho U^\dagger$ with $U\in SU(2)$.
\end{itemize}
The joint state space of two direction bits is then by necessity $AB=(A\otimes B, (AB)_+,\u^A\otimes\u^B,E_{AB},\G_{AB})$, where
\begin{itemize}
\item $A\otimes B$ is the real vector space of Hermitian $4\times 4$ matrices,
\item $(AB)_+$ is the set of positive semidefinite $4\times 4$ matrices,
\item the unit functional $\u^{AB}=\u^A\otimes \u^B$ is the map $\rho\mapsto \tr(\rho)$,
\item $E_{AB}$ is some subset of the quantum effects $\rho\mapsto\tr(\rho M)$ with $M$ a positive semidefinite $4\times 4$ matrix, $0\leq M \leq\mathbf{1}$,
\item $\G_{AB}$ is the projective unitary group, $\rho\mapsto U\rho U^\dagger$ with $U\in SU(4)$.
\end{itemize}
\end{theorem}
\proof
The standard Bloch representation of a qubit shows that the vector space $A$ as well as $A_+$ and $\u^A$ can be chosen in the claimed form.
In the ball representation, it is clear that the only possibilities for $\G_A$ are $SO(3)$ and $O(3)$. The noiseless version $A'$ from Lemma~\ref{LemNoisyToNoiseless}
will have the same transformation group. However, it is shown in~\cite{LocalToGlobalQT} that $O(3)$ is
impossible if we want to construct a global state space $A'B'$ with interaction out of noiseless $3$-balls (in a nutshell, the group $O(3)$ would introduce partial transpositions on
$A'B'$ which yield negative probabilities). Thus, the group must be $SO(3)$, which (in the chosen representation) is the projective unitary group.

It is easy to confirm that the special effects $\m_x$, $x\in\R^3$, $|x|=1$, can be represented in the following way: for every $x$, there is a complex unit vector
$|\varphi_x\rangle\in\C^2$ with
\[
   \m_x(\rho)=\tr(\rho M),\qquad\mbox{where}\qquad M=a|\varphi_x\rangle\langle\varphi_x|+\left(c-\frac a 2 \right)\mathbf{1}.
\]
These matrices have trace $\tr(M)=2c$. The convex hull of all these matrices is a subset of $E_A$:
\[
   \left\{M\,\,\left|\,\,\tr(M)=2c,M\leq \left(c+\frac a 2 \right)\cdot\mathbf{1}\right.\right\}\subset E_A.
\]

As it has been shown in~\cite{LocalToGlobalQT}, the noiseless composite state space $A'B'$ equals quantum theory on two qubits.
Since $\G_{A'B'}=\G_{AB}$ in the construction of Lemma~\ref{LemNoisyToNoiseless}, it follows that $\G_{AB}$ must be the projective
unitary group as claimed (that the vector space is $AB=A\otimes B$ and $\u^{AB}=\u^A\otimes \u^B$ follows directly from the definition of
a composite, Definition~\ref{DefTomoComposite}). Since the unitary group generates the set of all quantum states from a pure product state,
it follows that the quantum state space of two qubits, $\Omega^Q:=\{\rho\in A\otimes B\,\,|\,\, \tr(\rho)=1,\rho\geq 0\}$, is contained in $\Omega_{AB}$.
Suppose there was any $\sigma\in\Omega_{AB}\setminus\Omega^Q$, then this would be a Hermitian matrix with at least one negative eigenvalue.
Using an appropriate unitary $U$, this matrix can be diagonalized and be brought into the form $U\sigma U^\dagger=\sum_{i,j=0}^1 \lambda_{i,j} |i\rangle\langle i|\otimes |j\rangle\langle j|$
with $\lambda_{0,0}<0$, denoting by $\{|0\rangle,|1\rangle\}$ a basis of $\C^2$. Using the linear functional $\m(\rho):=\n(\rho):=\langle 0|\rho|0\rangle$,
we get $\m\otimes \n (U\sigma U^\dagger)=\lambda_{0,0}<0$. However, this contradicts ineq.~(\ref{ineqPositiveBack}), which shows that all noiseless
product quantum measurements $\m\otimes \n$ on all bipartite states $\omega^{AB}\in\Omega_{AB}$ must yield positive probabilities. Therefore $\Omega_{AB}=\Omega^Q$,
hence $(AB)_+$ is the set of positive semidefinite $(4\times 4)$-matrices.

We do not really know what $E_{AB}$ is: since all its elements must be non-negative on all quantum states, it must be a subset of the quantum effects.
Since there are no further conditions on $E_{AB}$ in Definition~\ref{DefTomoComposite}, it could possibly coincide with the set of quantum effects, or
be a proper subset. All we know is that it contains the unitary orbit of all allowed product effects.
\qed

\section{Proof of non-existence of frame bits (in a special case)}
\label{appendixX}
We now prove the claim made in Section~\ref{SecFrameBits} in the main text. Recall the ``frame bit'' setup as explained in Fig.~\ref{fig_alicebobX}.
We start by giving a formal definition of a frame bit, modifying and specializing Definition~\ref{DefDirectionBit} for direction bits.

\begin{definition}[Frame bits]
\label{DefFrameBits}
For $d\geq 2$, a dynamical state space $(A,A_+,\u^A,E_A,\G_A)$ together with a distinguished continuous representation of $SO(d)$ as a subgroup of
$\G_A$ (denoted $R\mapsto G_R$), and a distinguished effect $\m\in E_A$, will be called
a \emph{frame bit for spatial dimension $d$}, if the following conditions are satisfied:
\begin{itemize}
\item For all $Y\in SO(d)$, define $\m_Y:=\m\circ G_{Y^{-1}}$.
\item \textbf{Assumption 1':} There exists $\omega\in\Omega_A$ with $\m_{\mathbf{1}}(\omega)>\m_Y(\omega)$ for all $Y\neq \mathbf{1}$.
\item \textbf{Assumption 2':} Suppose that $\omega\neq\omega'\in\Omega_A$ are states with the property that the maps $Y\mapsto \m_Y(\omega)$
and $Y\mapsto \m_Y(\omega')$ both have a unique maximum for the same frame $Y_0$. Then there exist $0<\lambda_j< 1$, $\sum_j\lambda_j=1$,
and pairwise distinct rotations $R_j\in SO(d)$ such that
\[
   \m_Y(\omega')=\sum_j\lambda_j\m_{R_j Y}(\omega)\qquad\mbox{for all }Y\in SO(d)
\]
or vice versa (that is, with $\omega$ and $\omega'$ interchanged).
\end{itemize}
\end{definition}

Note that Assumption 1' implies that for every $X\in SO(d)$ there exists $\varphi\in\Omega_A$ with $\m_X(\varphi)>\m_Y(\varphi)$ for
all $Y\neq X$. This follows by setting $\varphi:=G_X \omega$ and using $\m_Y\circ G_{X^{-1}} = \m_{XY}$.

In analogy to Lemma~\ref{LemNoisySOd} for direction bits, we can prove the following:
\begin{lemma}
\label{LemNoisyFrame}
Under the premises of Assumption 2' in Definition~\ref{DefFrameBits}, we obtain
\[
   \omega'=\sum_j \lambda_j G_{R_j^{-1}}\omega.
\]
\end{lemma}
\proof
Let $\varphi:=\sum_j \lambda_j G_{R_j^{-1}}\omega$. Direct calculation shows that $\m_Y(\varphi)=\m_Y(\omega')$ for all $Y\in SO(d)$, hence
$Y\mapsto \m_Y(\varphi)$ also has a unique maximum in orientation $Y_0$. Denote the maximal value by $m:=\m_{Y_0}(\omega')=\m_{Y_0}(\varphi)$.
Suppose that $\varphi\neq\omega'$. Then Assumption 2' implies that one of the following two cases must be true:
\begin{itemize}
\item[1.] We have $\m_Y(\omega')=\sum_j \mu_j \m_{S_j Y}(\varphi)$ for all $Y$, where $0<\mu_j<1$, $\sum_j\mu_j=1$, and $S_j$ are pairwise
distinct rotations.

In this case, we obtain
\[
   m=\m_{Y_0}(\omega')=\sum_j \mu_j \m_{S_j Y_0}(\varphi)=\sum_j \mu_j \underbrace{\m_{S_j Y_0}(\omega')}_{\leq m}.
\]
This is only possible if $\m_{S_j Y_0}(\omega')=m$ for all $j$. By the unique maximum assumption, this implies that $S_j Y_0=Y_0$ for all $j$,
hence all $S_j=\mathbf{1}$ are equal, which is a contradiction.
\item[2.] We have $\m_Y(\varphi)=\sum_j \mu_j \m_{S_j Y}(\omega')$ for all $Y$, where $0<\mu_j<1$, $\sum_j\mu_j=1$, and $S_j$ are pairwise
distinct rotations. Then the argument is completely analogous to case 1.
\end{itemize}
This proves the claim.
\qed

In the following, it will turn out to be useful to introduce some abbreviations. Call any state $\omega$ with the property $\m_X(\omega)>\m_Y(\omega)$
for all $Y\neq X$ a \emph{codeword for $X$}.
Furthermore, for every state $\omega$, define
\[
   \Delta(\omega):=\max_{Y\in SO(d)}\m_Y(\omega)-\min_{Y\in SO(d)} \m_Y(\omega)=:\m_{\max}(\omega)-\m_{\min}(\omega).
\]
Note that $\Delta$ is continuous, but in general non-linear. Clearly $\Delta(G_R\omega)=\Delta(\omega)$ for all $R\in SO(d)$. Furthermore, we have
\begin{lemma}
\label{LemDeltaConvexity}
The map $\Delta$ is convex, i.e.\ if $\omega_i\in\Omega_A$ and $0\leq\lambda_i$, $\sum_i \lambda_i=1$, then
\[
   \Delta\left( \sum_i \lambda_i \omega_i\right) \leq \sum_i \lambda_i \Delta(\omega_i).
\]
If $0<\lambda_i$ for all $i$, we have equality if and only if there exists orientations $Y,Z\in SO(d)$ such that $\m_{\max}(\omega_i)=\m_Y(\omega_i)$ for all $i$, and
$\m_{\min}(\omega_i)=\m_Z(\omega_i)$ for all $i$.
\end{lemma}
\proof
Let $\psi:=\sum_i \lambda_i \omega_i$, let $Y$ be some frame with $\m_{\max}(\psi)=\m_Y(\psi)$, and let $Z$ be some frame
with $\m_{\min}(\psi)=\m_Z(\psi)$. Then
\begin{eqnarray*}
   \sum_i \lambda_i \Delta(\omega_i) &=& \sum_i\left(\strut \lambda_i \m_{\max}(\omega_i)-\lambda_i \m_{\min}(\omega_i)\right) 
   \geq \sum_i\left(\strut \lambda_i \m_Y(\omega_i)-\lambda_i \m_Z(\omega_i)\right)  \\
   &=& \m_Y\left(\sum_i \lambda_i\omega_i\right)-\m_Z\left(\sum_i \lambda_i \omega_i\right) = \m_{\max}(\psi)-\m_{\min}(\psi)=\Delta(\psi).
\end{eqnarray*}
Assume now $\lambda_i>0$ for all $i$. Inspecting the single inequality in the chain above proves the claimed condition for equality.
\qed

Assumptions 1' and 2' imply the following:

\begin{lemma}
\label{LemSame}
Suppose that $\omega$ and $\omega'$ are both codewords for the same $X\in SO(d)$, and $\Delta(\omega)=\Delta(\omega')$. Then $\omega=\omega'$.
\end{lemma}
\proof
Suppose that $\omega\neq\omega'$. Then Assumption 2' implies that $\omega'=\sum_j \lambda_j G_{R_j^{-1}}\omega$ for $0<\lambda_j<1$, $\sum_j\lambda_j=1$
and rotations $R_j\in SO(d)$ with $R_j\neq R_k$ for $j\neq k$ (if vice versa, rename $\omega\leftrightarrow \omega'$). Using Lemma~\ref{LemDeltaConvexity} we get
\[
   \Delta(\omega')\leq\sum_j\lambda_j \Delta(G_{R_j^{-1}}\omega)=\sum_j\lambda_j \Delta(\omega)=\Delta(\omega),
\]
with equality if and only if there exists $Y\in SO(d)$ with $\m_{\max}(G_{R_j^{-1}}\omega)=\m_Y(G_{R_j^{-1}}\omega)$ and there exists $Z\in SO(d)$
with $\m_{\min}(G_{R_j^{-1}}\omega)=\m_Z(G_{R_j^{-1}}\omega)$ for all $j$. Since we have equality by assumption, it follows that
\[
   \m_{\max}(\omega)=\m_{\max}(G_{R_j^{-1}}\omega) = \m_Y(G_{R_j^{-1}}\omega)=\m_{R_j Y}(\omega)\qquad\mbox{for all }j.
\]
But since $\omega$ by assumption has a unique maximizing direction, we must have $R_j Y=R_k Y$ for all $j,k$ and thus $R_j=R_k$, which
is a contradiction.
\qed

Now we prove the existence of a unique maximally mixed state, and a bit more:
\begin{lemma}
\label{LemUniqueMu}
There is a unique state $\mu$ such that $c:=\m_Y(\mu)$ is constant in $Y\in SO(d)$. Moreover, if $\omega$ and $\omega'$ are both codewords for $X$
with $\Delta(\omega')<\Delta(\omega)$, then $\omega'=\lambda\omega+(1-\lambda)\mu$ for $\lambda:=\Delta(\omega')/\Delta(\omega)$.
\end{lemma}
\proof
Let $\omega$ be any $X$-codeword.
There is at least one ``uniform noise'' state $\mu$ for which $\m_Y(\mu)$ is constant in $Y$: it is $\mu:=\int_{R\in SO(d)} G_R \omega\, dR$.
Let $\omega'$ be another $X$-codeword with $\Delta(\omega')<\Delta(\omega)$ (there exists at least one,
for example $\omega'=\alpha\omega+(1-\alpha)\mu$ for $0<\alpha<1$). Set $\lambda:=\Delta(\omega')/\Delta(\omega)$, and define
$\omega'':=\lambda\omega+(1-\lambda)\mu'$, where $\mu'$ is any ``uniform noise'' state, i.e.\ $\m_Y(\mu')$ is constant in $Y$.
It follows that $\omega''$ is a codeword for $X$, too, and we have $\Delta(\omega'')=\lambda\Delta(\omega)=\Delta(\omega')$. Thus,
Lemma~\ref{LemSame} implies that $\omega''=\omega'$, and so $\omega'=\lambda\omega+(1-\lambda)\mu'$. Since this equation is true for \emph{all}
uniform noise states $\mu'$, they must all be equal -- there is a unique state $\mu$ such that $\m_Y(\mu)$ is constant in $Y$.
\qed

The following lemma is the frame bit analogue of Lemma~\ref{LemHomeo}:
\begin{lemma}
\label{LemMaxb}
There is a constant $0<b\leq 1$ which we call \emph{intensity parameter} such that for all $X\in SO(d)$
\[
   b=\max\{\Delta(\omega)\,\,|\,\, \omega\mbox{ is a codeword for }X\}.
\]
Moreover, for every $X\in SO(d)$, there is a unique codeword $\omega_X$ for $X$ such that $\Delta(\omega_X)=b$, and we have $\omega_Y=G_{YX^{-1}}\omega_X$
for all $X,Y\in SO(d)$.
\end{lemma}
\proof
Fix any $X\in SO(d)$.
Lemma~\ref{LemUniqueMu} implies that all codewords for $X$ lie on the line which starts at the maximally mixed state $\mu$, crosses the state $\omega(X)$,
and extends to infinity. Since the state space is compact and convex, there is a unique state $\omega_X$ at which this line crosses the state space's boundary.
By construction, it has the maximal value of $\Delta(\omega)$ among all codewords for $X$. Set $b:=\Delta(\omega)$.

For all $Y\in SO(d)$, define $\omega_Y:=G_{YX^{-1}}\omega_X$. It is easy to check that $\omega_Y$ is a codeword for $Y$, and $\Delta(\omega_Y)=b$.
If there was any other codeword $\omega'_Y\neq\omega_Y$ for $Y$ with $\Delta(\omega'_Y)\geq b$, then $\omega'_X:=G_{X Y^{-1}}\omega'_Y$ would
be a codeword for $X$ with $\Delta(\omega'_X)\geq b$ and $\omega'_X\neq \omega_X$, which is impossible.
\qed

Now we obtain the key lemma:
\begin{lemma}
Every state $\omega$ can be written in the form $\omega=\lambda\omega_X+(1-\lambda)\mu$, where $\lambda\in[0,1]$ and $X\in SO(d)$.
\end{lemma}
\proof
Let $\omega$ be any state. By continuity and compactness of $SO(d)$, there is some $X\in SO(d)$ such that $\m_X(\omega)\geq \m_Y(\omega)$
for all $Y\in SO(d)$ (in general, $X$ is not unique -- choose one maximizer arbitrarily). For $0<\varepsilon<1$, set $\omega_\varepsilon:=(1-\varepsilon)\omega
+\varepsilon\omega_X$, then $\omega_\varepsilon$ is a codeword for $X$. If $\Delta(\omega)=b$, then $\omega=\omega_X$, and we
are done due to Lemma~\ref{LemMaxb}. Otherwise, $\Delta(\omega)<b$, and so $\Delta(\omega_\varepsilon)<b$ for all $\varepsilon$ small enough
since $\Delta$ is continuous, and Lemma~\ref{LemUniqueMu} implies that
$\omega_\varepsilon=\lambda_\varepsilon\omega_X+(1-\lambda_\varepsilon)\mu$ with $\lambda_\varepsilon:=\Delta(\omega_\varepsilon)/b$.
We have $\lim_{\varepsilon\to 0} \Delta(\omega_\varepsilon)=\Delta(\omega)$, and the claim follows for
$\lambda:=\lim_{\varepsilon\to 0}\lambda_\varepsilon=\Delta(\omega)/b$ by taking the limit $\varepsilon\to 0$.
\qed

Exactly the same argumentation as in Lemma~\ref{LemDirBitProps} -- including the introduction of ``Bloch vectors'' $\hat \omega$ for
states $\omega$ -- now proves the following:
\begin{lemma}
\label{LemFrameBallness}
The frame bit state space is equivalent to a Euclidean $D$-dimensional unit ball, and the map $X\mapsto\hat\omega_X$ is a homeomorphism
of $SO(d)$ to the unit sphere $S^{D-1}$.
\end{lemma}

Since $SO(d)$ is not simply connected for $d\geq 2$, but $S^{D-1}$ is simply connected for $D\geq 3$,
this is only possible if $D=2$ and thus (from dimension counting) $d=2$. But in this case, frames and directions coincide.

\begin{theorem}
\label{TheNoFrame}
``Frame bit'' state spaces allowing the protocol in Fig.~\ref{fig_alicebobX}, while at the same time satisfying Assumptions 1' and 2' above,
do not exist -- unless $d=2$, where they coincide with direction bits.
\end{theorem}

At first sight, this result may seem surprising, in particular Lemma~\ref{LemFrameBallness} which says that the frame bit state space must be
a Euclidean unit ball, exactly as the direction bit state space. The first obvious guess, before doing any calculations, would have been that
the pure normalized frame states $\omega_X$ with $X\in SO(d)$ can simply be parametrized by the matrix $X$ as their ``Bloch vector'', i.e. $\hat\omega_X=X$,
similarly as $\hat\omega_x=x$ for directions $x$ up to an orthogonal transformation (cf.\ Lemma~\ref{LemLinearity}).

We will now illustrate that this first guess does not work: it results in a state space that satisfies Assumption 1', but not Assumption 2', confirming
Theorem~\ref{TheNoFrame}. We only discuss the simplest non-trivial case $d=3$. Surprisingly, in this case, it turns out that our naive
guess reproduces $4$-level quantum theory over the real numbers.

\begin{example}
\label{ExOrbitope}
Suppose we define a state space $\Omega$ with the orthogonal matrices $X\in SO(3)$ as the pure states. As usual, we have to add a component for the normalization,
such that $\Omega$ becomes
\[
   \Omega:={\rm conv}\{(1,X)\,\,|\,\, X\in SO(3)\}.
\]
The vector space that carries the cone of unnormalized states is $10$-dimensional; by construction, for every $X\in SO(3)$, we have a pure state
$\omega_X=(1,X)\in\Omega$. Every (mixed) state $\omega\in\Omega$ is then of the form $\omega=(1,M)$ with $M\in\R^{3\times 3}$ some matrix
which, according to~\cite[Corollary 5.2]{Zietak}, has operator norm $\|M\|_\infty\leq 1$. We denote this state by $\omega_M$.

According to~\cite[Proposition 4.1]{Sanyal}, the full state space $\Omega$ is an \emph{orbitope} which can be parametrized in the following way. Denote the normalized
state space of $4$-level quantum theory over the reals by $\Omega_{QM}^{4,\R}$; that is,
\[
   \Omega_{QM}^{4,\R}:=\{\rho\in\R^{4\times 4}\,\,|\,\, \rho\geq 0,\tr(\rho)=1\},
\]
then the state space $\Omega$ can be written in the form
\[
   \Omega=\left\{\omega_M\,\,\left|\,\, M=\left(
      \begin{array}{ccc}
          u_{11}+u_{22}-u_{33}-u_{44} & 2 u_{23} - 2 u_{14} & 2 u_{13} + 2 u_{24} \\
          2 u_{23}+2 u_{14} & u_{11}-u_{22}+u_{33}-u_{44} & 2u_{34}-2 u_{12} \\
          2 u_{24}-2 u_{13} & 2 u_{12} +2 u_{34} & u_{11}-u_{22}-u_{33}+u_{44}
      \end{array}
   \right),\enspace U\in\Omega_{QM}^{4,\R}\right.\right\}.
\]
It is easy to check that the map $U\mapsto M$ is affine and invertible -- hence our candidate $\Omega$ of ``frame bit'' state space for dimension $d=3$
is equivalent to $4$-level real quantum theory, given that we define effects and transformations accordingly (cf.\ Definition~\ref{DefEquiv}).

The obvious choice to set up the representation of the rotation group $R\mapsto G_R$, is via $G_R\omega_X=\omega_{RX}$ for all $X\in SO(3)$.
Every special orthogonal matrix $X\in SO(3)$ has trace $\tr(X)\in[-1,3]$. Thus, the following analog of direction bits' noiseless spin measurements,
describing the probability of the ``yes''-outcome in Fig.~\ref{fig_alicebobX}, yields valid probabilities in the unit interval for $Z\in SO(3)$ (and then for
all $\omega_Z\in\Omega$ by convexity):
\[
   \m_Y(\omega_Z)=\frac 1 4 \left(\tr(Y^T Z)+1\right)\qquad (Y\in SO(3), \omega_Z\in\Omega).
\]
The expression $\tr(Y^T Z)=:\langle Y,Z\rangle$ is the Hilbert-Schmidt inner product between matrices; the corresponding Cauchy-Schwarz inequality
proves that $\m_Y(\omega_Z)=1$ if and only if $Z=Y$. Thus, the protocol given in the caption of Fig.~\ref{fig_alicebobX} works for this state space -- that is,
the state space satisfies Assumption 1'.

However, we will now show that Assumption 2' is violated.
Consider the $4\times 4$ positive semidefinite diagonal matrices $U=\frac 1 {13} {\rm diag}(6,4,2,1)$ and $U'=\frac 1{40}{\rm diag}(18,3,3,16)$; they satisfy
$\tr(U)=\tr(U')=1$. Thus, the corresponding matrices $M=\frac 1 {13}{\rm diag}(7,3,1)$ and $M'=\frac 1 {20} {\rm diag}(1,1,14)$ correspond to valid states
$\omega_M,\omega_{M'}\in\Omega$. If $Y\in SO(3)$, then $\tr(Y^T M)=\frac 1 {13}\left(7 Y_{1,1}+3 Y_{2,2}+1 Y_{3,3}\right)$.
But for the standard orthonormal basis $\{|i\rangle\}_{i=1}^3$, we have
$Y_{i,i}=\langle i|Y|i\rangle\leq \||i\rangle\|\cdot \|Y|i\rangle\| =1$, with equality if and only if $Y|i\rangle= |i\rangle$.
This proves that the identity matrix $Y=\mathbf{1}$ is the unique maximizer of the function $M\mapsto \tr(Y^T M)$, and thus of $\m_Y(\omega_M)$.

However, the same calculation applies to $M'$, and so we have two states $\omega_M$ and $\omega_{M'}$ that are both valid codewords for the
frame $Y=\mathbf{1}$. Thus, according to Lemma~\ref{LemNoisyFrame}, Assumption 2' implies that either $\omega_{M'}=\sum_j \lambda_j G_{R_j^{-1}} \omega_M$,
or with $\omega_M$ and $\omega_{M'}$ exchanged. Thus, one of the two following equations must be true:
\begin{eqnarray}
   M'&=&\sum_j \lambda_j R_j^{-1} M,\qquad \mbox{or}\label{eqAlt1}\\
   M&=&\sum_j \lambda_j R_j^{-1} M'.\label{eqAlt2}
\end{eqnarray}
For real matrices $X$, define the norms $\|X\|_1:=\tr\sqrt{X^T X}$ and $\|X\|_2:=\sqrt{\tr(X^T X)}$, then $\|X\|_1=\|RX\|_1$ and $\|X\|_2=\|RX\|_2$ if
$R$ is orthogonal. Now eq.~(\ref{eqAlt1}) implies that $\|M'\|_k\leq \|M\|_k$ for $k=1$ and $k=2$, while eq.~(\ref{eqAlt2}) implies that $\|M\|_k\leq\|M'\|_k$ for
$k=1$ and $k=2$. However, it turns out that $\|M\|_1>\|M'\|_1$, while $\|M\|_2<\|M'\|_2$. Thus, none of the two states is strictly noisier than the other, and
Assumption 2' is violated.
\end{example}

\section{Inferring spatial geometry from probability measurements}
\label{AppSpatialGeometry}
As mentioned in Section~\ref{SecSpatialGeometry} in the main text, we now describe an operational procedure that allows observers
to determine physical angles from probability measurements.

Imagine some observer (which we call Bob) in $d$-dimensional space who
holds a direction bit measurement device as in Fig.~\ref{fig_directionbit}. Suppose that Bob does not know how to measure lengths and angles in his local laboratory;
say he does not have the necessary tools (rulers etc.) to accomplish this.
In more detail, Bob may rotate his measurement device into some direction $x$, but he does not know what the resulting direction $x$ actually is, or what
rotation he actually performed.
He lacks the tools to determine angles between possible settings of his device, or between the orientations of several different
devices that he might hold.

However, suppose that Bob has access to several direction bit preparation devices. They might just lie around in his lab,
or they might be physical systems arising in nature, preparing systems in (generally mixed) direction bit states $\omega$.
Again, given any of these preparations, Bob has initially no idea what the prepared state (or its Bloch vector $\hat\omega$) is;
still, he may operate any of these devices as often as he likes, preparing many independent copies of the corresponding unknown state $\omega$.
Moreover, we assume that Bob knows the intensity and noise parameters $a$ and $c$ and the outcome $i_0$, as
defined before eq.~(\ref{eqomegaprimey}), in order to operate his measurement device properly (otherwise we may imagine that Bob
starts by testing his device on many different states $\omega$ to identify a useful outcome $i_0$, and to obtain estimates of the
corresponding parameters $a$ and $c$).

We will now show that the ballness of $\Omega_d$ allows Bob to perform a bootstrapped strategy which establishes a spatial
orthonormal coordinate frame. This allows him to determine the angle between any two possible orientations of his measurement device.
For simplicity, we first discuss the case of dimension $d=2$ here, and give the general protocol for $d\geq 2$ below.
We dismiss the trivial case $d=1$.

Bob starts by choosing two arbitrary preparation devices at random, preparing two unknown states $\omega_1,\omega_2$.
Generically, the corresponding Bloch vectors $\hat\omega_1,\hat\omega_2$ will be linearly independent (if, for some reason, $\hat\omega_1$
and $\hat\omega_2$ turn out to be (close to) linearly dependent, the protocol will fail and Bob will have to start again).

Now Bob determines $\m_x(\omega_1)=c+(a/2)\langle\hat\omega_x,\hat\omega_1\rangle$ for many different directions $x$ by repeated measurements. He never knows which direction $x$
he is currently actually measuring, but by trying out many different directions, he can determine a good estimate of $\max_x\m_x(\omega_1)=c+(a/2)|\hat\omega_1|$
and thus of $|\hat\omega_1|$, and he may
rotate his device in a direction which is very close to the actually maximizing direction $x$ that satisfies $\hat\omega_x=\hat\omega_1/|\hat\omega_1|$ (still,
without knowing any coordinate description of $x$ or $\hat\omega_x$).

Having done so, Bob leaves his device in direction $x$, and performs repeated measurements on $\omega_2$ to obtain an estimate of
$\m_x(\omega_2)=c+(a/2)\langle\hat\omega_x,\hat\omega_2\rangle$, and thus of $\langle\hat\omega_x,\hat\omega_2\rangle|\hat\omega_1|=
\langle\hat\omega_1,\hat\omega_2\rangle$. Moreover, Bob can also estimate $|\hat\omega_2|$ by repeating the strategy that he used
to determine $|\hat\omega_1|$. This is all the information he needs to determine the coordinates of $\hat\omega_1$ and $\hat\omega_2$ in some
orthonormal coordinate system. For example, he may choose the coordinates such that $\hat\omega_1=|\hat\omega_1|\cdot(1,0)^T$,
and $\hat\omega_2=|\hat\omega_2|\cdot(\cos x,\sin x)^T$, where $x$ must be chosen in accordance with $\langle\hat\omega_1,\hat\omega_2\rangle=
|\hat\omega_1|\cdot|\hat\omega_2|\cdot \cos x$.

Now suppose Bob rotates his devices in some unknown direction $y$. By measuring $\m_y(\omega_1)$ and $\m_y(\omega_2)$ to good accuracy,
he may determine the overlaps $\langle\hat\omega_y,\hat\omega_1\rangle$ and $\langle\hat\omega_y,\hat\omega_2\rangle$ and thus, since
$\hat\omega_1,\hat\omega_2$ is a basis, the coordinates of $\hat\omega_y$ in the given orthonormal frame.

If Bob holds a second measurement device which points in another unknown direction $z$, he may do the same thing, and altogether
compute the angle $\angle(\hat\omega_y,\hat\omega_z)$ between the two direction's Bloch vectors. While there was some freedom to assign
an orthonormal frame to establish coordinates for $\hat\omega_1$ and $\hat\omega_2$, this angle is independent of the specific choice of frame.

As shown in Lemma~\ref{LemLinearity} in Appendix~\ref{AppDirBit}, there exists some orthogonal transformation $O\in O(d)$ such that
$\hat\omega_x=O x$ for all directions $x$. Thus, if Bob's space carries a metric, such that there is an actual physical angle $\angle(y,z)$ between
the two devices' directions, we have $\angle(y,z)=\angle(\hat\omega_y,\hat\omega_z)$, and the angle that Bob determines by probability
measurements must agree with the actual physical angle.

Now we give a protocol which allows an observer to determine the angle between
different direction bit measurement devices (or different settings of the same device), by means of probability measurements, in arbitrary
dimensions $d\geq 2$.
The protocol will yield more or less accurate estimates of the corresponding angle, depending on the statistical effort that the observer spends to obtain probability estimates.
We assume that the observer knows the outcome $i_0$ as defined in the main text, as well as the visibility and noise parameters $a$ and $c$,
and the spatial dimension $d$.
\begin{protocol}
\label{Protocol}
In $d\geq 2$ spatial dimensions, an observer (called Bob) can estimate the angle $\angle(y,z)$ between two given
measurement devices $\m_y$ and $\m_z$ (acting on systems according to Postulates 1 and 2) by the following protocol:
\begin{enumerate}
\item Bob randomly selects $d$ direction bit preparation devices that he finds in his lab (or in nature), preparing (unknown)
direction bit states $\omega_1,\ldots,\omega_d$.

The protocol will assume that the corresponding Bloch vectors $\hat\omega_1,\ldots\hat\omega_d$ are linearly independent, which is generically the case.
Otherwise, the protocol fails and has to be repeated.
\item For every $i=1,\ldots, d$, Bob measures $\omega_i$ in many different (unknown) directions $x\in \R^d$, $|x|=1$. This way, he determines
$\max_x\m_x(\omega_i)=c+(a/2)|\hat\omega_i|$, and he can rotate his device close to the (unknown) maximizing direction $x_i$ where $\hat\omega_{x_i}=\hat\omega_i/|\hat\omega_i|$,
setting the device up to perform the measurement $\m_{x_i}$.
\item By measuring the probabilities $\m_{x_i}(\omega_j)=c+(a/2)\langle\hat\omega_i,\hat\omega_j\rangle/|\hat\omega_i|$ for all $j\neq i$, he can determine the matrix
$X_{ij}:=\langle\hat\omega_i,\hat\omega_j\rangle$.
\item Bob computes any matrix $S$ that solves the equation $S^T S=X$. A solution of this kind exists: any matrix $S$ with columns $\hat\omega'_1,\ldots,\hat\omega'_d$
is a solution, if $\hat\omega'_i$ is the coordinate representation of $\hat\omega_i$ in any orthonormal basis. Conversely, it follows from the polar decomposition
that every solution is of this form.

Hence, in this step of the protocol, Bob obtains the coordinates of the $\hat\omega_1,\ldots,\hat\omega_d$ in some orthonormal basis.
\item For any pair of measurement devices pointing in directions $y$ and $z$, Bob can determine the coordinates of $\hat\omega_y$ and $\hat\omega_z$
in the previously obtained orthonormal basis
by measuring $\m_y(\omega_i)=c+(a/2)\langle\hat\omega_y,\hat\omega_i\rangle$ and $\m_z(\omega_i)$ for $i=1,\ldots, d$, and therefore compute $\angle(\hat\omega_x,\hat\omega_y)$.
But according to Lemma~\ref{LemLinearity}, there is some orthogonal matrix $O$ such that $\hat\omega_x=Ox$ and $\hat\omega_y=Oy$, hence
this angle equals $\angle(x,y)$.
\end{enumerate}
\end{protocol}

As announced in Section~\ref{AppSpatialGeometry}, we now give a modification of the direction bit setup, showing that physical space
can in some situations inherit its linear and Euclidean structure from state space. The following example is not meant to describe actual physics in our
universe; it is simply a ``proof of principle'' demonstrating the mechanism under very specific conditions.

\begin{example}
\label{ExTopMf}
Imagine an observer Bob in $d$-dimensional space, which is simply a topological manifold $M$. Bob's local laboratory is assumed to reside in a (small)
part of this manifold, in the vicinity of some point $p\in M$. We assume that there are systems $C$ (say, internal degrees of freedom of particles) described
by a convex state space $\Omega_C$ which is also $d$-dimensional, but not necessarily a Euclidean ball.

We also assume that there is an analogue of a direction bit measurement device which can ``point in different directions'' and can be ``rotated''. However,
since $M$ does not carry a metric tensor, Bob's local laboratory space does not carry an inner product (there is not even the notion of a tangent space
to begin with). Thus, there is no literal notion of direction vectors or rotations, and we have to define what we mean by these notions in a generalized sense.

We do this by assuming that there is
a special (small) open neighborhood $U$ of $p$ that is homeomorphic to a $d$-dimensional Euclidean ball, with a topological boundary $\partial U$ homeomorphic
to the $(d-1)$-sphere $S^{d-1}$, such that
for every $x\in\partial U$ there is an effect $\mathcal{E}_x\in E_C$ which describes the first outcome of
the ``device pointing in direction $x$''. Formally, we only assume that the measurement device can be in different macroscopic states indexed by $x\in\partial U$.
The concrete physical interpretation will be left completely open, with the wording ``pointing in direction $x$'' chosen only to supply a more concrete mental picture.

For simplicity, let us assume that we have a $2$-outcome device, with outcomes labelled ``yes'' and ``no'', such that $\mathcal{E}_x(\omega)$
yields the probability of outcome ``yes'' if the device ``points in direction $x$'' and is applied to the state $\omega\in\Omega_C$.
For obvious physical reasons, $\mathcal{E}_x$ should be continuous in $x$. A sketch is given in Fig.~\ref{fig_topmf}.
We make an additional important assumption, namely that the effects determine the space points; that is, if $x\neq y$, then $\mathcal{E}_x\neq\mathcal{E}_y$.

\begin{figure*}[!hbt]
\begin{center}
\includegraphics[angle=0, width=6cm]{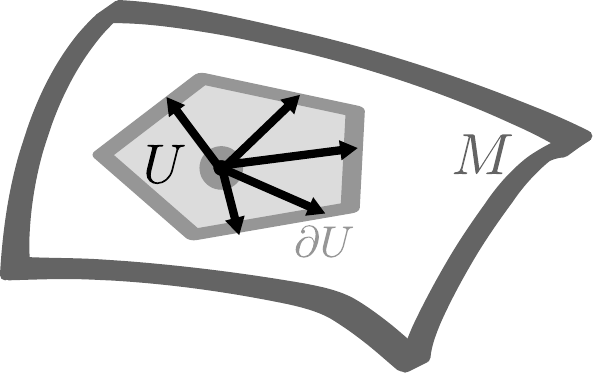}
\caption{The topological manifold $M$ and a neighborhood $U$ of a point $p$ with a boundary that is homeomorphic to a $(d-1)$-sphere. The different ``directions''
$x\in\partial U$ in which a measurement device may be oriented are sketched as black arrows, illustrating the intuition that a measurement device
``points'' in the corresponding direction. However, the elements $x\in\partial U$ are in general not vectors in any mathematically well-defined sense;
they are only meant to label the different possible states of the macroscopic measurement device, leading to different types of measurements.}
\label{fig_topmf}
\end{center}
\end{figure*}

The analog of a ``rotation'' is then any physical transformation which takes a measurement device pointing in direction $x\in\partial U$ to point
in some other direction $y=H(x)\in\partial U$. Which transformations with corresponding maps $H$ are actually possible depends on the physics in Bob's universe.
To comply with some of our intuition on rotations, we only consider those transformations that are continuous and can be physically reversed by some inverse transformation
of this kind. We will assume that the relevant physical quantities (measurement devices, particles etc.) are exactly as before if $H$ and then $H^{-1}$ is applied
(however, there may be parts of the universe that have changed in this process; for example, a distant observer may have noticed the applications
of $H$ and then of $H^{-1}$ and kept some memory of this).

Since these transformations map $\partial U$ continuously onto itself, and $\partial U$ is homeomorphic to the $(d-1)$-sphere $S^{d-1}$, we obtain
a subgroup $\mathcal{H}$ of homeomorphisms of $\partial U$, which can also be seen as a subgroup of homeomorphisms of the unit sphere $S^{d-1}$.
So far, there is no reason why the transformations $H\in\mathcal{H}$ should act linearly;
this notion does not even have any meaning at this point. We assume that these transformations allow Bob to collimate his device in any ``direction''
$x\in\partial U$ that he likes; in other words, $\mathcal{H}$ acts transitively on $\partial U$.

But now suppose that some of the transformations $H\in\mathcal{H}$ have impact on the measured outcome probabilities: the probability to see
the first outcome may change if the measurement device is transformed via $H$. If it makes sense that Bob
undergoes the transformation $H$ together with the measurement device such that the device has not changed from his perspective (that is, a ``joint
rotation''), he will model
the net effect on the probabilities by some other transformation $H'$ that acts on the state $\omega$ instead. It must satisfy the equation
\begin{equation}
   \label{DefHPrime}
   \mathcal{E}_{H(x)}(\omega)=\mathcal{E}_x\left(H'(\omega)\right)\qquad\mbox{for all }x\in\partial U,\enspace\omega\in\Omega_C.
\end{equation}
We will assume that this is possible. Due to the probabilistic interpretation of states, the map $H'$ must be linear.
Because of the assumed reversibility of $H$, it must be a reversible transformation, i.e.\ $H'\in\mathcal{G}_C$.
We do not yet know whether this equation and linearity determine $H'$ uniquely. Let $\omega\in\Omega_C$ be any pure state,
and define $\mu:=\int_{\mathcal{G}_C} G(\omega)\, dG$ (which may well depend on the choice of $\omega$).
Then in particular $H'\mu=\mu$ for all possible $H'$, since $H'\in\mathcal{G}_C$. Thus, if $x,y\in\partial U$ are arbitrary, there is some $H\in\mathcal{H}$ such that $H(x)=y$, and
\[
   \mathcal{E}_y(\mu)=\mathcal{E}_{H(x)}(\mu)=\mathcal{E}_x(H' \mu)=\mathcal{E}_x(\mu)=:m\geq 0.
\]
Thus, all $\mathcal{E}_x$ lie in the $d$-dimensional affine subspace $A:=\{\mathcal{E}\,\,|\,\,\mathcal{E}(\mu)=m\}$ of the $(d+1)$-dimensional dual space $C^*$.
Let $A'\subseteq A$ be the affine span of all $\mathcal{E}_x$, and $d':=\dim A'$.
Let $h:\partial U \to S^{d-1}$ be a homeomorphism and $j:A'\to\R^{d'}$ an invertible affine map. Then the map $s\mapsto j\left(\mathcal{E}_{h^{-1}(s)}\right)$
is a continuous injective map from $S^{d-1}$ to $\R^{d'}$. Due to Lemma~\ref{LemSphereDim}, we must have $d'\geq d=\dim A$, and so $A'=A$.
Suppose that $m=0$; in this case, relabel the two
outcomes of the device ``yes$\leftrightarrow$no'', such that the new $\mathcal{E}_x$ satisfy $\mathcal{E}_x(\mu)=1$ for all $x\in\partial U$.
Thus, we may assume that $m>0$, such that $A$ is not a linear subspace.
Consequently, the $\mathcal{E}_x$ linearly span $C^*$, and so eq.~(\ref{DefHPrime}) determines $H'$ uniquely.

If $H$ does not alter the outcome probabilities,
the corresponding map $H'$ will be the identity map; in particular, $H\mapsto H'$ need not be injective. Again, since the $\mathcal{E}_x$ span $C^*$, we obtain
\[
   \mathcal{E}_x\mapsto \mathcal{E}_{H(x)}=\mathcal{E}_x\circ H'=:L_H(\mathcal{E}_x)\quad\mbox{extends to a linear invertible map $L_H$ from the dual space }C^*\mbox{ to itself.}
\]
Let $\mathcal{H}'$ be the topological closure of the group of all $H'$, where $H\in\mathcal{H}$. Since it is a subset of the compact group
of reversible transformations of $C$, it must itself be compact. 
Let $y_1,\ldots,y_{d+1}\in\partial U$ be any set of points such that $\mathcal{E}_{y_1},\ldots,\mathcal{E}_{y_{d+1}}$ is a basis of $C^*$.
Define the subspace $S$ of $C^*$ by
\[
   S:=\{f\in C^*\,\,|\,\, f(\mu)=0\}.
\]
Then the functionals $\mathcal{E}_{y_i}-m\,\u^C$ span $S$, and we can find $d$ of them which constitute a basis of $S$. Call these
functionals $\mathcal{F}_1,\ldots,\mathcal{F}_d$ (in some arbitrary order). Now we can define a coordinate map $\Lambda:S\to\R^d$ via
\[
   \mathcal{E}=\lambda_1 \mathcal{F}_1+\ldots+\lambda_d\mathcal{F}_d\quad\Leftrightarrow\quad  \Lambda(\mathcal{E}) = (\lambda_1,\ldots,\lambda_d)^T .
\]
This allows us to define coordinates $\vec\lambda(x)$ of space points $x\in\partial U$ via
\[
   \vec\lambda(x):=\Lambda(\mathcal{E}_x-m\,\u^C).
\]
What is the action of $H$ in these coordinates? Since $L_H(\u^C)=\u^C$, we have
\begin{equation}
   \vec\lambda\left(\strut H(x)\right) = \Lambda (\mathcal{E}_{H(x)}-m\,\u^C) = \Lambda\circ L_H(\mathcal{E}_x-m\,\u^C)
   =\Lambda\circ L_H\circ\Lambda^{-1}\,\vec\lambda(x).
   \label{eqOrtho}
\end{equation}
In other words, we have constructed a fictitious
$d$-dimensional linear space such that all $x\in\partial U$ can be represented
as elements of this linear space, and all transformations $H\in\mathcal{H}$ act linearly (represented by $\Lambda\circ L_H\circ \Lambda^{-1}$).
This vector space structure is inherited from the convexity of probabilities.

Define the group $\mathcal{L}$ as the topological closure of $\{\Lambda\circ L_H\circ \Lambda^{-1}\,\,|\,\, H\in\mathcal{H}\}$.
Due to $L_H(\mathcal{E})=\mathcal{E}\circ H'$, compactness of $\mathcal{H}'$ implies compactness of $\mathcal{L}$.
Thus, there is an inner product on $\R^d$ such that $\langle v,w\rangle=\langle L v,L w\rangle$ for all $v,w\in \R^d$ and $L\in \mathcal{L}$.
With respect to this inner product, $\mathcal{L}$ is a subgroup of $SO(d)$. Hence eq.~(\ref{eqOrtho}) implies that $\|\vec\lambda\left(\strut H(x)\right)\| = \|\vec\lambda(x)\|$
for the corresponding norm, and since $H$ is transitive on $\partial U$, we obtain that
\[
   \mbox{there exists }r>0\mbox{ such that }\|\vec\lambda(x)\|=r\qquad\mbox{for all }x\in\partial U.
\]
Let $h:\partial U \to S^{d-1}$ be a homeomorphism, then $(1/r)\cdot\vec\lambda\circ h^{-1}$ is a continuous injective map from the sphere $S^{d-1}$ into
itself. According to Lemma~\ref{LemTop}, it must be surjective -- in other words, the set $\{\vec \lambda(x)\,\,|\,\,x\in\partial U\}$ is the full sphere
of radius $r$ in $\R^d$.
Since $\mathcal{L}$ is transitive on this sphere, we see that $\mathcal{L}$ acts irreducibly on $\R^d$; hence the inner product $\langle\cdot,\cdot\rangle$
is in fact unique.

In other words, we have obtained a unique Euclidean structure on our vector space representation of $\partial U$, inherited from the group of reversible transformations
on state space.

We can say more about the set of states $\Omega_C$. Let us introduce coordinates on the $(d+1)$-dimensional space $C^*$:
for $\mathcal{E}\in C^*$, set
\[
   \bar\mathcal{E}:=\left(\strut \mathcal{E}(\mu),\Lambda(\mathcal{E}-\mathcal{E}(\mu)\,\u^C)\right)\in\R^{d+1}.
\]
Basically, this extends the coordinate map $\Lambda$ from the subspace $S$ to all of $C^*$. In particular, we have $\bar\u^C=(1,0,\ldots,0)$.
For $H\in\mathcal{H}$, set
$\bar H:=\mathbf{1}\oplus \Lambda\circ L_H\circ\Lambda^{-1}$, where $\mathbf{1}$ acts as the identity on the first entry of the vector
to which it is applied. We introduce an inner product on $\R^{d+1}$: if $a,b\in\R$ and $v,w\in\R^d$, set $\langle (a,v),(b,w)\rangle:=ab+\langle v,w\rangle$,
where the inner product on the right-hand side is the $\mathcal{L}$-invariant inner product constructed above. It follows that
$\langle \bar H x,\bar H y\rangle=\langle x,y\rangle$ for all $x,y\in\R^{d+1}$. Now we can represent elements $\omega\in C$ (for example states)
by vectors $\bar\omega\in\R^{d+1}$, defined by the equation $\mathcal{E}(\omega)=\langle\bar\mathcal{E},\bar\omega\rangle$.
For $\omega\in\Omega_C$, we have $1=\u^C(\omega)=\langle\bar\u^C,\bar\omega\rangle=\bar\omega_1$, where $\bar\omega_1$ is the first
component of the vector $\bar\omega$. Thus, we can represent every $\omega\in\Omega_C$ via $\bar\omega=(1,\hat\omega)$. We get
\begin{eqnarray*}
   \langle \bar\mathcal{E},\bar H^{-1}\bar\omega\rangle &=& \langle\bar H \bar\mathcal{E},\bar\omega\rangle
   =\langle(\mathcal{E}(\mu),\Lambda L_H(\mathcal{E}-\mathcal{E}(\mu)\,\u^C)),\bar\omega\rangle
   =\langle (\mathcal{E}(\mu),\Lambda((\mathcal{E}-\mathcal{E}(\mu)\,\u^C)\circ H')),\bar\omega\rangle
   =\langle\overline{\mathcal{E}\circ H'},\bar\omega\rangle\\
   &=&(\mathcal{E}\circ H')(\omega)=\mathcal{E}(H'(\omega))=\langle\bar\mathcal{E},\overline{H'(\omega)}\rangle.
\end{eqnarray*}
Thus, we obtain $\overline{H'(\omega)}=\bar H^{-1}\bar\omega$. In summary, for two states $\varphi,\omega\in\Omega_C$, we have
\[
   \varphi=H'(\omega)\Leftrightarrow \bar\varphi=\bar H^{-1}\bar\omega\Leftrightarrow \hat\varphi=\Lambda\circ L_{H^{-1}}\circ \Lambda^{-1}\hat\omega.
\]
In other words, $\mathcal{H}$ acts on the subspace containing the $\hat\omega$ as the group $\mathcal{L}$; since $\mathcal{L}$ is transitive on the unit
sphere, this implies that the set of ``Bloch vectors'' $\{\hat\omega\,\,|\,\,\omega\in\Omega_C\}$ is a Euclidean ball (of some radius), and equivalent
to a $d$-dimensional Euclidean unit ball, exactly as direction bits are.
Thus, Bob may use Protocol~\ref{Protocol} to determine angles between different orientations of his measurement device.
In retrospect, we also see that the maximally mixed state $\mu$ is unique (it is the center of the ball); hence the linear structure that we have constructed
is unique as well.
\end{example}

The following lemmas have been used in Example~\ref{ExTopMf}.

\begin{lemma}
\label{LemSphereDim}
There is no continuous injective map $f:S^{d-1}\to\R^n$ if $n<d$.
\end{lemma}
\proof
It is sufficient to consider the case $n=d-1$; otherwise, we can compose $f$ with an embedding of $\R^n$ into $\R^{d-1}$.
The result for $n=d-1$ follows from an application of the invariance of domain theorem~\cite[Exercise 7.6]{Dold}.
\qed

\begin{lemma}
\label{LemTop}
Every continuous injective map from the sphere $S^{d-1}$ to itself is surjective.
\end{lemma}
\proof Suppose there was a non-surjective continuous injective map $f$ from $S^{d-1}$ to itself. Let $s\in S^{d-1}$ be any point which
is not attained by $f$; then $f$ can be interpreted as a continuous injective map from $S^{d-1}$ to the punctured sphere
$S^{d-1}\setminus\{s\}$, which is well-known~\cite{Jaenich} to be homeomorphic to $\R^{d-1}$. Let $g:S^{d-1}\setminus\{s\}\to\R^{d-1}$
be a corresponding homeomorphism, then $g\circ f$ is a continuous injective map from $S^{d-1}$ to $\R^{d-1}$, contradicting Lemma~\ref{LemSphereDim}.
\qed

\section{Simplified proof of the result of Section 4.3 in Ref.~\cite{EntDynBeyondQT} for SO(d)}
\label{AppSimpler}
In~\cite{EntDynBeyondQT}, it has been shown that two noiseless $d$-dimensional Euclidean ball state spaces can be combined into an interacting, joint state space
if and only if $d=3$. In that paper, we have considered the general case of ball state spaces with any compact group of reversible transformations which
is transitive on the unit sphere. However, here, we are only interested in the special case that the group of reversible transformations on a direction bit
contains the full orthogonal group $SO(d)$, as established in Lemma~\ref{LemDirBitProps}. It turns out that this simplifies the proof of a key lemma in~\cite{EntDynBeyondQT}
significantly.

Here, we give the simplified proof, as a reference for readers who would like to follow the argumentation in~\cite{EntDynBeyondQT}. Therefore, we do not introduce
the relevant notation here in the appendix, but refer the reader to the introductory chapters of~\cite{EntDynBeyondQT}, and just use the notation that has
been introduced there.

\begin{lemma}[Section 4.3 in~\cite{EntDynBeyondQT}, special case of $SO(d)$, $d\geq 2$]
If a generator $W\in\tilde\mathfrak{g}$ is of the block-diagonal form
\[
   W=\left[
      \begin{array}{cccc}
         0 & \mathbf{0} & \mathbf{0} & \mathbf{0} \\
         \mathbf{0} & Y & \mathbf{0} & \mathbf{0} \\
         \mathbf{0} & \mathbf{0} & X & \mathbf{0} \\
         \mathbf{0} & \mathbf{0} & \mathbf{0} & Z
      \end{array}
   \right],
\]
then $Z=X\otimes\mathbf{1}+\mathbf{1}\otimes Y$, i.e.\ $W$ generates non-interacting dynamics.
\end{lemma}
\proof
Since $W$ is antisymmetric, so are $X$ and $Y$, which are thus generators of rotations. By assumption, we can perform the
rotations $\exp(t\hat X)\otimes \exp(t\hat Y)$ on the joint system, which are generated by
\[
   W'=\left[
      \begin{array}{cccc}
         0 & \mathbf{0} & \mathbf{0} & \mathbf{0} \\
         \mathbf{0} & Y & \mathbf{0} & \mathbf{0} \\
         \mathbf{0} & \mathbf{0} & X & \mathbf{0} \\
         \mathbf{0} & \mathbf{0} & \mathbf{0} & X\otimes\mathbf{1}+\mathbf{1}\otimes Y
      \end{array}
   \right]\in\tilde\mathfrak{g}.
\]
Since $\tilde\mathfrak{g}$ is a Lie algebra, it also contains the element
\[
   W'':=W'-W={\rm diag}(0,\mathbf{0},\mathbf{0},V),\qquad\mbox{where }V:=X\otimes\mathbf{1}+\mathbf{1}\otimes Y-Z\mbox{ is antisymmetric}.
\]
Applying constraint~(35) from~\cite{EntDynBeyondQT} in the special case $\mathbf{x}=-\mathbf{a}$ implies
\[
   \left[\begin{array}{c} 1 \\-\mathbf{a}\end{array}\right]\otimes 
   \left[\begin{array}{c} 1 \\-\mathbf{b}\end{array}\right] M (W'')^2 M^{-1}
   \left[\begin{array}{c} 1 \\\mathbf{a}\end{array}\right] \otimes
   \left[\begin{array}{c} 1 \\\mathbf{b}\end{array}\right]\geq 0
\]
for all $\mathbf{a},\mathbf{b}\in\R^d$ with $|\mathbf{a}|=|\mathbf{b}|=1$, which simplifies to $(\mathbf{a}\otimes\mathbf{b})\cdot NV^2N^{-1}(\mathbf{a}\otimes\mathbf{b})\geq 0$.
Summing over all $\mathbf{a}$ and all $\mathbf{b}$ in an orthonormal basis yields $0\leq\tr(NV^2N^{-1})=\tr(V^2)$. But $V^T V\geq 0$, and since $V^T=-V$, we have
$V^2\leq 0$, hence $\tr(V^2)\leq 0$. Both inequalities together give $\tr(V^2)=0$, which is only possible if $V^2=0$, and thus $V=0$.
\qed

\end{document}